\documentclass[11pt]{article} 

\usepackage[letterpaper,margin=1in]{geometry}

\newif\ifred
\redtrue 

\usepackage[T1]{fontenc}
\usepackage{microtype}

\usepackage{titlesec}
\titleformat*{\paragraph}{\bfseries}

\usepackage{amsthm,amsmath,amssymb,amsfonts,amssymb,mathtools}
\usepackage{amsmath,amssymb,amsfonts,amssymb,mathtools}
\usepackage{xcolor}
\usepackage{empheq}
\usepackage{dsfont}
\usepackage{mathrsfs}
\usepackage{graphicx}
\usepackage{enumitem}
\usepackage{aliascnt} 
\usepackage{xspace}
\usepackage{bbm}

\usepackage{caption}
\usepackage{tikz}
\usetikzlibrary{patterns}
\usetikzlibrary{patterns}
\usetikzlibrary{arrows,shapes,automata,backgrounds,petri,positioning}
\usetikzlibrary{shadows}
\usetikzlibrary{calc}
\usetikzlibrary{spy}
\usetikzlibrary{angles, quotes}
\usetikzlibrary{matrix}
\usepackage{pgf,pgfplots}
\usepackage{pgfmath,pgffor}
\pgfplotsset{compat=1.17}

\usepackage{framed}
 \usepackage[noend]{algorithm2e}
\usepackage[noend]{algpseudocode}

\definecolor[named]{ACMBlue}{cmyk}{1,0.1,0,0.1}
\definecolor[named]{ACMYellow}{cmyk}{0,0.16,1,0}
\definecolor[named]{ACMOrange}{cmyk}{0,0.42,1,0.01}
\definecolor[named]{ACMRed}{cmyk}{0,0.90,0.86,0}
\definecolor[named]{ACMLightBlue}{cmyk}{0.49,0.01,0,0}
\definecolor[named]{ACMGreen}{cmyk}{0.20,0,1,0.19}
\definecolor[named]{ACMPurple}{cmyk}{0.55,1,0,0.15}
\definecolor[named]{ACMDarkBlue}{cmyk}{1,0.58,0,0.21}

\usepackage[colorlinks,citecolor=blue,linkcolor=magenta,bookmarks=true]{hyperref}
\usepackage[nameinlink]{cleveref}
\crefname{ineq}{Inequality}{Inequality}
\creflabelformat{ineq}{#2{\upshape(#1)}#3}
\crefname{sub}{Subsection}{Subsection}
\creflabelformat{Subsection}{#2{\upshape(#1)}#3}
\crefname{sdp}{SDP}{SDP}
\creflabelformat{sdp}{#2{\upshape(#1)}#3}
\crefname{lp}{LP}{LP}
\creflabelformat{lp}{#2{\upshape(#1)}#3}
\usepackage{aliascnt}
\usepackage{cleveref}
\crefname{ineq}{Inequality}{Inequality}
\creflabelformat{ineq}{#2{\upshape(#1)}#3}
\crefname{sub}{Subsection}{Subsection}
\creflabelformat{Subsection}{#2{\upshape(#1)}#3}
\crefname{sdp}{SDP}{SDP}
\creflabelformat{sdp}{#2{\upshape(#1)}#3}
\crefname{lp}{LP}{LP}
\creflabelformat{lp}{#2{\upshape(#1)}#3}

\makeatletter
\newenvironment{Ualgorithm}[1][htpb]{\def\@algocf@post@ruled{\kern\interspacealgoruled\hrule  height\algoheightrule\kern3pt\relax}\def\@algocf@capt@ruled{under}\setlength\algotitleheightrule{0pt}\SetAlgoCaptionLayout{centerline}\begin{algorithm}[#1]}
{\end{algorithm}}
\makeatother

\DeclarePairedDelimiter\ceil{\lceil}{\rceil}

\def\colorful{0}

\ifnum\colorful=1
\newcommand{\inote}[1]{\footnote{{\bf [[Ilias: {#1}\bf ]] }}}
\newcommand{\dnote}[1]{\footnote{{\bf [[Daniel: {#1}\bf ]] }}}
\newcommand{\nnote}[1]{\footnote{{\bf [[Nikos: {#1}\bf ]] }}}
\newcommand{\vnote}[1]{\footnote{{\bf [[Vasilis: {#1}\bf ]] }}}
\newcommand{\snote}[1]{\footnote{{\bf [[Sihan: \textcolor{blue}{#1}\bf ]] }}}

\newcommand{\new}[1]{\textcolor{red}{#1}}
\newcommand{\snew}[1]{\textcolor{blue}{#1}}

\else
\newcommand{\inote}[1]{}
\newcommand{\dnote}[1]{}
\newcommand{\nnote}[1]{}
\newcommand{\vnote}[1]{}
\newcommand{\snote}[1]{}

\newcommand{\new}[1]{{#1}}
\newcommand{\snew}[1]{{#1}}

\fi

\newtheorem{theorem}{Theorem}[section]

\newtheorem{lemma}[theorem]{Lemma}
\newtheorem{informal theorem}[theorem]{Theorem (informal statement)}

\newtheorem{proposition}[theorem]{Proposition}
\newtheorem{corollary}[theorem]{Corollary}
\newtheorem{claim}[theorem]{Claim}
\newtheorem{fact}[theorem]{Fact}

\newtheorem{remark}[theorem]{Remark}

\newtheorem{definition}[theorem]{Definition}

\newcommand{\lp}{\left}
\newcommand{\rp}{\right}

\newcommand\snorm[2]{\left\| #2 \right\|_{#1}}
\renewcommand\vec[1]{\mathbf{#1}}
\DeclareMathOperator*{\pr}{\mathbf{Pr}}
\DeclareMathOperator*{\E}{\mathbf{E}}

\newcommand{\normal}{\mathcal{N}}

\newcommand{\samplesize}{\mathcal V}

\newcommand{\bw}{\mathbf{w}}

\newcommand{\R}{\mathbb{R}}

\newcommand{\Z}{\mathbb{Z}}

\newcommand{\eps}{\epsilon}
\newcommand{\dtv}{d_{\mathrm TV}}
\newcommand{\poly}{\mathrm{poly}}

\newcommand{\sgn}{\mathrm{sign}}
\newcommand{\sign}{\mathrm{sign}}

\newcommand{\OPT}{\mathrm{opt}}
\newcommand{\D}{D}

\newcommand{\Ind}{\mathds{1}}
\newcommand{\1}{\Ind}

\newcommand{\wt}{\widetilde}
\newcommand{\wh}{\widehat}

\newcommand{\wstar}{\bw^{\ast}}
\newcommand{\x}{\vec x}

\newcommand{\Exx}{\E_{\x\sim \D_\x}}

\newcommand{\Ey}{\E_{(\x,y)\sim \D}}

\newcommand{\opt}{\mathrm{opt}}

\newcommand{\iid}{{i.i.d.}\ }
\newcommand{\abs}[1]{\lp| #1 \rp|}

\renewcommand\Pr{\pr}

\DeclareMathOperator{\Jac}{Jac}
\DeclareMathOperator{\coe}{coe}

\newcommand{\LC}[1]{ \text{MC}(#1) }

\title{Super Non-singular Decompositions of Polynomials and their Application to Robustly Learning Low-degree PTFs}

\author{
Ilias Diakonikolas\thanks{Supported by NSF Medium Award CCF-2107079,
NSF Award CCF-1652862 (CAREER), a Sloan Research Fellowship, and
a DARPA Learning with Less Labels (LwLL) grant.}\\
UW Madison\\
{\tt ilias@cs.wisc.edu}\\
\and
Daniel M. Kane\thanks{Supported by NSF Award CCF-1553288 (CAREER) and a Sloan
 Research Fellowship.}\\
UC San Diego\\
{\tt dakane@ucsd.edu }\\
\and
Vasilis Kontonis\thanks{Supported in part by NSF Award CCF-2144298 (CAREER).}\\
UW Madison\\
{\tt kontonis@wisc.edu } \\
\and
Sihan Liu\\
UC San Diego\\
{\tt sil046@ucsd.edu}
\and
Nikos Zarifis\thanks{Supported in part by NSF Award CCF-1652862 (CAREER) 
and a DARPA Learning with Less Labels (LwLL) grant.}\\
UW Madison\\
{\tt zarifis@wisc.edu}\\
}

\begin{document}

\maketitle

\begin{abstract}
We study the efficient learnability of low-degree polynomial threshold functions
(PTFs) in the presence of a constant fraction of adversarial corruptions. 
Our main algorithmic result is a polynomial-time PAC learning algorithm for this 
concept class in the strong contamination model under the Gaussian distribution 
with error guarantee $O_{d, c}(\opt^{1-c})$, for any desired constant $c>0$, 
where $\opt$ is the fraction of corruptions. 
In the strong contamination model, an omniscient adversary 
can arbitrarily corrupt an $\opt$-fraction 
of the data points and their labels. 
This model generalizes the malicious noise model and the adversarial label noise 
model. Prior to our work, known polynomial-time algorithms 
in this corruption model (or even in the weaker adversarial label noise model) 
achieved error $\tilde{O}_d(\opt^{1/(d+1)})$, which 
deteriorates significantly as a function of the degree $d$. 

Our algorithm employs an iterative approach inspired by localization techniques 
previously used in the context of learning linear threshold functions. 
Specifically, we use a robust perceptron algorithm to compute 
a good partial classifier and then iterate on the unclassified points. 
In order to achieve this, we need to take a set defined by a number 
of polynomial inequalities and partition it into several well-behaved subsets. 
To this end, we develop new polynomial decomposition techniques 
that may be of independent interest.
\end{abstract}


\setcounter{page}{0}
\thispagestyle{empty}
\newpage

\section{Introduction} \label{sec:intro}

A degree-$d$  polynomial threshold function (PTF) 
is any Boolean function $f: \R^n \to \{ \pm 1\}$ 
of the form $f(x) = \sign(p(\x))$\footnote{The function $\sign: \R \to \{ \pm 1\}$ is defined as $\sgn(t)=1$ if $t \geq 0$ 
and $\sgn(t)=-1$ otherwise.}, 
where $p : \R^n \to \R$ is a degree-$d$ polynomial 
with real coefficients. For $d=1$, we obtain 
Linear Threshold Functionss (LTFs) or halfspaces. 
PTFs are a fundamental class 
of Boolean functions that have been extensively studied 
in many contexts for at least the past five decades~\cite{Dertouzos:65, MinskyPapert:68, Muroga:71}. 
\new{Over the past two decades, low-degree PTFs have been the focus 
of renewed research interest in various fields of theoretical computer science, including complexity theory~\cite{Sherstov:09, DGJ+:10,MZstoc10, DKN10, Kane11ccc, Kane12, DeS14, SherstovW19, ODonnellST20} 
and learning theory~\cite{KKMS:08, DHK+:10, DOSW:11, DeDFS14, DKS18a, DiakonikolasK19}.}

In this paper we study the problem of PAC learning degree-$d$ PTFs 
in the presence of a constant fraction of adversarially corrupted data. 
\new{More concretely,}
we define the following data contamination model considered in the current work.

\begin{definition}[Nasty Noise or Strong Contamination Model] \label{def:nasty-learning}
{\em Let $\mathcal{C}$ be a class of Boolean functions on $\R^n$, 
$D_{\x}$ a distribution over $\R^n$, and $f$ an unknown target function 
$f \in \mathcal{C}$. For $0< \opt < 1/2$,
we say that a set $T$ of $m$ labeled examples is an {\em $\opt$-corrupted}
set of examples from $\mathcal{C}$
if it is obtained using the following procedure: 
First, we draw a set $S = \{ (\x^{(i)}, y_i)\}$
of $m$ labeled examples, $1 \leq i \leq m$, where for each $i$
we have that $\x^{(i)} \sim D_{\x}$, $y_i = f(\x^{(i)})$, and the $\x^{(i)}$'s are independent.
Then an omniscient adversary, upon inspecting the set $S$,
is allowed to remove an  $\opt$-fraction of the examples
and replace these examples by the same number of arbitrary examples of its choice.
The modified set of labeled examples is the $\opt$-corrupted set $T$.

A learning algorithm in the \new{strong contamination}/nasty noise model 
is given as input an $\opt$-corrupted set of examples
from $\mathcal{C}$ and its goal is to output a hypothesis $h$ such 
that with high probability the error $\Pr_{\x \sim D} [h(\x) \neq f(\x)]$ is small, 
\new{as compared to the information-theoretically optimal error of $\opt$}.}
\end{definition}

That is, in the nasty noise model~\cite{BEK:02}, an omniscient adversary can
arbitrarily corrupt a small constant fraction of both the data points and their labels. The nasty noise model is equivalent to the strong contamination model
studied in the field of robust statistics~\cite{DKKLMS16, DK23-book} and 
generalizes well-studied corruption models, including 
the agnostic (adversarial label noise) model~\cite{Haussler:92, KSS:94} 
and the malicious noise model~\cite{Valiant:85, KearnsLi:93}. 
In the adversarial label noise (agnostic) model, the adversary can corrupt an 
$\opt$-fraction of the labels, but cannot change the distribution of the 
unlabeled points. In the malicious model, the adversary can {\em add} 
an $\opt$-fraction of corrupted labeled examples, but is not allowed to 
adversarially remove clean labeled examples.

\new{
The goal of this work is to understand the efficient learnability of degree-$d$ PTFs under the Gaussian distribution in the presence of nasty noise. 
Our main algorithmic result is the following 
(see \Cref{thm:final-thm} for a more detailed formal statement):}

\begin{theorem}[Main Learning Result]\label{intro-thm:final-thm}
There exists an algorithm that, given  any $c, \eps \in (0,1)$, has
sample and computational complexity $n^{O(d)}\poly_{d,c}(1/\eps)$, and 
learns the class of degree-$d$ PTFs on $\R^n$ \new{in the nasty noise model}
under the Gaussian distribution within 0-1 error
$O_{c,d}(1) \, \opt^{1-c} + \eps$. 
\end{theorem}

\paragraph{Discussion} 
Some comments are in order. 
We start by noting that our learning algorithm is not proper. 
Specifically, the output hypothesis is 
a decision list whose leaves are degree-$d$ PTFs.

It is instructive to quantitatively compare the 
complexity and error guarantee of~\Cref{intro-thm:final-thm} with prior work. 
The $L_1$-polynomial regression algorithm of~\cite{KKMS:08} achieves 
the optimal error of $\opt+\eps$ in the (weaker) adversarial label noise model 
with sample and computational complexity $n^{\poly(d/\eps)}$. Moreover, the 
exponential complexity dependence in $1/\eps$ is inherent~\cite{DKR23, Tiegel22}. 
The latter computational lower bounds motivate the design of 
faster (ideally, fully-polynomial time) algorithms 
with relaxed error guarantees. 
When restricting to fully-polynomial time algorithms (i.e., with 
runtime $\poly_d(n/\eps)$),~\cite{DKS18a} gave a \new{robust} learner with error 
guarantee $\tilde{O}_d(\opt^{1/(d+1)}) +\eps$. \new{For $d>1$, this was 
the best previously known error guarantee (for $\poly_d(n/\eps)$ time algorithms) 
even in the weaker adversarial label noise model.} 
(See the following subsection for a detailed summary of prior work.)

\new{The latter error guarantee deteriorates dramatically 
as a function of the degree $d$. A natural question that motivated this work 
is whether it is possible to qualitatively nearly-match the $d=1$ 
case --- where polynomial-time algorithms with error $O(\opt)+\eps$ 
are known~\cite{ABL17, DKS18a} --- for any constant degree $d$ (or even for $d=2$!).} 
More concretely:
\begin{quote}
{\em Is there a $\poly_d(n/\eps)$ time algorithm that, 
for any constant $d$, robustly learns \\ 
degree-$d$ PTFs with error $O_{c,d}(1) \, \opt^{c}$, where $c>0$ is {\em independent} of $d$?} 
\end{quote}
Our main result answers this question in the affirmative.
Moreover, we can take the parameter $c$ above 
to be {\em any constant less than $1$}. 
Achieving error $\tilde{O}_d(\opt)$ or $O_d(\opt)$ is left as an open question. 

\new{Finally, we reiterate that our algorithm is the 
first algorithm with this error guarantee {\em even in the weaker 
model of adversarial label noise}.}


Interestingly, to obtain our algorithmic result, we 
generalize the {\em localization} technique~\cite{BlumFKV96, ABL17}, 
developed in the context of learning {\em linear} threshold functions,
for the problem of learning degree-$d$ PTFs.  
To achieve this goal,
we develop the algorithmic theory of {\em super non-singular polynomial 
decompositions}, which we believe is of broader interest 
beyond learning theory.



\subsection{Prior Work} \label{ssec:prior}

In the realizable PAC learning model (i.e., with clean/consistent labels), 
low-degree PTFs are known to be efficiently learnable 
in the distribution-free setting via a reduction to linear 
programming~\cite{MT:94}. Specifically, the class of degree-$d$ PTFs on $\R^n$ 
can be learned to 0-1 error $\eps$ with sample size $m = \tilde{O}(n^d/\eps)$ 
in $\poly(m)$ time. By standard VC-dimension arguments, this sample size is 
information-theoretically necessary for any learning algorithm.

In the presence of adversarial noise 
in the data (the focus of the current work), 
the learning problem becomes significantly 
more challenging computationally.  
Specifically, in the distribution-free setting, the agnostic 
learning problem (i.e., in the presence of adversarial label noise) 
is known to be computationally intractable, 
even for the special case of 
$d=1$ and constant accuracy~\cite{Daniely16, DKMR22-massart, Tiegel22}. 
As a result, research in this area has focused on 
the {\em distribution-specific} setting,
i.e., with respect to specific natural distributions on the domain, 
such as the Gaussian distribution. 

In the distribution-specific agnostic model, 
the $L_1$-polynomial regression algorithm~\cite{KKMS:08} 
learns degree-$d$ PTFs within error $\opt+\eps$ 
with sample and computational complexity $n^{O(d^2/\eps^4)}$ 
under the Gaussian distribution 
(and the uniform distribution on the hypercube)~\cite{KOS:08, DHK+:10, DSTW:10, DRST14, HKM14, Kane:10}. 
Importantly, the exponential dependence in $1/\eps$ is inherent in the complexity of the 
problem, both in the Statistical Query model~\cite{DKPZ21} 
and under standard cryptographic assumptions~\cite{DKR23, Tiegel22} 
\new{(even under the Gaussian distribution)}.

\new{The aforementioned hardness results motivated the design of 
faster algorithms with relaxed error guarantees.} 
Over the past fifteen years, substantial progress has been made 
in this direction, in particular for the special case of 
Linear Threshold Functions (corresponding to $d=1$).
Specifically, a sequence of works~\cite{KLS:09jmlr, ABL17, Daniely15, DKS18a}
developed $\poly(n/\eps)$ time robust learners for LTFs 
in the malicious/nasty model (thus, also in the adversarial label noise model)
under the Gaussian distribution and, in some cases, for isotropic 
(i.e., zero-mean, identity covariance) log-concave distributions.
In more detail, \cite{ABL17} gave a malicious learning 
algorithm for {\em homogeneous} LTFs (i.e., halfspaces whose separating hyperplane goes through the origin) with near-optimal error guarantee 
of $O(\opt)+\eps$ under all isotropic log-concave distributions\footnote{It 
turns out that the homogeneity assumption is important here. Specifically, the 
underlying algorithm does not extend to arbitrary LTFs with the same error 
guarantees.}. Subsequently,~\cite{DKS18a} gave an efficient algorithm that 
achieves error of $O(\opt)+\eps$ for arbitrary LTFs and succeeds under the 
Gaussian distribution. \new{At the technical level, \cite{ABL17} developed a 
{\em localization method} (see also~\cite{balcan2013active} for a precursor) 
which is crucial to obtain the near-optimal error guarantee of $O(\opt)$. In 
fact, the algorithm of ~\cite{DKS18a} for general halfspaces proceeds by a 
refinement of this idea.}

For the case of degree-$d$ PTFs, progress in this direction has been slow. 
The only prior algorithmic work on the topic is due to~\cite{DKS18a}. 
That work gave a $\poly(n^d/\eps)$ time algorithm that succeeds in the 
presence of nasty noise under the Gaussian distribution\footnote{We note that their algorithm works under a slightly more general class of distributions, whose moments up to order $2d$ are known a priori.} 
and attains an error guarantee of $O_d(\opt^{1/(d+1)})+\eps$. 

\subsection{Organization} \label{ssec:org}
The structure of this paper is as follows: In \Cref{sec:techniques} we give a detailed technical overview of our algorithm and the structural results used in its analysis.
\Cref{sec:prelims} contains the notation, basic facts, and technical lemmas that will be used throughout the remaining analysis.
In \Cref{sec:efficient-decomposition}, we present an efficient algorithm (\Cref{prop:super-non-singular-extension}) for constructing super non-singular polynomial decompositions.
\Cref{sec:proof-anti-concentration} is used to establish our main structural result (\Cref{thm:(anti-)concentration}) that demonstrates the (anti)-concentration properties of Gaussian distributions conditioned on sets of the form $\vec q(\x) \in R$, where $\vec q$ is some super non-singular polynomial transformation, and $R$ is an axis-aligned rectangle.
In \Cref{sec:polynomial-set-cutter} we use the tools developed in \Cref{sec:efficient-decomposition,sec:proof-anti-concentration} to construct a localization sub-routine (\Cref{prop:set-cutter}) which partitions sets defined by polynomial inequalities into subsets such that the Gaussian distribution conditioned on each one of them satisfy good (anti-)concentration properties.
In \Cref{sec:combine} we present the robust margin-perceptron algorithm, and combine it with the set partitioning routine in \Cref{prop:set-cutter} to produce our final algorithm for learning PTFs under adversarial corruptions. 

\section{Technical Overview} 
\label{sec:techniques}

\paragraph{Prior Work: Learning via \new{Degree-$d$} Chow-Parameters}
A polynomial threshold function \new{(PTF)} can be thought 
of as a linear threshold function \new{(LTF)} 
composed with the Veronese map $\x \mapsto \x^{\otimes d}$. 
Thus, we can think of the question of (robustly) learning a \new{PTF} 
of Gaussian inputs as the problem of learning an LTF with input 
$\wt \x \coloneqq \x^{\otimes d}$. 

A common approach to learning PTFs \new{(and other related geometric concept classes)}
in the literature 
is \new{via (low-degree)} Chow-parameter fitting \footnote{The Chow parameters of a Boolean function $f(\x)$ are defined as the vector $\E_{\x}[f(\x) \x]$.
Similarly, the degree-$d$ Chow-parameter tensor is defined as
$\E_{\x}[f(\x) \x^{\otimes d}]$.}; see, e.g.~\cite{TTV09,DDS12stoclong, DeDFS14,DKS18a,DGKKS20}.
More precisely, in \cite{DKS18a}, the approach
is to find an LTF (as a function of the tensor feature $\wt \x$) 
such that $\E_{\wt \x}[\sgn(\vec w \cdot \wt \x) \wt \x ] =
\E_{(\wt \x, y)}[y \wt \x]$.

The Chow-parameter fitting approach requires two crucial assumptions 
about the distribution of $\wt \x = \x^{\otimes d}$.
First, it requires concentration bounds in order to show that 
the \new{adversarial} noise cannot affect significantly the relevant 
Chow parameters $\E_{(\wt \x, y)}[y \wt \x ]$. 
Gaussian hypercontractivity indeed implies that polynomials of Gaussian 
random variables enjoy strong concentration.  
In addition to this, showing that a small error in the Chow parameters 
translates to a small error in the total variation distance 
of the corresponding threshold function, requires some 
anti-concentration properties of the underlying distribution. 
More precisely, it requires showing that any linear function 
is not-too-small with high probability. We can obtain this 
using results of \cite{CW:01}; but, unfortunately, 
the anti-concentration provided is weak, 
showing that $\Pr(|p(\x)| < \eps \||p\|_{L_2}) < O_d(\eps^{1/d})$ 
for any degree-$d$ polynomial $p(\cdot)$. 
This translates quantitatively to an algorithm that robustly 
learns degree-$d$ PTFs to error $O(\OPT^{1/O(d)})$ --- 
a far cry from our goal of error $O(\OPT^{1-c})$, 
especially when $d$ is large.

\paragraph{Our Approach: Learning PTFs Using Perceptron and Localization} 
Our \new{high-level} plan to improve upon \new{the error guarantee of  $O(\OPT^{1/O(d)})$}
is via \new{the method of} localization, 
a powerful approach for learning with corrupted labels; 
see~\cite{BlumFKV96, balcan2013active, ABL17}.  
\new{For technical reasons, our starting point is an early instantiation 
of this technique~\cite{BlumFKV96} 
developed in the context of learning LTFs with random label noise.}
At a high-level, localization consists of first learning some LTF 
that achieves good error for all large-margin points, 
and then conditioning on low-margin examples 
to learn a new (or improve the current) hypothesis. 
\new{Importantly,} all localization-based algorithms require that, 
\emph{after conditioning on the low-margin region} $|\vec w \cdot \wt \x| < \eps$, 
the resulting distribution satisfies strong anti-concentration properties.
While this property is true for learning LTFs under the Gaussian distribution,
it completely fails to hold under the conditional distribution 
of low-margin points with respect to a PTF, i.e., $|p(\x)| \leq \eps$.  
Our approach consists of two \new{new} ingredients: 
(i) a robust version of the localized margin-perceptron algorithm for 
learning PTFs under weaker (anti-)concentration assumptions, 
and (ii) a localization process for PTFs so that the corresponding conditional 
distributions satisfy (anti-)concentration.  
In the following presentation, we first focus on the 
localization process for PTFs, 
and then present our robust margin-perceptron learning algorithm.

\subsection{PTF Localization via Partitioning}

\begin{figure}
    \centering
    \includegraphics[width=5cm]{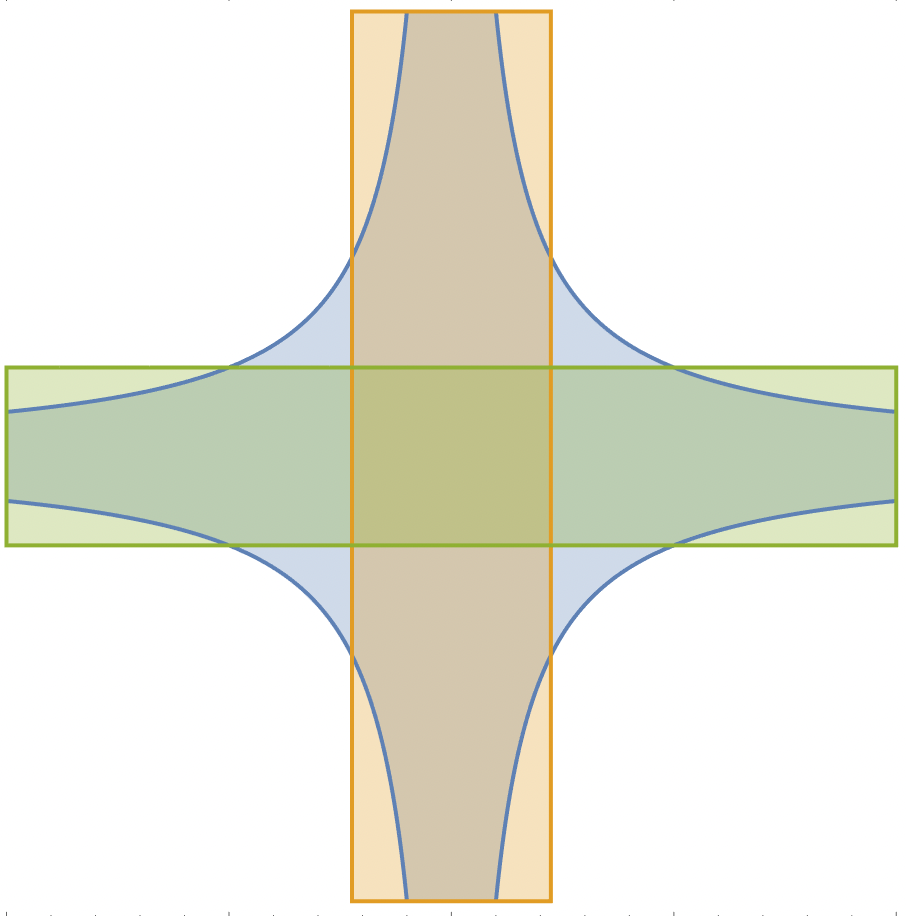}
    ~~~
    \includegraphics[width=5cm]{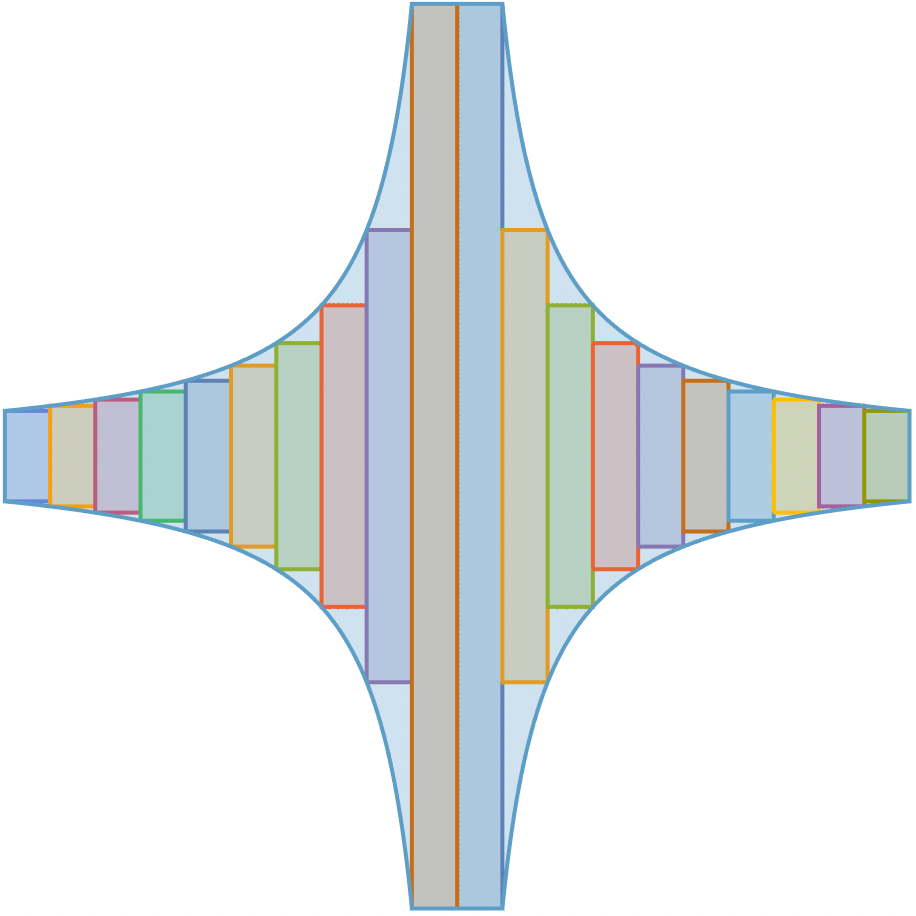}
    \caption{The localization region $|p(\x_1, \x_2)| = |\x_1^2 \x_2^2| \leq \eps$ is shown in blue. It is essentially a union of two rectangles (shown in the left figure) of width roughly $\sqrt{\eps}$.
    It is easy to see that (i) the total mass of the union is roughly $\sqrt{\eps})$;
    (ii) the expected value of $\x_1^2$ conditioned on the union is roughly $\Theta(1)$ (due to the contribution of the green rectangle).
    If the conditional distribution were a Gaussian, Carbery-Wright anti-concentration would imply that the conditional probability of  $\abs{\x_1^2} < \eps$ 
    should be at most $\poly(\eps)$. 
    In sharp contrast, the mass of the set $\abs{\x_1^2} < \eps$ 
    conditioned on the union is roughly $\Theta(1)$ 
    (due to the contribution of the orange rectangle).
    \newline
    To mitigate the issue, we will partition the low-margin set $|p(\x_1,\x_2)| \leq \eps$ into
    multiple rectangles as in the right figure.
    Since the Gaussian conditioned on each rectangle is a log-concave distribution, we have the desirable (anti-)concentration properties by \cite{CW:01}.} \label{fig:pain} 
    \end{figure}

\paragraph{Naive localization fails}
We first investigate why naively conditioning in the low-margin region $|p(\x)| < \eps$ 
fails to satisfy the required anti-concentration property, 
when $p(\x)$ has degree larger than $1$.
In particular, consider the polynomial $p(\x_1,\x_2) = \x_1^2 \x_2^2$.
To simplify the calculations, we first observe that the set $\abs{\x_1^2 \x_2^2} < \eps$
is similar to the union of two intervals 
$R = \{|\x_1^2| \leq \eps \} \cup \{|\x_2^2| \leq \eps\}$; see \Cref{fig:pain}.
The probability of $R$ under the standard Gaussian distribution is roughly $\sqrt{\eps}$, 
so we still need to learn a classifier inside it (note that we could simply 
ignore a region of mass $O(\epsilon)$). We examine the anti-concentration property 
of a different polynomial $q(\x_1, \x_2) = \x_1^2$ under the Gaussian distribution 
conditioned on the union of two rectangles $R$.
It is not hard to see that the $L_2$-norm of $q$ is $\Theta(1)$ under the conditional 
distribution (the points \new{within the} green rectangle in \Cref{fig:pain}
have large $\x_1$ coordinate with constant probability).
To give some intuition of what would be ``good'' anti-concentration, 
we remark that for a polynomial $q$ whose $L_2$-norm is constant, 
we would like the probability of $\abs{q(\x_1, \x_2)} < \eps$ 
to be roughly $\poly(\eps)$ 
(as is indeed the case for the Gaussian, by~\cite{CW:01}).
However, $q$ turns out to have much worse anti-concentration 
conditional on $R$, as we have 
\(
\pr[ \x_1^2 \leq \eps \mid \x_1^2 \x_2^2 \leq \eps]
\geq 
\pr[ \x_2^2 \leq 1] \pr[\x_1^2 \leq \eps]/  \pr[R]
= \Omega(1) 
\,.
\)
Thus, a naive localization procedure --- which tries to reapply 
a learner on the low-margin conditional distribution directly --- 
is unlikely to work as long as the learner requires 
\emph{any non-trivial} anti-concentration property.

\paragraph{Localization via Partitioning}
 A way to preserve the (anti)-concentration properties in the previous example 
 is to (approximately) partition the region where $|\vec x_1^2 \vec x_2^2| < \eps$ 
 into small (axis-aligned) rectangles (see the right figure in \Cref{fig:pain}). 
 The Gaussian distribution conditioned on each rectangle is a log-concave distribution, 
 and thus has good concentration and anti-concentration. Hence, we could attempt to use the margin-perceptron learner on each of the conditional distributions.  
 As one of our main contributions, we give an efficient algorithm that finds such a 
 partition for {\em any} low-degree polynomial. In particular, for a degree-$d$ 
 polynomial $p$, we show that the low-margin area $|p(\x)| \leq \eps$ can be partitioned 
 into $O_d(1)$ many subsets such that the Gaussian distribution conditioned 
 on each of them satisfies strong (anti-)concentration properties.
\begin{theorem}[Informal -- Partitioning the Low-Margin Region of Polynomials]
\label{intro-prop:set-cutter}
Fix $\eps \in (0, 1)$ and let $p: \R^n \mapsto \R$ be a polynomial of degree at most $d$. There exists an efficient algorithm that (approximately) decomposes 
\new{the set} $\{\vec x \in \R^n: \abs{p(\vec x)} < \eps\}$ into $m = \poly_d(1/\eps)$
sets $S^{(1)}, \cdots, S^{(m)} \subset \R^n$ such that $\normal(\vec 0, \vec I)$ conditioned on $S^{(i)}$ \new{satisfies} good anti-concentration.    
\end{theorem} 

\paragraph{Super Non-Singular Decomposition}
To get some intuition of how the partition routine operates, 
we revisit the example $p(\x_1, \x_2) = \x_1^2 \x_2^2$.
 What made this possible in this example is that the function 
 can be decomposed into the \emph{linear terms} $\x_1$ and $\x_2$. 
 Conditioning the Gaussian density on a rectangle of the form 
 $\x_1 \in I_1, \x_2 \in I_2$ yields a log-concave distribution 
 with good anti-concentration.
 More generally, if we could always decompose a polynomial $p$ 
 into \emph{a small number of} linear polynomials $q_1, \ldots, q_\ell$, 
 then we would still have anti-concentration
 in the resulting conditional distributions after partitioning 
 the region $\abs{p(\vec x)} < \eps$ into rectangles 
 defined by the linear terms, i.e., every $q_i(\x)$ lies in an interval $I_i$.
 Unfortunately, this is not possible for a general polynomial $p$.
 However, such a decomposition will exist if we allow the set of 
 polynomials $q_1,\ldots, q_\ell$ (or more precisely the polynomial mapping 
 $\x \mapsto \vec q(\x) := (q_1(\x), \ldots, q_\ell(\x))$) to only 
 {\em resemble a linear transformation locally}.
 To achieve this, we will need to \new{leverage} and generalize the results 
 of~\cite{Kane12subpoly} on \emph{non-singular decompositions}, 
 which itself builds on the techniques of diffuse decompositions from \cite{Kane12}.
In particular, we say that a collection of polynomials $q_1, q_2, \ldots, q_{\ell}$ is 
{\em non-singular} if there is only a negligible probability that the Jacobian 
of the (vector-valued) polynomial transformation $\vec q(\x)$ 
(i.e., the matrix $[\nabla q_1(\x) \nabla q_2(\x) \ldots \nabla q_\ell(\x)]$) 
has small singular values.
\snew{Intuitively, when this is the case, the polynomial transformation $\vec q$ 
will locally resemble a non-singular linear transformation.}
In \cite{Kane12subpoly}, it is shown that for any polynomial of degree at most $d$, 
there exists a non-singular set of of $O_d(1)$ polynomials $q_1, \ldots, q_\ell$ 
so that $p$ can approximately be written as a polynomial in the $q_i$'s. 
It turns out that having a non-singular decomposition 
is not enough to establish the anti-concentration properties that we require. 
We introduce the notion of \emph{super non-singularity}, 
which enforces ``local linearity'' by restricting the high-order derivatives 
of the polynomials. 
In particular, we \new{establish} two structural results on 
super non-singular sets of polynomials.
\begin{enumerate}[leftmargin=*]
    \item \emph{Anti-concentration via Super Non-Singular Decompositions.} The Gaussian distribution conditioned on a super non-singular set of polynomials, each lying in some interval, \new{satisfies} good anti-concentration and concentration properties (see \Cref{thm:(anti-)concentration}).
    \item \emph{Efficient Super Non-Singular Decomposition Algorithm.} Given a general degree-$d$ polynomial $p: \R^n \mapsto \R$, 
    there exists a computationally efficient algorithm that finds a super non-singular set of $m = O_d(1)$ polynomials $q_1,\ldots, q_m$ such that there exists a polynomial $h(\cdot)$ of degree at most $d$ which satisfies $p(\x) \approx h(q_1(\x),\ldots, q_m(\x))$ (see \Cref{prop:super-non-singular-extension}).
\end{enumerate}
Equipped with the above structural and algorithmic results, 
\new{obtaining} an efficient partition algorithm is relatively straightforward. 
After computing a super non-singular decomposition $q_1,\ldots, q_m$ of $p$ using 
\Cref{prop:super-non-singular-extension}, we have that since $p$ can be approximately 
expressed as a polynomial $h(q_1(\x),\ldots, q_m(\x)$, 
the value of $p(\vec x)$ is (approximately) determined by the values of $q_i(\vec x)$.
Therefore, we can show that the set $\{\vec x \in \R^n: \abs{p(\vec x)} < \eps\}$ 
can be approximately covered by sets of the form 
$\{\vec x:  (q_1(\vec x), \cdots, q_{\ell}(\vec x) \in R\}$, 
where $R$ is an $m$-dimensional axis-aligned rectangle.  
Hence, anti-concentration properties of the conditional distributions 
on these sets follow.

This allows us to perform at least one round of localization 
by partitioning
the set $L = \{\x:  |p(\x)| \leq \eps\}$.  
Assuming that we have obtained a polynomial $p'$ 
that achieves good error in the region $R$,
we then need to ``localize'' on the region 
$L' = \{ \x: |p(\x)| \leq \eps, |p'(\x)| \leq \eps \}$.  
We show that this is possible by a subtle 
``extendibility'' property of our super non-singular decomposition algorithm.
Specifically, assuming that we have a \new{super} non-singular decomposition 
$q_1, \cdots, q_{\ell}$ of $p$ (which was used to compute $p'$), 
we can then extend it into a larger super non-singular set 
$\{q_1, \cdots, q_{\ell}, q_{\ell+1}, \cdots, q_m\}$ 
such that we can still approximately 
express $p'(\vec x)$ as a polynomial in $q_1, \cdots, q_m$. 
For each rectangle $R$ of the original partition of $p(\x)$, 
we can now further cover the region 
$\{\vec x \in \R^n: \abs{p'(\vec x)} < \eps,  (q_1(\vec x), \cdots, q_{\ell}(\vec x)) \in R \}$ 
with sets of the form 
$\{\vec x:  (q_1(\vec x), \cdots, q_{\ell}(\vec x) \in R, (q_{\ell+1}(\vec x), \cdots, q_{m}(\vec x) \in R')\}$, 
where $R'$ is some other $(m-\ell)$-dimensional axis-aligned rectangle.
\snew{As a slight digression,}
we note that this extendibility property is also what makes 
the proof of \Cref{thm:(anti-)concentration} possible.

\subsection{Anti-concentration via Extendible Super Non-Singular Decomposition}
We have already seen (see \Cref{fig:pain}) that the Gaussian distribution 
conditioned on sets of the form 
$\abs{p_1(\vec x)} < \eps, \abs{p_2(\vec x)} < \eps, \cdots$ 
for generic polynomials $p_i$ does not satisfy good anti-concentration.
To mitigate this issue, we need the polynomials appearing in the conditioning 
to collectively satisfy a strong non-singularity condition 
concerning their high-order derivatives.
In the following definition, we denote by $\nabla_x$ 
the standard gradient operator and by $D_{\vec y}$ 
the derivative in the direction $\vec y$.
\begin{definition}[Super Non-Singular Polynomial Transformation (SNPT)]
\label{def:super-non-singular}
Let $\eps \in (0, 1)$ and $N \in \mathbb Z_+$.
Let $S := \{q_1, \cdots, q_m\}$, where $q_i: \R^n \mapsto \R$ 
is a set of harmonic (see \Cref{def:harmonic-components}) real-valued polynomials 
of degree at most $d$.
For $1 \leq k \leq d$,
let $S_k \subseteq S$ be the set of degree-$k$ harmonic polynomials (contained in $S$).
The set of polynomials $S$ is $(\eps, N)$-super non-singular 
if for any integer 
$1 \leq k \leq d$ it holds
$$
\Pr_{\vec x \sim \normal(\vec 0, \vec I), \vec y^{(i)}\sim \normal(\vec 0, \vec I) \text{ for } 1 \leq i \leq k-1}
\lp [
\snorm{2}{
\nabla_{\vec x}
D_{\vec y^{(k-1)}}
\cdots
D_{\vec y^{(1)}}
\lp(  \sum_{i \in S_k} \vec a_i \; q_i(\vec x) \rp)}  <\eps  \rp]
< \eps^{N}
$$
for all $\vec a \in \R^m$ such that $\sum_{i \in S_k} \vec a_i^2  = 1$.
We will also call $(\eps, N)$-super non-singular a polynomial transformation 
$\vec q(\x) = (q_1(\x),\ldots, q_m(\x))$ defined by 
\new{an} $(\eps, N)$-super non-singular set $S$.
\end{definition} 
We remark that \Cref{def:super-non-singular} resembles the definition 
of non-singular polynomials in \cite{Kane12subpoly}, 
but imposes additional requirements on the high-order derivatives of the polynomials. This additional structure turns out to be crucial 
in proving \Cref{prop:super-non-singular-extension}, 
which is itself an important building block 
to establish \Cref{thm:informal-anticoncentration}. 
As one of our main contributions, we show that \new{the distribution} 
$\normal(\vec 0, \vec I)$, 
conditioned on a set of super non-singular polynomials each lying in some interval (satisfying some mild conditions), 
satisfies good polynomial concentration and anti-concentration properties.
For brevity, we \new{henceforth} refer to both concentration 
and anti-concentration as (anti-)concentration.
\begin{theorem}[Informal -- Conditional (anti-)concentration for SNPT, see \Cref{thm:(anti-)concentration}] \label{thm:informal-anticoncentration}
Let $\vec q$ be a degree $d$, ``sufficiently'' super non-singular polynomial transformation 
(i.e., for large enough $\eps, N$ in \Cref{def:super-non-singular}). Let $R \subseteq \R^m$ be an axis-aligned rectangle that is not too far from the origin
and let $D$ be $\normal(\vec 0, \vec I)$ conditioned on the set $\{\x: \vec q(\x) \in R\}$.
For any unit variance, mean-zero polynomial $p$ of degree at most $d$ we have:
\begin{itemize}
\item   (Anti-Concentration) $\Pr_{\x \sim D}[|p(\x)| \leq t ] \leq t^{1/(2 d)}$.
\item (Concentration)  For all $K\in \mathbb Z_+$ up to some constant \footnote{The constant depends on ``how'' super non-singular the set is; see \Cref{thm:(anti-)concentration}.}, it holds
$\Pr_{\x \sim D}[|p(\x)| > t ] \leq t^{-1/K}$.
\end{itemize}
\end{theorem}
We now provide a sketch of the high-level ideas behind the proof of the above theorem.  
In what follows, 
we denote by $p(D)$ the distribution of the random variable 
$p(\x)$ when $\x \sim \D$.
Let $D$ be the distribution of $\vec x \sim \normal(\vec 0, \vec I)$ 
conditioned on $\vec q (\vec x) \in R $ for a rectangle $R$.
Our goal is to show that for any low-degree polynomial $p$, 
the distribution $p(D)$ has good anti-concentration.

\paragraph{Constructing a Low-Dimensional Surrogate Distribution} 
As our first step,
instead of directly analyzing the (anti-)concentration properties of $p$ 
under the $n$-dimensional distribution $D$, which is challenging, 
we construct low-dimensional ``surrogates'' for $D$ and $p$.
Specifically, we consider a low-dimensional distribution $Q$ 
together with a polynomial $f$, such that
the outcome of $p(D)$ enjoys roughly the same concentration and 
anti-concentration properties as $f(Q)$.  
Given the construction, we can in turn focus on analyzing 
this low-dimensional surrogate pair.

The fact that super non-singular decompositions are ``extendible'' 
will play a critical role in this construction. 
In particular, the super non-singular polynomials $\{q_1, \cdots, q_{\ell}\}$ 
appearing in the conditioning of $D$ will first be extended into 
a super non-singular decomposition for the target polynomial $p$ \new{such that} 
there exists a set of super non-singular polynomials 
$\{q_1, \cdots, q_{\ell}, q_{\ell+1}, \cdots, q_m\}$, 
where $m = O_{d,\ell}(1)$, 
and a composition polynomial $f:\R^m \mapsto \R$ such that
\(
p( \vec x ) \approx  f( q_1(\vec x), \cdots, q_m(\vec x) ) .
\)
Define $\vec q:\R^n \mapsto \R^m$ to be the vector-valued polynomial 
whose $i$-th coordiniate is $q_i$, 
and $Q$ to be the $m$-dimensional distribution of 
$\vec q(\normal(\vec 0, \vec I))$ conditioned on
the set $\{ \vec y \in \R^m: \vec y_i \in I_i ~ \forall i =1,\ldots, \ell\}$.
Then, subject to the intervals appearing in the conditioning satisfying 
some mild conditions and the polynomials appearing in the conditioning 
being sufficiently super non-singular, one can verify that 
$p(D)$ enjoys roughly the same (anti-)concentration properties as $f(Q)$.
For the details of this argument, we refer to \Cref{sec:transfer-properties}, 
where we conclude the proof of \Cref{thm:(anti-)concentration}. 

Given such a construction, we can now shift our focus from the $n$-dimensional 
distribution $D$ to the $m=O_{d,\ell}(1)$-dimensional conditional distribution $Q$ 
defined by a set of super non-singular polynomials.
In particular, if we use $\vec q:\R^n \mapsto \R^m$ to denote the vector-valued 
polynomial whose $i$-th coordinate is $q_i$, 
we are interested in the distribution of $\vec q \lp( \normal(\vec 0, \vec I) \rp)$ 
conditioned on the event $\{ \vec q_i(\vec x) \in I_i \}_{i=1}^{\ell}$, 
where $I_i$ is some interval. 
To build some intuition as to why such conditional distributions 
may have desirable properties, we can start with the simple case 
where all $q_i$ are linear functions. In that case, 
the distribution of $\vec q( \normal(\vec 0, \vec I) )$ 
is simply some other Gaussian distribution $\normal'$. 
Then, even if we condition on that the $i$-th coordinate of 
$\vec q( \normal(\vec 0, \vec I) )$ is equal to $\vec a_i$, 
the resulting distribution will simply be some lower-dimensional 
Gaussian distribution. 
Under mild conditions on $\vec a_i$ and $q_i$, 
the resulting low-dimension Gaussian will be not too different 
from a standard Gaussian, in the sense that its mean is not very far 
from the origin and its covariance is bounded above and below 
by multiples of the identity. 
For more details, we refer to \Cref{sec:reasonable-gaussian}.
\begin{definition}[$(\delta, \kappa)$-Reasonable Gaussian]
\label{intro-def:reasonable-gaussian}
Let $\normal(\vec \mu, \vec \Sigma)$ be a Gaussian distribution.
Given $\delta \in (0, 1)$ and $\kappa > 1$,
we say $\normal(\vec \mu, \vec \Sigma)$ is a $(\delta, \kappa)$-reasonable Gaussian if 
$ \snorm{2}{\vec \mu} \leq  \kappa  $ and
$
\delta \vec I \preceq \vec \Sigma \preceq \kappa \vec I.
$
\end{definition}
When the polynomials are of degree more than $1$, 
it becomes hard to characterize the exact form of $\vec q(\normal(\vec 0, \vec I))$. 
Nonetheless, the hope is that we can still compare its probability density function 
to that of some other more structured distribution family.
\begin{definition}[Distribution Comparability]
\label{def:comparability}
Let $Q, Q'$ be probability distributions with the same support. 
We say that $Q$ and $Q'$ are comparable if for all $\vec x$ in their common support, 
it holds
\(
 1/2 \; Q'(\vec x) \leq Q(\vec x) \leq 2 \; Q'(\vec x).
 \)
 \footnote{For readers who are familiar with the notion of R\'enyi divergence, 
 this is equivalent to stating that the symmetrized R\'enyi divergence 
 of infinite order between the two distributions is bounded by some constant.}
\end{definition}
We show that if two distributions are comparable to each other, 
then \new{they} will have similar (anti-)concentration properties --- 
even under an arbitrary conditioning. The formal statement of this \new{fact}
and its proof can be found in \Cref{sec:transfer-properties}.


\paragraph{Super Non-Singular Polynomial Transformations of Gaussians are Reasonable} 
Given a polynomial transformation $\x \mapsto \vec q(\x)$, 
we will say that $\vec q$ is super non-singular if the set 
of its polynomial coordinates $\vec q_i(\x)$ is super non-singular.
We show that a super non-singular transformation $\vec q$ 
behaves similarly to a linear transformation, 
in the sense that the distribution
$q(\normal(\vec 0, \vec I)\new{)}$ is comparable to a mixture of reasonable Gaussians.
\begin{proposition}[Informal -- Super Non-Singular Polynomial Transformations are Reasonable]
\label{intro-prop:Gaussian-Comparability}
Let $\vec q$ be a $(\delta^{1/(3d)}, C)$-super non-singular polynomial transformation for some sufficiently small $\delta$ and large $C$\footnote{See \Cref{prop:Gaussian-Comparability} for the precise constants.}.
Then,  $\vec q( \normal(\vec 0, \vec I) )$ is $O( \delta^N )$-close in total variation distance to 
some distribution that is comparable to the mixture distribution
$\int \normal_{\theta} d\theta$, where each $\normal_{\theta}$
is a $(\delta, \log^{ O(d) }(1/\delta  ) )$-reasonable Gaussian.
\end{proposition}

We now provide a proof sketch for \Cref{intro-prop:Gaussian-Comparability}.
We first note that $\vec q( \normal(\vec 0, \vec I) )$ has a distribution identical 
to $\vec q(\sqrt{1-\delta^2} \vec x + \delta \vec z)$ for $\delta \in (0, 1)$ an 
appropriately chosen real number and $\vec x, \vec z$ distributed as 
two \iid Gaussians. 
Fixing the value of $\vec x$ and Taylor expanding around $\vec z$, 
we find that $\vec q(\sqrt{1-\delta^2} \vec x + \delta \vec z)$ 
is approximately $\vec g_{\vec x} + \delta \Jac_{\vec q}(\vec x) \; \vec z  + O(\delta^2) \vec e_{\vec x}(\vec z) $, 
where $\Jac_{\vec q}$ represents the Jacobian of the transformation 
$\vec q$, $\vec g_{\vec x}$ is some vector that depends only on $\vec x$, 
and $\vec e_{\vec x}$ is some degree-$d$ polynomial.
It turns out if $\vec q$ consists of super non-singular polynomials, 
$\Jac_{\vec q}(\vec x)$ will have no small singular values 
with high probability (see \Cref{lem:non-singular-jacobian}).  
Conditioned on some fixed value of $\vec x$ that makes $\Jac_{\vec q}(\vec x)$ 
non-singular, the distribution of 
$\vec g_{\vec x} + \delta \Jac_{\vec q}(\vec x)  \vec z$ where $\vec z \sim \normal(\vec 0, \vec I)$ will be a reasonable Gaussian. 
We remark that the transformation still has the high-order error term 
$O(\delta^2) \vec e_{\vec x}(\vec z)$ that we have to bound.
We notice that the coefficient in front of the high-order term 
is significantly smaller than the minimum singular value of the linear component. 
As a result, the distribution produced by the transformation will still be close 
in total variation distance to some distribution \emph{comparable} 
to a reasonable Gaussian distribution. 
We refer to \Cref{sec:proof-anti-concentration} for more details.

We remark that all of the above analysis is done for a fixed value of $\vec x$ 
that ensures non-singularity of $\Jac_{\vec q}(\vec x)$. 
Hence, to conclude the proof, we simply need to take a mixture 
over the values of $\vec x$ following the standard Gaussian distribution. 
By super non-singularity, $\Jac_{\vec q}(\vec x)$ has no small singular values 
with high probability. 
Consequently, most of the distributions within the mixture 
will be comparable to a reasonable Gaussian distribution. 
\Cref{prop:Gaussian-Comparability} thereby follows.

Given \Cref{intro-prop:Gaussian-Comparability}, and the definition of comparability, 
we conclude that the transformation $\vec q$ conditioned 
on an axis-aligned rectangle enjoys good (anti)-concentration properties.
By our construction, the target polynomial $p$ under the target distribution $D$ 
enjoys roughly the same (anti)-concentration properties 
as some polynomial $h$ under $\vec q(\normal(\vec 0, \vec I))$ 
conditioned on an axis-aligned rectangle. 
The proof of \Cref{thm:(anti-)concentration} follows.


\subsection{Efficiently Extending a Super Non-Singular Decomposition}
\label{sec:extendible-decomposition-technique}
In this section, we discuss our efficient algorithm for 
obtaining and extending a super non-singular decomposition.
\cite{Kane12subpoly} shows that any polynomial of degree at most $d$ 
can be approximately decomposed into a non-singular polynomial set 
of size at most $O_d(1)$.
In \Cref{intro-prop:super-non-singular-extension}, we show that this 
is also true for the notion of super non-singularity. 
 \Cref{intro-prop:super-non-singular-extension} extends \new{and strengthens} 
 the result of \cite{Kane12subpoly} in two ways: 
 (i) we \new{are} able to decompose \emph{multiple} 
 (as opposed to just one, as in \cite{Kane12subpoly}) generic polynomials 
 into a \emph{common} set of super non-singular polynomials, \new{and} 
 (ii) we are able to do so when the generic polynomials arrive 
 in an \emph{online} fashion. In particular, given a super non-singular set 
 of polynomials $\mathcal Q$ obtained while decomposing 
 some polynomials $p_1, \cdots, p_t$ in the past rounds, 
 after receiving the new polynomial $p_{t+1}$, 
 we are able to extend $\mathcal Q$ into a larger set of 
 super non-singular polynomials $\mathcal Q'$ 
 and decompose $p_{t+1}$ in terms of $\mathcal Q'$.
We remark that the fact \new{that} we can keep extending 
a super non-singular set of polynomials 
to ensure it can be used to represent increasingly more polynomials 
is a unique characteristic of super non-singular decomposition 
(compared to its ``non-super'' counterpart).
Crucially, this additional \new{``extendibility''} property 
of the decomposition is what makes the (anti-)concentration 
result (\Cref{thm:informal-anticoncentration}) 
and the polynomial set partitioning routine (\Cref{intro-prop:set-cutter}) possible.
In the following result, we present our efficient algorithm for extending 
a super non-singular decomposition.
\begin{theorem}[Informal -- Extendible Super Non-singular Decomposition]
\label{intro-prop:super-non-singular-extension}
Let $\ell,d, N'\in \mathbb Z_+$ and $\eps > 0$ be sufficiently small given $\ell,d$.
Let $S:=\{q_1, \cdots, q_{\ell}\}$, where $q_i: \R^n \mapsto \R$, be a set of  polynomials of degree at most $d$ and  
$p: \R^n \mapsto \R$ be another polynomial of degree at most $d$.
Suppose that $S$ is $(\eps^{1/3}, N)$ super non-singular, 
where $N$ is sufficiently large given $d, \ell, N'$.
Then there exists an algorithm which can
extend $S$ into a set of $m = O_{d,\ell}(1)$ polynomials
$\bar S:= \{q_1, \cdots, q_\ell, q_{\ell+1}, \cdots, q_m \}$ such that
\begin{itemize}
    \item $\bar S$ is $(\eps, N')$-super non-singular. 
    \item There exists a polynomial $h:\R^m \mapsto \R$ of degree at most $d$
    such that
    $
    \snorm{L_2}{ p(\vec x) - h(\vec q(\x)) } 
    \leq \eps^2$, where $\vec q = (q_1(\x),\ldots, q_m(\x))$.
\end{itemize}
Moreover, the algorithm runs in time $\poly(n) \;
\poly_{\ell, d, N'}(1/\eps)$.
\end{theorem}
Suppose we only want to compute a non-singular decomposition for a polynomial $p$.
The process given in \cite{Kane12subpoly} maintains a data-structure 
to which we refer as a partial decomposition.
Informally, the data-structure keeps track \new{of} 
a list of polynomials $q_1, \cdots, q_{\ell}$ 
(which is not necessarily non-singular), 
a coefficient vector $\vec b \in \R^{\ell}$, 
and a composition polynomial $h$ such that 
$p(\vec x) = h( \vec b_1 q_1(\vec x), \cdots, \vec b_{\ell} q_{\ell}(\vec x) )$.
If the list of polynomials is already non-singular, we are done.
Otherwise, following the definition of non-singularity, 
there exists a linear combination of the polynomials 
$q^*(\vec x) = \sum_{i} \vec b_i q_i(\vec x)$
such that the gradient of the combined polynomial $q^*$ is small 
with non-trivial probability under the Gaussian distribution.
In the second case, we show that we can approximately decompose 
the combined polynomial $q^*$ into a set of lower-degree polynomials 
$\alpha_1, \cdots, \alpha_m$. 
Hence, we can rewrite one of the polynomials $q_i$ 
which has non-trivial weight in the linear combination 
with the set of newly obtained lower-degree polynomials $\alpha_i$ 
and the remaining polynomials in the linear combination. 
We then end up with a new partial decomposition
consisting of the polynomials 
$q_1, \cdots, q_{i-1}, q_{i+1}, \cdots, q_{\ell}, \alpha_1, \cdots, \alpha_m$, 
a new coefficient vector $\vec b'$ and a new composition polynomial $h'$.
It turns out such a rewriting strategy will always decrease 
the total weights of the polynomials that have the same degree 
as $q_i$ in $\vec b'$, but may end up increasing the weights 
of the other polynomials in the linear combination.
However, if we always choose to rewrite the highest (or one of the highest) 
degree among the polynomials in the linear combination, 
it is then guaranteed that we will have fewer and fewer high-degree polynomials 
in the decomposition.
This process must then eventually terminate 
and give us a super non-singular set of polynomials.

In order to adapt the above strategy for extendible 
super non-singular decomposition, 
a caveat here is that for the additional extendibility property to hold, 
we are now allowed to 
rewrite any of the initial polynomials $q_1, \cdots, q_{\ell}$.
Fortunately, if a set of polynomials does not satisfy super non-singularity, 
we are always capable of finding a linear combination 
of the polynomials \emph{of the same degree} such 
that the combined polynomial has small gradients with non-trivial probability. 
Therefore, no matter which polynomial $q_i$ we choose to rewrite, 
it will always have the highest degree among the polynomials 
in the linear combination (since all of them have the same degree!).
The induction argument for showing the termination of the process 
will then go through, 
giving us a super non-singular decomposition algorithm.

\subsection{Learning via Localization and Margin-Perceptron}
At a high-level, our learning algorithm \new{can be viewed as} 
a robust version of the margin-perceptron algorithm of \cite{DunaganV04}. 
In \Cref{prop:perceptron}, we give a robust version of the perceptron algorithm 
(with a slightly different step-size) \new{and establish the following}:  
given sample access to a distribution corrupted by $\opt$-nasty noise, 
the margin-perceptron is a semi-agnostic LTF learner 
when the underlying (uncorrupted) $\x$-marginal 
is ``sufficiently'' (anti)-concentrated.  Our modified perceptron algorithm 
of \Cref{prop:perceptron} uses a robust sub-routine of \cite{DKS18a} 
for estimating the Chow-parameters of the LTF under nasty noise. 
For example, in the first round, we learn a PTF that achieves error 
$\opt^{1 - c}$ for all high-margin points in the set 
$\{\x : |p(\x)| \geq \eps^{c} \|p\|_2 \}$. 
We then localize (condition) on the set of low-margin points
$\{\x : |p(\x)| \leq \eps^{c} \|p\|_2 \}$, 
and use the partitioning algorithm of \Cref{intro-prop:set-cutter} 
to partition the above region into sets $S^{(1)},\ldots, S^{(m)}$ 
such that the standard normal conditional on those sets 
has good (anti)-concentration.  We then \new{again} condition on each set 
of this partition and use the robust-perceptron algorithm of \Cref{prop:perceptron} 
to learn a PTF inside each set; and continue recursively until the probability 
mass of the ``unclassified'' low-region is at most $O(\eps)$. 
Our final hypothesis is therefore a decision list of \new{degree-$d$} PTFs: 
one PTF \new{for} each set of the partition. 
For more details, we refer \new{the reader} to \Cref{sec:combine}.

\section{Preliminaries} \label{sec:prelims}
\subsection{Notation and Terminology}

\paragraph{Asymptotic Notation}
We will use the notation $O_{\alpha, f}(N)$, where $\alpha$ is some variable and $f: \mathbbm Z_+ \mapsto \mathbbm Z_+$ is some computable function. 
We use this notation to denote a quantity whose absolute value is bounded above by $N$ times some number that can be computed given the input $\alpha$ and oracle access to the function $f$.
The notation $\poly(n)$ is used to represent a quantity whose absolute value is bounded above by $O(n^C)$ for some large enough constant $C$.
We also write $\poly_{\alpha, f}(n)$. In that case, the quantity is bounded above by $O(n^C)$, 
where $C = O_{\alpha, f}(1)$.
In statements of lemmas/theorems, we often write  ``$\alpha$ is sufficiently large given 
$\beta$'', where $\alpha$ is some variable and $\beta$ can be either a variable or a function. 
By that, we mean there exists a number $C_{\beta} = O_{\beta}(1)$ such that the guarantees stated in the lemma/theorem will be true as long as $\alpha > C_{\beta}$.
The phrase ``$\alpha$ is sufficiently small given $\beta$'' means  $\alpha^{-1}$ is sufficiently large given $\beta$.

\paragraph{Vectors, Matrices and Tensors} 
We use lower case bold letters to denote a vector. Given a vector $\vec x \in \R^n$, 
we define its $\ell_2$-norm 
as $\snorm{2}{ \vec x }^2 
= \sum_{i=1}^n \vec x_i^2$.
We use upper case bold letters to denote matrices and tensors.
Given a matrix or a tensor $\vec A \in \R^{n_1 \times \cdots \times n_d }$, we define its Frobenious norm as~$\snorm{F}{ \vec A }^2 = \sum_{i_1 \in [n_1], \cdots, i_d \in [n_d]} \vec A_{i_1, \cdots, i_d}^2$.
When $\vec A$ is a matrix,  
we write $\sigma_i(\vec A)$ to denote the $i$-th largest singular value of $\vec A$.
Following this convention, $\sigma_{\max}(\vec A)$ is used to denote the induced vector norm of $\vec A$ and $\sigma_{\min}(\vec A)$ is used to denote the smallest singular value of $\vec A$.
We sometimes need to study how a matrix's singular values change after a small perturbation. For that, we will use Weyl's Inequality for singular values.
\begin{fact}[Weyl's Inequality for Singular Values]
\label{fact:weyl}
Let $\vec A, \vec B$ be matrices of size $\R^{m \times n}$.
Then it holds
$$
\abs{\sigma_i( \vec A + \vec B ) - \sigma_i(\vec A)} \leq \sigma_{\max}(\vec B).
$$
\end{fact}

\paragraph{$L_p$ Norms of functions}
Let $f:\R^n \mapsto \R$ be a real-valued function.
Given a distribution $D$ on $\R^d$, we define the $L_p$-norm of $f$ with respect to $D$ as
$$
\snorm{D, L_p}{f}
= \lp(\E_{\vec x \sim D}
\lp[ \abs{f(\vec x)}^p \rp] \rp)^{1/p}.
$$
Let $\vec f: \R^n \mapsto \R^m$ be a vector-valued function.
We use $\vec f_i$ to denote 
\snew{the $i$-th coordinate of the vector-valued function $\vec f$.}
We define the $L_p$-norm of $\vec f$ 
with respect to $D$ as
$$
\snorm{D, L_p}{\vec f}
= \lp( \sum_{i=1}^m \snorm{D, L_p}{\vec f_i}^p \rp)^{1/p}.
$$
If the distribution $D$ is the standard Gaussian $\normal(\vec 0, \vec I)$, we omit the subscript $D$ and write just $\snorm{L_p}{f}$ and $\snorm{L_p}{\vec f}$.

\paragraph{Multi-index Notation}
We will use the following multi-index notation.
Let $\vec s \in \mathbb Z^{n}$ and $\vec x \in \R^n$. 
We write $\vec x^{\vec s}$ to denote the degree-$\abs{\vec s}$ monomial $  \prod_{i=1}^n \vec x_i^{\vec s_i} $.
Given $\vec k, \vec s \in \mathbb Z^{n}$, 
we write $\vec k \leq \vec s$ when $\vec k_i \leq \vec s_i$ for all $i \in [n]$.
Given $\vec k \leq \vec s$, we write
\begin{align*}
\binom{\vec s}{\vec k}:= \prod_{i=1}^n \binom{\vec s_i}{\vec k_i}.    
\end{align*}
Let $f: \R^n \mapsto \R$ be a degree-$d$ polynomial and $\vec T$ be a tensor indexed by $\vec a \in [d]^n$.
We say that $\vec T$ is the coefficient tensor of $f$ if $\vec T$ satisfies
$$
f(\vec x) = \sum_{ \vec a \in [d]^n }
\vec T_{\vec a} \; \vec x^{\vec a}.
$$
Notice that since $f$ is of degree $d$, $\vec T_{\vec a}$ must be non-zero only if $\snorm{1}{\vec a} \leq d$.
We will write $\coe(f)$ to denote this coefficient tensor $\vec T$.
For a low-degree \snew{polynomial whose inputs have low dimensions}, an important quantity related to its $L_p$-norm is the largest of the absolute values of the entries in $\coe(f)$.
We will write $\LC{f}$ to denote this quantity.
\paragraph{Derivatives}
Given a real-valued function $f: \R^n \mapsto \R$, the gradient of $f$ is a vector valued function and we denote it by $\nabla f$.
Given a vector $\vec y \in \R^n$, we write $D_{\vec y} f$ to denote its directional derivative along the direction of $\vec y$, i.e., 
$D_{\vec y} f(\vec x) = \vec y \cdot \nabla \vec x$.
Let $g(\vec x, \vec y) = D_{\vec y} f(\vec x)$. For such a function $g$ with two input arguments, we write $\nabla_{\vec x} g$ or $\nabla_{\vec y} g$ to denote its gradient with respect to the corresponding argument.
Given a vector-valued function $\vec f: \R^m \mapsto \R^n$, the Jacobian of $\vec f$ is a matrix-valued function and we denote it by $\Jac_{\vec f}$.

\paragraph{Probabilities} We denote by $\normal(\vec g, \vec M)$ the Gaussian distribution with mean $\vec g$ and covariance $\vec M$.
Given $\vec x \in \R^n$, we denote by $\normal(\vec x; \vec g, \vec M)$ the probability density function of $\normal(\vec g, \vec M)$ evaluated at the point $\vec x$.
Let $S$ be a subset of points in $\R^n$ and $D$ be a distribution in $\R^n$.
We denote by $D(S)$ the mass of the set $S$ under $D$, and denote by $D \mid S$ the conditional distribution of $\vec x \sim D$ conditioned on  $\vec x \in S$.
\subsection{Orthogonal Polynomials}
In this subsection, we will need the definition of the Hermite polynomials and harmonic components of a polynomial.
The univariate Hermite polynomials that we use will be ``probabilist's Hermite polynomials'' normalized to have $L_2$-norms $1$.
\begin{definition}[Univariate Hermite Polynomial]
The degree-$0$ Hermite polynomial is $H_0(x) = 1$.
We define the degree-$d$ Hermite Polynomial recursively as
$$
H_d(x) = \frac{1}{\sqrt{d}} \lp( x \; H_{d-1}(x) -  H_{d-1}'(x) \rp) \, ,
$$
where $H_{d-1}'$ denotes the derivative of $H_{d-1}$.
\end{definition}
We have the following natural recurrence relation of the coefficients in univariate Hermite polynomials.
\begin{fact}
Assume that $H_d(x) = \sum_{k=0}^d a_{d, k} \; x^k$.
Then we have the recurrence relation
\begin{align*}
a_{d+1, k} = 
\begin{cases}
    &-d \;  \sqrt{ \frac{d-1}{d+1} } \; a_{d-1, k} \, , \text{   if } k = 0 \, , \\
    & \sqrt{ \frac{d}{d+1} } a_{d, k-1} -d \; 
    \sqrt{ \frac{d-1}{d+1} } \; 
    a_{d-1, k} \, , \text{   if } k > 0 .
\end{cases}
\end{align*}
\end{fact}
Using the above recurrence relation, we can derive a bound on the largest coefficient of Hermite polynomials.
\begin{fact}
For the degree-$d$ univariate Hermite polynomial $H_d$,
we have that
$
\LC{ H_d } \leq 2^d.
$
\end{fact}
We can use the univariate Hermite polynomials as building blocks to construct multivariate Hermite polynomials.
\begin{definition}[Multivariate Hermite Polynomials]
Given a vector $\vec a \in \mathbbm N^n$, we denote the multivariate Hermite polynomials of degree $\snorm{1}{\vec a}$ as 
$$
H_{\vec a}(\vec x)
= \prod_{i=1}^n
H_{\vec a_i} (\vec x_i).
$$
\end{definition}
Similar to the univariate Hermite polynomials, 
the coefficients in a multivariate Hermite polynomial can also be bounded by the input dimension and the degree of it.
\begin{fact}
\label{fact:hermite-coefficient-bound}
Let $H_{\vec a}:\R^m \mapsto \R$ be a multivariate Hermite polynomial with $\snorm{1}{\vec a} = d$.
We have that
$
\LC{ H_{\vec a} } \leq  (d+1)^d \; 2^{d^2}.
$    
\end{fact}
For any positive integer $d$, the set of multivariate Hermite polynomials $H_{\vec a}$ where $\snorm{1}{\vec a} \leq d$ forms an orthogonal basis of all polynomials of degree $d$.
Hence, we can always write a polynomial as a (unique) linear combination of Hermite polynomials. 
If we keep only the degree-$k$ Hermite polynomials in the linear decomposition, we obtain the $k$-th harmonic components of the polynomial.
\begin{definition}[Harmonic Components]
\label{def:harmonic-components}
Given a polynomial $p: \R^n \mapsto \R$ of degree $d$, suppose it admits the linear decomposition
$$
p(\vec x) = \sum_{ \snorm{1}{\vec a} \leq d } \vec T_{\vec a} \; H_{\vec a} (\vec x).
$$
For positive integer $k \leq d$, 
we use $p^{[k]}$ to denote the $k$-th harmonic component of $p$ defined as 
$$
p^{[k]}(\vec x) = \sum_{ \snorm{1}{\vec a} = k }
\vec T_{\vec a} \; H_{\vec a} (\vec x).
$$
We say that $p$ is harmonic of degree $k$ if
$p = p^{[k]}$.
\end{definition}
The main reason we are interested in harmonic polynomials is that the $L_2$-norms of their gradients is comparable to the $L_2$-norms of themselves.
\begin{fact}
\label{fact:gradient-norm-bound}
Let $p$ be a polynomial of degree $d$.
Then it holds
$$
\snorm{L_2}{ \nabla p }
\leq d \; \snorm{L_2}{p}
$$
with equality if and only if $p$ is harmonic.
\end{fact}
We will use the following addition formula for multivariate Hermite polynomials.
\begin{fact} \label{fact:hermite-addition-formula}
Given $a, b > 0$ such that $a^2 + b^2 = 1$ 
and the Hermite polynomial $H_{\vec s}: \R^n \mapsto \R$, we have
\begin{align*}
H_{\vec s}( a \vec x + b \vec y )  
= 
\sum_{ \vec k \leq \vec s }
\sqrt{ \binom{\vec s}{\vec k} }
a^{ \abs{\vec s - \vec k} }
b^{\abs{\vec k}}
H_{\vec s - \vec k}(  \vec x)
\; H_{\vec k}(\vec y).
\end{align*}
\end{fact}
\begin{proof}
For $x,y \in \R$ and a univariate Hermite polynomil $H_d$, by the Taylor-expansion of $H_d$, we have
\begin{align}
\label{eq:univariate-addition}
H_d( ax + by ) = \sum_{ k \leq d } \sqrt{ \binom{d}{k} } a^{ d - k }  b^{ k } H_{d - k}(x) H_k(y).    
\end{align}
For the multivariate Hermite polynomial $H_{\vec s}$, we can first express it as a product of univariate Hermite polynomials, i.e., $H_{\vec s}(\vec x) = \prod_{i=1}^n H_{\vec s_i}(\vec x_i)$, and apply \Cref{eq:univariate-addition} for each of the univariate Hermite polynomial:
$$
H_{\vec s}(a \vec x + b \vec y)
= \prod_{i=1}^n \sum_{ \vec k_i \leq \vec s_i }
\sqrt{ \binom{\vec s_i}{\vec k_i} } a^{ \vec s_i - \vec k_i }
b^{\vec k_i} H_{ \vec s_i - \vec k_i }(\vec x_i) H_{\vec k_i}(\vec y_i)
= 
\sum_{ \vec k \leq \vec s }
\sqrt{ \binom{\vec s}{\vec k} }
a^{ \abs{\vec s - \vec k} }
b^{\abs{\vec k}}
H_{\vec s - \vec k}(  \vec x)
\; H_{\vec k}(\vec y).
$$
This concludes the proof of \Cref{fact:hermite-addition-formula}.
\end{proof}
     
\subsection{Norms of Polynomials}
One useful fact is that the $L_p$-norm is monotonically increasing as $p$ increases.
\begin{fact}[Monotonicity of $L_p$-Norms]
\label{fact:Lp-monotone}
Let $f:\R^n \mapsto \R$ be a real-valued function.
For $1 \leq t < k$, we have $\snorm{L_t}{f} \leq\snorm{L_k}{f}$.
\end{fact}
Under a Gaussian or log-concave distribution, hypercontractivity gives us the ``reverse'' of the above $L_p$-norm inequality.
\begin{lemma}[Hypercontractivity] \label{lem:hyper}
Let $p: \R^n \mapsto \R$ be a degree-$d$ polynomial and $t \geq 2$. 
Then it holds
$$
\snorm{L_t}{p} \leq \sqrt{t-1}^d \snorm{L_2}{p}.
$$
Let $D$ be a log-concave distribution on $\R^n$. Then, it holds
$$
\snorm{D, L_t}{p} \leq O(t)^d \snorm{D, L_2}{p}.
$$
\end{lemma}
\begin{proof}
The first inequality follows from Theorem 2 of \cite{nelson1973free}.
The second inequality follows by applying Theorem 7 from \cite{CW:01} with $q = d\; t$ and $r = 2d$.
\end{proof}
As a simple corollary, $L_2$-norm itself can be further upper bounded by the $L_1$-norm.
\begin{claim}[Upper bound of $L_2$-norm in $L_1$-norm]
\label{clm:l2-l1-bound}
Let $f$ be a degree-$d$ polynomial. Then it holds
$$
\snorm{L_2}{f} \leq 3^{4d/3} \snorm{L_1}{f}.
$$
\end{claim}
\begin{proof}
Given two functions $h, g: \R^n \mapsto \R$, define $w (\vec x) = h(\vec x) \; g(\vec x)$.
Hölder's inequality then says
$$
\snorm{L_1}{w} \leq \snorm{L_t}{h} \snorm{L_k}{g}
$$
for any $t, k$ satisfying $1/t + 1/k \leq 1$.
Using Hölder's inequality with  $1/t = 1/3$ and $1/k = 2/3$ yields
\begin{align*}
\snorm{L_2}{f}^2 
&= 
\E_{ \vec x \sim \normal(\vec 0, \vec I) }    
\lp[ 
\abs{f(\vec x)}^{4/3} \; \abs{f(\vec x)}^{2/3}
\rp]
\leq 
\E_{ \vec x \sim \normal(\vec 0, \vec I) }    
\lp[ f(\vec x)^4 \rp]^{1/3}
\; 
\E_{ \vec x \sim \normal(\vec 0, \vec I) }    
\lp[ \abs{f(\vec x)}\rp]^{2/3} \\
&= \snorm{L_4}{f}^{4/3} \; \snorm{L_1}{f}^{2/3}.
\end{align*}  
Applying \Cref{lem:hyper} then gives
\begin{align*}
\snorm{L_2}{f}^2 
\leq 
3^{d/2} \; 
\snorm{L_2}{f}^{4/3} \; \snorm{L_1}{f}^{2/3},
\end{align*}
which implies our claim.
\end{proof}
We will often be dealing with the gradient of a real-valued polynomial. As it turns out, the $L_2$-norm of the gradient (as a vector-valued polynomial) can be bounded above by the $L_2$-norm of the real-valued polynomial up to a factor of the degree of the polynomial.
\begin{fact}[Lemma 9 of \cite{Kane12}]
\label{lem:derivative-norm-bound}
Let $p: \R^n \mapsto \R$ be a degree-$d$ polynomial.
We then have the bound
$
\E_{\vec x \sim \normal(\vec 0, \vec I)}
\lp[ 
\snorm{2}{\nabla p(\vec x)}^2
\rp]
\leq d \snorm{L_2}{p}^2 
$
with equality if and only if $p$ is harmonic.
\end{fact}

\subsection{(Anti-)Concentration Properties of Polynomials} 
\label{sec:polynomial-concentration}
We will use the following anti-concentration theorem about polynomials under Gaussian and log-concave distributions.
\begin{lemma}[Anti-concentration of Real-valued Polynomials from \cite{CW:01}]
\label{lem:CW-concentration}
Let $p:\R^n \mapsto \R$ be a degree-$d$ polynomial.
Let $D$ be either the standard Gaussian or a log-concave distribution.
Then for any $t > 0$, we have
$$
\Pr_{ \vec x \sim D }
\lp[ \abs{ p(\vec x) > t \snorm{D, L_2}{p} } \rp]
\leq O \lp( d t^{1/d} \rp).
$$
\end{lemma}
Low-degree polynomials under Gaussian and log-concave distributions satisfy fairly strong tail bounds. This is one of the fundamental properties leveraged by our algorithm.
\begin{lemma}[Concentration of Real-valued Polynomials]
\label{lem:polynomial-concentration}
Let $p:\R^n \mapsto \R$ be a degree-$d$ polynomial.
Under the Gaussian distribution,
for any $t > 0$, we have
$$
\Pr_{ \vec x \sim \normal(\vec 0, \vec I) }
\lp[ \abs{ p(\vec x) > t \snorm{L_2}{p} } \rp]
\leq O \lp( 2^{ - (t/2)^{2/d} } \rp).
$$
Let $D$ be a log-concave distribution.
Under $D$, for any $t > 0$, we have the following slightly weaker tail bound
$$
\Pr_{ \vec x \sim D }
\lp[ \abs{ p(\vec x) > t \snorm{D, L_2}{p} } \rp]
\leq O \lp( \exp\lp(  - \Omega(t)^{1/d} \rp) \rp).
$$
\end{lemma}
\begin{proof}
Both of the above follow from the corresponding hypercontractivity property (\Cref{lem:hyper}) and Markov's inequality.
\end{proof}
\Cref{lem:polynomial-concentration} can also be generalized to vector-valued polynomials.
As vector-valued polynomials only appear in our analysis under the Gaussian distribution, the rest of the lemmas in this section will only be stated with respect to the Gaussian distribution (though similar statements can be derived under log-concave distributions as well).
\begin{lemma}[Norm Concentration for Vector-valued Polynomials]
\label{lem:vector-norm-concentraion}
Let $\vec f: \R^n \mapsto \R^n$ be a vector-valued polynomial of degree $d$.
Then we have
$$
\Pr_{ \vec x \sim \normal(\vec 0, \vec I) }
\lp[ 
\snorm{2}{ \vec f(\vec x) } < 
O_d(1)  \log^{d/2}(1/\delta)
\snorm{L_2}{ \vec f }
\rp]
\geq 1 - \delta.
$$
\end{lemma}
\begin{proof}
Define the real-valued polynomial $g: \R^n \mapsto \R$ as
$g(\vec x) = \snorm{2}{ \vec f(\vec x) }^2$.
We then have
$$
\snorm{L_1}{g} = \E_{\vec y \sim \normal(\vec 0, \vec I)}
\lp[ \snorm{2}{ \vec f(\vec x) }^2 \rp].
$$
By \Cref{clm:l2-l1-bound}, we have that
$\snorm{L_2}{g} \leq O_d(1) \snorm{L_1}{g}$.
As $g$ is a degree-$2d$ polynomial, 
\Cref{lem:polynomial-concentration} gives that
\begin{align*}
    \Pr_{ \vec x }\lp[ 
    \abs{g(\vec x)} < O \lp( \log^{d}(1/\delta) \rp) \snorm{L_2}{g}
    \rp]
    \geq 1 - \delta \, ,
\end{align*}
which implies that
$$
    \Pr_{ \vec x }\lp[ 
    \snorm{2}{ \vec f(\vec x) }^2 < O_d(1) \;  \log^{d}(1/\delta)
    \E_{\vec y \sim \normal(\vec 0, \vec I)}
    \lp[ \snorm{2}{ \vec f(\vec x) }^2 \rp]
    \rp]
    \geq 1 - \delta.
$$
Taking the square root of the inequality describing the probability event then concludes the proof.
\end{proof}
Consequently, gradients of real-valued polynomials satisfy good tail bounds as well.
\begin{lemma}[Norm Concentration of Gradients]
\label{lem:gradient-norm-concentration}
Let $f: \R^n \mapsto \R$ be a real-valued polynomial of degree $d$ and $\delta > 0$ be sufficiently small given $d$.
Then, we have
\begin{align*}
\Pr_{ \vec x \sim \normal(\vec 0, \vec I) }    \lp[ 
\snorm{2}{ \nabla f(\vec x) } < \log^{d/2}(1/\delta) \snorm{L_2}{f}
\rp] \geq 1 - \delta.
\end{align*}
\end{lemma}
\begin{proof}
Notice that $\nabla f(\vec x)$ is a degree at most $(d-1)$ vector-valued polynomial.
When $\delta$ is sufficiently small given $d$, we have $O_d(1) \; \log^{d-1}(1/\delta) \leq \log^d(1/\delta)$.
The proof then follows from \Cref{lem:vector-norm-concentraion} and \Cref{lem:derivative-norm-bound}.
\end{proof}
In fact, we can take a step further: derivatives of a vector-valued polynomial $\vec f: \R^n \mapsto \R^m$, i.e., the Jacobian $\Jac_{\vec f}: \R^n \mapsto \R^{n \times m}$, are also well-concentrated.
In particular, if $\vec f$ has bounded $L_2$-norm,  the Frobenious norm of its Jacobian matrix is also bounded with high probability.
\begin{lemma}
\label{lem:Jacobian-Frob-concentration}
Let $\vec f: \R^n \mapsto \R^m$ be a vector-valued polynomial of degree $d$ and $\delta > 0$ be sufficiently small given $d$.
Then it holds
$$
\Pr_{ \vec x \sim \normal(\vec 0, \vec I) }
\lp[ 
\snorm{F}{\Jac_{\vec f}(\vec x)}
\leq  
\log^{d/2}(1/\delta) \; \snorm{L_2}{\vec f}
\rp] \geq 1 - \delta.
$$
\end{lemma}
\begin{proof}
Notice that the $i$-th row of $\Jac_{\vec f}(\vec x)$ is exactly the gradient vector $\nabla \vec f_i(\vec x)$.
Using \Cref{lem:derivative-norm-bound}, we have
$$
\E_{\vec x \sim \normal(\vec 0, \vec I)}\lp[ \snorm{2}{ \nabla \vec f_i(\vec x) }^2 \rp]
\leq d \snorm{L_2}{\vec f_i}^2.
$$
Define the real-valued polynomial $g: \R^d \mapsto \R$ as 
$g(\vec x) = \snorm{F}{\Jac_{\vec f}(\vec x)}^2$.
It follows that
$$
\snorm{L_1}{ g }
= \sum_{i=1}^m 
\E_{\vec x \sim \normal(\vec 0, \vec I)}\lp[ \snorm{2}{ \nabla \vec f_i(\vec x) }^2 \rp]
\leq d \sum_{i=1}^m \snorm{L_2}{\vec f_i}^2 = 
d \snorm{L_2}{ \vec f }^2.
$$
Notice that each entry in $\Jac_{\vec f}(\vec x)$ is a degree at most $(d-1)$ polynomial since it is some partial derivative of some entry of $\vec f$.
Hence, $g$ must be a degree at most $(2d-2)$ polynomial.
Using \Cref{clm:l2-l1-bound}, we have
$$
\snorm{L_2}{ g } \leq O_d(1) \; \snorm{L_2}{ \vec f }^2.
$$
Hence, using \Cref{lem:polynomial-concentration}, 
with probability at least $1 - \delta$, we have
\begin{align*}
\snorm{F}{\Jac_{\vec q}(\vec x)}^2
= g(\vec x)
\leq O_d(1) \; \log^{d - 1}(1/\delta)  \; \snorm{L_2}{\vec f}^2
\leq \log^{d}(1/\delta) 
\; \snorm{L_2}{\vec f}^2
\, ,
\end{align*}
where the last inequality holds as long as $\delta$ is sufficiently small given $d$.
Taking the square root of both sides then concludes the proof.
\end{proof}

We will sometimes define new polynomials by restricting a subset of the input variables to some fixed values.
We show that the $L_2$-norm of the restricted polynomial with high probability will not increase by too much if we choose the values of the restricted inputs randomly.
\begin{lemma}
\label{lem:restricted-polynomial-norm}
Let $\vec f: \R^n \times \R^n \mapsto \R^m$ be a vector-valued polynomial of degree $d$. 
For each $\vec x \in \R^n$, 
define the restricted function $\vec f_{\vec x}: \R^n \mapsto \R^m$ as $\vec f_{\vec x}(\vec y) = \vec f(\vec x, \vec y)$.
Then we have
$$
\Pr_{ \vec x \sim \normal(\vec 0, \vec I) }
\lp[ 
\snorm{L_2}{ \vec f_{\vec x} } \leq O_d(1) \; \log^{d/2}(1/\delta) 
\snorm{L_2}{\vec f}
\rp] \geq 1- \delta.
$$
\end{lemma}
\begin{proof}
Define the degree-$2d$ real-valued polynomial
$g(\vec x) = \E_{ \vec y \sim \normal(\vec 0, \vec I) } \lp[ 
\snorm{2}{\vec f_{\vec x}(\vec y)}^2
\rp]$.
We have
$$
\snorm{L_1}{g}
= \E_{\vec x, \vec y \sim \normal(\vec 0, \vec I)}
\lp[ \snorm{2}{ \vec f(\vec x, \vec y) }^2 \rp].
$$
Using \Cref{clm:l2-l1-bound}, we have
$$
\snorm{L_2}{ g }
\leq O_d(1) \; \E_{\vec x, \vec y \sim \normal(\vec 0, \vec I)}
\lp[ \snorm{2}{ \vec f(\vec x, \vec y) }^2 \rp].
$$
Hence, by \Cref{lem:polynomial-concentration}, 
with probability at least $1-\delta$ over the randomness of $\vec x \sim \normal(\vec 0, \vec I)$,
it holds
\begin{align*}
\E_{ \vec y \sim \normal(\vec 0, \vec I) } \lp[ 
\snorm{2}{\vec f_{\vec x}(\vec y)}^2
\rp] 
&= g(\vec x) 
\leq O_d(1) \; \log^{d}(1/\delta) \; 
\E_{\vec x, \vec y \sim \normal(\vec 0, \vec I)}
\lp[ \snorm{2}{ \vec f(\vec x, \vec y) }^2 \rp]
= O_d(1) \; \log^{d}(1/\delta) \snorm{L_2}{ \vec f }^2.
\end{align*}
Taking the square roots of both sides concludes the proof.
\end{proof}

\subsection{Low-dimensional, Low-degree Polynomials}
In \Cref{sec:polynomial-concentration}, we show that for a low-degree polynomial, its values and derivatives are not too much larger than its $L_2$-norm with high probability under the standard Gaussian distribution (and more generally under log-concave distributions).
This phenomenon has a much more intuitive explanation if the input space of the polynomial is small as well.
In particular, in low-dimensional space, most of the points under the Gaussian distribution will have relatively small $\ell_2$-norms.
In this subsection, we will directly bound the values and derivatives of a polynomial at these points in terms of the $\ell_2$-norms of the points and the $L_2$-norm of the polynomial. The arguments for these bounds are usually easy to establish for univariate polynomials and in general not much harder when the input space is low-dimensional.

Lying in the heart of these bounds is the following lemma that relates the $L_2$-norm of a polynomial to its maximum coefficient. This result will also be useful when we try to analyze the effect of a linear transformation on the $L_2$-norms of polynomials.
\begin{lemma}[Maximum Coefficient and $L_2$-norm]
\label{fact:coefficeint-norm}
Let $f:\R^m \mapsto \R$ be a real-valued polynomial of degree $d$.
Then it holds
$$
\LC{f} = \Theta_{d,m}(1) \; \snorm{L_2}{f}.
$$
\end{lemma}
\begin{proof}
We first give an upper bound of $ \snorm{L_2}{f}$ in terms of $ \LC{f} $.
Suppose $f(\vec x) = \sum_{ \vec s : \snorm{1}{\vec s} \leq d } \vec T_{\vec s} \vec x^{\vec s} $.
Using the definition and applying the triangle inequality gives us
\begin{align*}
\snorm{L_2}{ f }
= \snorm{L_2}{ \sum_{ \vec s : \snorm{1}{\vec s} \leq d } \vec T_{\vec s} \vec x^{\vec s}  }
\leq  \sum_{ \vec s : \snorm{1}{\vec s} \leq d } 
 \abs{\vec T_{\vec s}} \; \snorm{L_2}{  \vec x^{\vec s} }.
\end{align*}
Next notice that $\vec x^{\vec s}$ is a degree at most $d$ monomial. Therefore, its $L_2$-norm is simply the square root of some degree at most $2d$ moment of $\normal(\vec 0, \vec I)$, which is at most $O_d(1)$.
Besides there are at most $(m+1)^d$ non-zero terms in the coefficient tensor $\vec T$ (by counting the number of $\vec s$'s satisfying $\snorm{1}{\vec s} \leq d$).
Therefore, we have
$$
\snorm{L_2}{ f }
\leq O_d(1) \; \snorm{1}{ \coe(f) }
\leq O_{d,m}(1) \; \LC{f}.
$$
Then we give an upper bound of $ \LC{f}$ in terms of $\snorm{L_2}{f}$.
We start by showing this for some Hermite polynomial $H_{\vec a}$, where $\snorm{1}{\vec a} = d$.
By \Cref{fact:hermite-coefficient-bound}, we have that $\LC{H_{\vec a}} \leq O_d(1)$.
Since the $L_2$-norms of the (normalized) Hermite polynomials are all $1$, it follows that 
$\LC{H_{\vec a}} \leq O_d(1) \; \snorm{L_2}{H_{\vec a}}$.
Now for a general polynomial $f$, we can first expand it in the basis of Hermite polynomials:
$$
f(\vec x) = \sum_{ \vec a: \snorm{2}{\vec a} \leq d } \vec W_{\vec a}  H(\vec a)(\vec x) \, ,
$$
for some tensor $\vec W$ indexed by $\vec a$.
Then the coefficient tensor satisfies
$$
\coe(f) = \sum_{ \vec a: \snorm{2}{\vec a} \leq d } \vec W_{\vec a}  \coe(H_{\vec a}).
$$
Hence, using the triangle inequality for the $\snorm{\infty}{\cdot}$ norm on the equation above, it holds
\begin{align}
\label{eq:MC-upper-bound}
\LC{f}
\leq \sum_{ \vec a: \snorm{2}{\vec a} \leq d } \abs{\vec W_{\vec a}} \LC{H_{\vec a}}
\leq 
O_d(1) \;  \snorm{1}{ \vec W }.    
\end{align}
Using the fact that the Hermite polynomials are orthonormal under $\normal(\vec 0, \vec I)$, 
we note that the $L_2$-norm of $f$ can be written as
\begin{align}
\label{eq:L_2-f-W-equality}
\snorm{L_2}{f}^2
=
\E_{ \vec x \sim \normal(\vec 0 , \vec I) } \lp[ 
\lp( \sum_{  \vec a: \snorm{2}{\vec a} \leq d } \vec W_{\vec a} H_{\vec a}(\vec x) \rp)^2
\rp]
=  \snorm{2}{\vec W}^2.
\end{align}
Since $\vec W$ has at most $(m+1)^d$ non-zero terms (due to the fact that $f$ has degree $d$), we must have
\begin{align}
\label{eq:W-Cauchy}
\snorm{1}{\vec W} \leq (m+1)^{d/2} \snorm{2}{\vec W}.    
\end{align}
Combining Equations~\eqref{eq:MC-upper-bound}, \eqref{eq:L_2-f-W-equality} and \eqref{eq:W-Cauchy}, we then obtain 
$$
\LC{f} \leq O_{d,m}(1) \; \snorm{L_2}{f}.
$$
This completes the proof of \Cref{fact:coefficeint-norm}.
\end{proof}

Since the $L_2$-norm allows us to control the maximum coefficient of a polynomial, it is not hard to see that the value of the polynomial evaluated at points with bounded $\ell_2$-norm can be controlled by its $L_2$-norm as well.
\begin{corollary}
\label{lem:bounded-real-evaluation}
Let $f: \R^m \mapsto \R$ be a polynomial of degree $d$, where $m \geq d \geq 2$.
Then we have
$$
\abs{f(\vec x)} \leq  
O_{d,m}(1) \; 
\snorm{L_2}{f} \snorm{2}{\vec x}^d.
$$
\end{corollary}
\begin{proof}
Let $\vec T$ be the coefficient tensor of $f$. We can write
\begin{align*}
f^2(\vec x)
&=
\lp( \sum_{ \vec a \in \mathbbm N^m: \snorm{1}{\vec a} \leq d } \vec T_{\vec a}  \vec  x^{\vec a} \rp)^2
\leq 
\sum_{ \vec a \in \mathbbm N^m: \snorm{1}{\vec a} \leq d } \vec T_{\vec a}^2
\sum_{ \vec a \in \mathbbm N^m: \snorm{1}{\vec a} \leq d } \lp( \vec x^{ \vec a} \rp)^2 \\
&\leq O_{d,m}(1) \; \LC{f}^2 \; \sum_{ \vec a \in \mathbbm N^n: \snorm{1}{\vec a} \leq d } \lp( \vec x^{ \vec a} \rp)^2
\leq O_{d,m}(1) \;  \LC{f}^2 \; \snorm{2}{\vec x}^{2d} \, ,
\end{align*}
where we use Cauchy's inequality in the first inequality, in the second inequality we upper bound each $\vec T_{\vec a}^2$ by $\LC{f}^2$ and use the fact that there are at most $O_{d,m}(1)$ non-zero entries in $\vec T$, 
and in the third inequality we use the fact that there are at most $O_{d,m}(1)$ $\vec a \in \mathbbm N^m$ satisfying $\snorm{1}{\vec a} \leq d$ and that
$ \abs{ \vec x^{\vec a} } \leq \snorm{\infty}{\vec x}^d \leq \snorm{2}{\vec x}^d $.
By \Cref{fact:coefficeint-norm}, we have $\LC{f} \leq O_{d,m}(1) \snorm{L_2}{f}$. The result hence follows.
\end{proof}
\Cref{lem:bounded-real-evaluation} can also be generalized to vector-valued polynomials and their derivatives.
\begin{corollary}
\label{lem:bounded-vector-evaluation}
Let $\vec f: \R^m \mapsto \R^k$ be a vector-valued polynomial of degree $d$.
Then we have
$$
\snorm{2}{ \vec f(\vec x) } \leq  
O_{d,m}(1) \; 
\snorm{L_2}{\vec f} \; \snorm{2}{\vec x}^d.
$$
\end{corollary}
\begin{proof}
We have
\begin{align*}
    \snorm{2}{\vec f(\vec x)}^2
    &= \sum_{i=1}^k
    \vec f_i^2(\vec x)
    \leq O_{d,m}(1)  \; 
    \snorm{2}{\vec x}^{2d} \; \sum_{i=1}^k \; \snorm{L_2}{\vec f_i}^2 \\
    &= O_{d,m}(1) \; 
    \snorm{2}{\vec x}^{2d} \; 
    \snorm{L_2}{\vec f}^2 \, ,
\end{align*}
where in the inequality we use \Cref{lem:bounded-real-evaluation}.
Taking the square roots of both sides concludes the proof.
\end{proof}

\begin{corollary}
\label{lem:bounded-Jacobian-evaluation}
Let $\vec f: \R^m \mapsto \R^k$ be a vector-valued polynomial of degree $d$.
Then we have
$$
\snorm{F}{ \Jac_{\vec f}(\vec x) } \leq  
O_{d,m}(1) \; \snorm{L_2}{ \vec f } \; \snorm{2}{\vec x}^d.
$$
\end{corollary}
\begin{proof}
We have
\begin{align*}
\snorm{F}{ \Jac_{\vec f}(\vec x) }^2
&= \sum_{i=1}^k 
\snorm{2}{ \nabla \vec f_{i}(\vec x) }^2
\leq 
O_{d,m}(1) \; \snorm{2}{\vec x}^{2d} \; \sum_{i=1}^k
 \snorm{L_2}{ \vec \nabla \vec f_i }^2  \\
&\leq 
O_{d,m}(1) \; \snorm{2}{\vec x}^{2d} \; \sum_{i=1}^k
 \snorm{L_2}{ \vec f_i }^2 \\
 &=
O_{d,m}(1) \; \snorm{2}{\vec x}^{2d} \; d \; \snorm{L_2}{\vec f}^2 \;,
\end{align*}
where in the first inequality we use \Cref{lem:bounded-vector-evaluation}, 
and in the second inequality we use
\Cref{lem:derivative-norm-bound}.
\end{proof}
\Cref{lem:bounded-Jacobian-evaluation} allows us to control the derivatives of a polynomial evaluated at points inside an origin-centered ball of bounded radius, i.e., $\snorm{2}{\vec x} \leq \ell$. 
As a result, the polynomial is Lipschitz continuous with a reasonable Lipschitz constant inside the area.
\begin{corollary}
\label{lem:bounded-lipchitz}
Let $\vec f: \R^m \mapsto \R^k$ be a vector-valued polynomial of degree $d$ and $\ell > 0$.
Then, for $\vec x, \vec y$ satisfying $\max \lp( \snorm{2}{\vec x}, \snorm{2}{\vec y} \rp) \leq \ell$, 
we have
$$
\snorm{2}{ \vec f(\vec x) - \vec f(\vec y) }
\leq O_{d,m}(1) \; \ell^d \; 
\snorm{L_2}{\vec f} \; 
\snorm{2}{ \vec x - \vec y }.
$$    
\end{corollary}
\begin{proof}
We start by bounding the derivatives of $\vec f$.
For any $\vec z$ with $\snorm{2}{\vec z} \leq \ell$, we have
\begin{align*}
\sigma_{\max} \lp( \Jac_{\vec f}(\vec z) \rp)
\leq \snorm{F}{\Jac_{ \vec f }(\vec z) }
\leq O_{d,m}(1) \; \snorm{L_2}{\vec f} \; 
\ell^d \, ,
\end{align*}
where in the first inequality we use the fact that the induced vector of a matrix is upper bounded by the Frobenious norm of itself, and in the last inequality we use \Cref{lem:bounded-Jacobian-evaluation}.
Hence, $\vec f$
is $O_{d,m}(\snorm{L_2}{\vec f} \; \ell^d)$-Lipschitz-continuous within the domain $\{\vec z \in \R^m: \snorm{2}{\vec z} \leq \ell \}$.
The lemma hence follows.
\end{proof}

\section{Efficient Super Non-singular Extendible Decomposition}
\label{sec:efficient-decomposition}
In this section, we present the formal statement of \Cref{intro-prop:super-non-singular-extension} and its detailed proof.
\begin{theorem}[Extendible Super Non-singular Decomposition]
\label{prop:super-non-singular-extension}
Let $n,\ell,d,M$ be positive integers, 
$f:\mathbb Z^+ \mapsto \mathbb Z^+$ be some function, and $\eps > 0$ be sufficiently small given $\ell,d,M,f$.
Let $S:=\{q_1, \cdots, q_{\ell}\}$, where $q_i: \R^n \mapsto \R$, be a set of  degree at most $d$ harmonic polynomials and  $p: \R^n \mapsto \R$ be some other degree-$d$ polynomial.
Suppose $S$ is $(\eps^{1/3}, C_{d,\ell,f,M})$ super non-singular, where $C_{d,\ell,f,M}$ is sufficiently large given $d, \ell, f, M$.
Then \snew{there exists an algorithm which can
extend $S$} into $\bar S:= \{q_1, \cdots, q_\ell, q_{\ell+1}, \cdots, q_m \}$ such that
\begin{itemize}
    \item $\bar S$ is of size $m = O_{d,\ell,f,M}(1)$.
    \item $\bar S$ is 
    $(\eps, f(m))$-super non-singular.
    \item There exists a polynomial $h:\R^m \mapsto \R$ 
    with $\snorm{L_2}{ h } \leq \eps^{-3d-1}$ such that
    $$
    \snorm{L_2}{ p(\vec x) - h( q_1(\vec x), \cdots, q_m(\vec x) ) } 
    \leq \eps^M.
    $$
\end{itemize}
Moreover, the algorithm runs in time $\poly(n) \;
\poly_{d, \ell, M, f}(1/\eps)$.
\end{theorem}

We first give an outline of our algorithm.
The algorithm will start with the trivial decomposition 
$S = \{ q_1, \cdots, q_{\ell}, q_{\ell+1} = p \}$ and $h(q_1(\x), \cdots,q_{\ell}(\x), p(\x)) = p(\x)$.
For convenience, we will refer to $S$ as the \emph{primitive polynomials} and $h$ as the \emph{composition polynomial} in the decomposition.
If $S$ is already super non-singular, we are done.
Otherwise, we want to find a linear combination of the polynomials in $S$ such that the combined polynomial has small high-order derivatives with non-trivial probability.
Then we can decompose this combined polynomial into a number of lower-degree polynomials using tools developed in \cite{Kane12} (see \Cref{lem:polynomial-decomposition}).  
We next rewrite $p$ in terms of the other polynomials in the linear combination and the new lower-degree polynomials found.
This process will continue until a super non-singular decomposition is found.

We will now walk through one iteration of the algorithm in more detail.
Let $S$ be the current set of primitive polynomials that are not super non-singular.
As discussed in \Cref{sec:extendible-decomposition-technique}, in order for the decomposition to be extendible, 
one subtlety here is that we do not want to rewrite any of the polynomials $q_1, \cdots, q_{\ell}$. 
We will refer to them as the \emph{initial polynomials} and the others as the \emph{non-initial polynomials}.
Hence, we want to find a linear combination of the polynomials in $S$ such that there is some non-initial polynomial $q_j \in S$ which has non-trivial weight in the linear combination, i.e., the absolute value of the coefficient in front of $q_j$ is non-trivial.
Fortunately, by our assumption that the set of initial polynomials are super non-singular,
we know any linear combination of them has non-trivial high-order derivatives with high probability.
Besides, we also have that the derivatives of polynomials are well-concentrated under the Gaussian distribution.
Combining the two observations then allows us to conclude that, 
if a linear combination of polynomials from $S$ has small high-order derivatives, then there exists some non-initial polynomial from $S$ that has non-trivial weight in the linear combination.
The above discussion is summarized in the following lemma.
\begin{lemma}
\label{lem:small-gradient-partial-combination}
Let $N,m,\ell,d$ be positive integers and $\eps \in (0,1)$ such that $\eps$ is sufficiently small in $N,m,\ell,d$.
Let $S:=\{q_1, \cdots,q_\ell, q_{\ell+1}, \cdots, q_m\}$ be a set of harmonic polynomials that are of degree at most $d$.
Suppose that $S$ is not $(\eps, N)$-super non-singular but 
$\{ q_1, \cdots, q_\ell \}$ is $(\eps^{1/3}, 3(N+1))$-super non-singular.
Furthermore, suppose 
$\snorm{L_2}{ q_i } \leq 1$ for all $i \in [m]$.
Denote by $S_k$ the set of degree-$k$ harmonic polynomials in $S$.
Then there exists an algorithm that 
runs in time $\poly(n^d) \; \poly_{m,N}(1/\eps)$ and finds
a positive integer $k \in [d]$ and a vector $\vec a \in \R^m$ with $\sum_{i \in S_k} \vec a_i^2 \in (0.9, 1.1)$ such that 
\begin{itemize}
    \item 
The polynomial
$q(\vec x) = \sum_{i=1}^m \vec a_i \; q_i(\vec x)$ satisfies
\begin{align}
\label{eq:good-linear-combination}
\Pr_{\vec x \sim \normal(\vec 0, \vec I), \vec y^{(i)}\sim \normal(\vec 0, \vec I) \text{ for } 1 \leq i \leq k-1}
\lp [
\snorm{2}{
\nabla_{\vec x}
D_{\vec y^{(k-1)}}
\cdots
D_{\vec y^{(1)}}
q(\x)}  <\eps/2  \rp]
> \eps^{N}/2.    
\end{align}
\item There exists $j$ such that $j \in S_k$, $j > \ell$, and $\vec a_j > \sqrt{\eps}$.
\end{itemize}
\end{lemma}
\begin{proof}
Since $S$ is not super non-singular, 
there exists 
an integer $k$ and a vector $\vec a \in \R^m$ with $\sum_{i \in S_k} \vec a_i^2 = 1$ such that  
\begin{align}
\label{eq:not-super-nonsingular-def}
&\Pr_{\vec x \sim \normal(\vec 0, \vec I), \vec y^{(i)}\sim \normal(\vec 0, \vec I) \text{ for } 1 \leq i \leq k-1}
\lp [
\snorm{2}{
\nabla_{\vec x}
D_{\vec y^{(k-1)}}
\cdots
D_{\vec y^{(1)}}
\lp( \sum_{i \in S_k} \vec a_i q_i(\vec x) \rp)}  < \eps  \rp]
> \eps^{N}.
\end{align}
It remains to show that there exists some $j$ satisfying $j \in S_k$, $j > \ell$, and $\vec a_j > \sqrt{\eps}$.
In fact, we will show that there exist $j \in S_k$ and $j > \ell$ such that $\vec a_j > 2 \; \sqrt{\eps}$. 
Suppose that there is no such $j$ for the sake of contradiction.
Then we claim that 
\begin{align}
\label{eq:small-weight-small-contribution}   
\snorm{2}{
\nabla_{\vec x}
D_{\vec y^{(k-1)}}
\cdots
D_{\vec y^{(1)}}
\lp( \sum_{i \in S_k, i > \ell} 
\vec a_i q_i(\vec x)
\rp)
} \leq O_{d,m,N}( \sqrt{\eps} )
\end{align}
with probability at least $1 - \eps^{N+1}$.
To see this, we note that
$\snorm{L_2}{ \sum_{i \in S_k, i > \ell } \vec a_i q_i(\vec x) } \leq 2 \; (m - \ell) \; \sqrt{\eps}$ by our assumption that $\snorm{L_2}{q_i} \leq 1$ for all $i$ and $\vec a_j < 2 \; \sqrt{\eps}$.
By iteratively applying \Cref{fact:gradient-norm-bound},
this further implies that 
$\snorm{L_2}{ 
D_{\vec y^{(k-1)}}
\cdots
D_{\vec y^{(1)}}
\sum_{i \in S_k, i > \ell } \vec a_i q_i(\vec x) } \leq O_{d,m}\lp( \sqrt{\eps} \rp)$.
Then Equation~\eqref{eq:small-weight-small-contribution} follows from \Cref{lem:gradient-norm-concentration}.

On the other hand, since the set $\{ q_1, \cdots, q_{\ell} \}$ is $(\eps^{1/3}, 3(N+1))$-super non-singular, we have
\begin{align}
\label{eq:def-of-super-nonsingular}
\snorm{2}{
\nabla_{\vec x}
D_{\vec y^{(k-1)}}
\cdots
D_{\vec y^{(1)}}
\lp( \sum_{i \in S_k, i \leq \ell} 
\vec a_i q_i(\vec x)
\rp)
} > \eps^{1/3}    
\end{align}
with probability at least $1 - \eps^{N+1}$.

By the union bound, 
\Cref{eq:small-weight-small-contribution} and \Cref{eq:def-of-super-nonsingular} are both satisfied with probability at least $1 - \eps^{N+1} - \eps^{N+1} \geq 1 -\eps^N$.  
Conditioning on \Cref{eq:small-weight-small-contribution} and \Cref{eq:def-of-super-nonsingular}, 
we have
\begin{align*}
&\snorm{2}{
\nabla_{\vec x}
D_{\vec y^{(k-1)}}
\cdots
D_{\vec y^{(1)}}
\lp( \sum_{i \in S_k} \vec a_i q_i(\vec x) \rp)
} \\ 
&\geq
\snorm{2}{
\nabla_{\vec x}
D_{\vec y^{(k-1)}}
\cdots
D_{\vec y^{(1)}}
\lp( \sum_{i \in S_k, i \leq \ell} \vec a_i q_i(\vec x) \rp)
}
-
\snorm{2}{
\nabla_{\vec x}
D_{\vec y^{(k-1)}}
\cdots
D_{\vec y^{(1)}}
\lp( \sum_{i \in S_k, i > \ell} \vec a_i q_i(\vec x) \rp)
} \\
&\geq \eps^{1/3} - O_{d,m,N}(\eps^{1/2})
> \eps \, ,
\end{align*}
where in the first inequality we use the triangle inequality, in the second inequality we use our conditioning, and the last inequality is true as long as $\eps$ is sufficiently small given $d,m,N$.
This then contradicts \Cref{eq:not-super-nonsingular-def}.

The above shows the existence of some $\vec a$ and $k$ which satisfy the properties required by the lemma. Though it may be hard to find such an $\vec a$ exactly, we can nonetheless find some $\tilde {\vec a}$ that is $\eps^3$-close to $\vec a$ in $\ell_{\infty}$-norm by searching over all vectors whose coordinates are multiples of $\eps^3$ and lie in the range $[-2, 2]$. One can then verify that such a vector still satisfies the properties required by the lemma.
Moreover, for each $\tilde {\vec a}$, checking whether it satisfies the condition in \Cref{eq:good-linear-combination} can be done in time $\poly(n^d \; m/\eps^N)$.
Since there are at most $\poly_{m}(1/\eps)$ such vector $\tilde{\vec a}$, the overall runtime hence follows.
This concludes the proof of \Cref{lem:small-gradient-partial-combination}.
\end{proof}
As we have discussed, a critical step in obtaining a super non-singular decomposition is to decompose a high-degree polynomial into a set of lower-degree polynomials. To achieve this, it suffices to apply Proposition 10 shown in Section 3.2 of \cite{Kane12}.
Roughly speaking, the lemma shows that such a polynomial $q$ with small high-order derivatives can be approximately written as the sum of products of lower-degree polynomials such that the 
highest degree harmonic part of the error term has a small $L_2$ norm.
\begin{lemma}[Proposition 10 from \cite{Kane12}]
\label{lem:polynomial-decomposition}
Let $p(\vec x)$ be a degree-$d$ polynomial with $\snorm{L_2}{p} \leq 1$ and let $\eps, N >0$ be real numbers such that
$$
\Pr_{\vec x \sim \normal(\vec 0, \vec I), \vec y^{(i)}\sim \normal(\vec 0, \vec I) \text{ for } 1 \leq i \leq k-1}
\lp [
\snorm{2}{
\nabla_{\vec x}
D_{\vec y^{(k-1)}}
\cdots
D_{\vec y^{(1)}}
p(\vec x)}  <\eps  \rp]
> \eps^{N}.
$$
Let $c \in (0,1)$ be a real number. 
Then there exist $k = O_{N, c, d}(1)$ many harmonic polynomials 
$\{\alpha_i(\vec x),  \beta_i(\vec x)\}_{i=1}^k$ of degree 
at most $d-1$ such that
\begin{itemize}
\item 
Define $g(\x) = p(\x) - \sum_{i=1}^k \alpha_i(\x) \; \beta_i(\x)$.
Then it holds
$$
\snorm{L_2}{ g^{[d]} }
\leq O_{N, c, d} \lp( \eps^{1-c}\rp) \, ,
$$
where $g^{[d]}$  denotes the degree-$d$ harmonic component of $g$ (see \Cref{def:harmonic-components}).
\item The degrees of $\alpha_i, \beta_i$ are at most $d-1$ and they satisfy $\deg(\alpha_i) + \deg(\beta_i) = d$.
\item The norms of the lower-degree polynomials satisfy $ \snorm{L_2}{\alpha_i} \; \snorm{L_2}{\beta_i} \leq O_{N, c, d} \lp(  \eps^{-c} \;  \rp) \; \snorm{L_2}{p^{[d]}}$ for all $i \in [k]$.
\end{itemize}
Finally, such a decomposition can be computed in time $\poly(n^d) \; \poly_{c,d,N}(1/\eps)$.
\end{lemma}
Our algorithm throughout its execution maintains a data structure that we will call a \emph{partial decomposition}.
\begin{definition}[Partial Decomposition]
    A partial decomposition consists of the following objects:
\begin{itemize}
    \item A positive integer $m$.
    \item A sequence of polynomials $Q = (q_1, \cdots, q_m)$, to which we refer as the primitive polynomials, with $\snorm{2}{ q_i } = 1$ for all $i$. We also require each polynomial to be non-constant.    
    Over these primitive polynomials, we say $q_1, \cdots, q_{\ell}$ are the initial polynomials and they are left untouched throughout the process.
    \item A set of rational numbers $(b_1, \cdots, b_m)$, where each $b_i$ is 
    a multiple of $1/6$ between $0$ and $M + 3d$.
    We say that $b_i$ is the \emph{magnitude coefficient} of the primitive polynomial $q_i$.
    For the initial polynomials, their magnitude coefficients are always fixed to be $0$.
    \item A composition polynomial $h: \R^m \mapsto \R$. We require that for each monomial $\prod_{ i=1 }^m \vec x_i^{ \vec \alpha_i } $ in $h$, it holds
    $\sum_{i=1}^m \alpha_i \text{deg}(q_i) \leq d$. The purpose of the composition polynomial is that one can approximately recover the target polynomial $p$ by computing
    $$
    h( \eps^{ b_1} q_1(\x), \cdots, \eps^{ b_m} q_m(\x)   ).
    $$
\end{itemize}    
\end{definition}
When the primitive polynomials in the current partial decomposition are not already super non-singular, we will employ \Cref{lem:small-gradient-partial-combination} and \Cref{lem:polynomial-decomposition} to
replace some non-initial primitive polynomials, i.e., some $q_j$ for $j > \ell$,  with a set of new polynomials with lower degrees.
Such a replacement may create a number of complications. 
Specifically, the number of the primitive polynomials may increase, the composition polynomial $h$ may become more complex, and the error between the target polynomial $p$ and $h( \eps^{b_1} q_1, \cdots, \eps^{b_m} q_m )$ may grow larger.
Throughout the iterative process, we need these extra ``complexities'' to be bounded (otherwise, even storing the decomposition would be intractable). 
We hence define an extra complexity parameter for the decomposition to keep track of the above complications. 
\begin{definition}[Complexity of Partial Decomposition]
We say that the partial decomposition has complexity $C > 0$ if the following holds:
\begin{itemize}
    \item $m \leq C$.
    \item $\snorm{L_2}{h} \leq C \eps^{-3d}$.
    \item $p(\vec x) = h( \eps^{b_1} q_1(  \vec x), \cdots, \eps^{b_m} q_m(\vec x) ) + \xi(\vec x)$, where $\xi: \R^b \mapsto \R$ is an error polynomial with $\snorm{L_2}{\xi} \leq C \eps^M$. We say that the $L_2$-norm of $\xi$ is the error of the partial decomposition.
\end{itemize}    
\end{definition}  
We now give some intuition of why we will make progress after one iteration of replacement.
The high-level idea is that if we repeat the process, we will eventually run out of high-degree polynomials (at least their weights become increasingly small). Hence, the process must eventually terminate and give us a super non-singular decomposition.
To formalize this intuition, we need to carefully define the ``potential'' of a decomposition. Then we show that this potential decreases monotonically.

Naively, one may want to define the potential of a decomposition as the sum of the magnitude coefficients $b_i$, since these numbers intuitively represent the fraction of each polynomial within the decomposition.  
However, decreasing the magnitude of one high-degree polynomial may introduce many additional low-degree polynomials. 
To make sure the potential is monotonically decreasing after such changes, we need to define the potential to be a weighted sum of $b_i$'s.
Specifically, the weights of $b_i$'s associated with high-degree polynomials will be much larger than those of $b_i$'s associated with low-degree polynomials.

Unfortunately, defining the potential as a sum weighted by \emph{any finite numbers} is still insufficient for the argument to go through.
Similar to the argument for diffuse decompositions in \cite{Kane12},
a technical issue is that 
we are not able to bound from above the number of additional polynomials introduced by a fixed number throughout the process.
In fact, the best bound we can put on the increment scales with the complexity parameter $C$ of the current decomposition. 
To show such a process will eventually terminate, we need to define the weights to be \emph{transfinite ordinal numbers}. 
In particular, for $b_i$ associated with a degree-$k$ primitive polynomial, we will weight it by $\omega^{ k }$, where $\omega$ is the ordinal number defined to be larger than any finite number.
\begin{definition}[Potential of Partial Decomposition]
 The potential of a partial decomposition is defined to be a transfinite number --- a function of the ordinal number $\omega$ (which is bigger than all finite numbers but smaller than $\omega + 1$), specifically
\begin{align*}
\sum_{i=1}^m \omega^{ \text{deg}(q_i) }
\lp( M + 3d - b_i \rp).
\end{align*}    
\end{definition}
Then each replacement indeed causes the potential to become a smaller ordinal number and we can conclude via transfinite induction that the process will eventually terminate.
As a result, the number of iterations needed may only be bounded by some Ackerman-type function of the maximum degree and the number of initial polynomials.
\begin{proof}[Proof of \Cref{prop:super-non-singular-extension}]
Our goal is to find a decomposition 
of the target polynomial $p$ such that the primitive polynomials in the decomposition form an $(\eps, f(m))$-super non-singular set, where $m$ is the size of the set, that
extends the original set of polynomials $\{q_1, \cdots, q_{\ell}\}$, which is already $(\sqrt{\eps}, C_{d,\ell,f,M}(1))$-super non-singular. 
We start with an initial partial decomposition with the primitive polynomials being $\{ q_1, \cdots, q_{\ell}, q_{\ell+1} = p\}$, the composition polynomial being $h( \vec y_1, \cdots, \vec y_{\ell}, \vec y_{\ell+1} ) =  \vec y_{\ell+1}$, and the sequence of magnitude coefficients $b_i$ being
all $0$.

Given a partial decomposition of potential $w$ and complexity $C$, we will argue that either the set of primitive polynomials is sufficiently super non-singular already or we can modify it to obtain another partial decomposition with complexity bounded from above by $O_{d,C,f,M}(1)$ and strictly smaller potential.
Moreover, the original primitive polynomials $q_1, \cdots, q_\ell$ are left untouched in the new decomposition. 
Then we can conclude via transfinite induction that this process will terminate in $O_{d, \ell, f, M}(1)$ many steps and the final decomposition obtained will have its complexity bounded by $O_{d, \ell, f, M}(1)$.
Consequently, the size of the final decomposition will be at most some number $m \leq C = O_{d, \ell, f, M}(1)$.
We will argue that each modification can be done in time $
\poly(n^d) \; \poly_{d, C, f, M}(1/\eps)$, from which our overall runtime follows.

We proceed to show the inductive step. 
In the new decomposition, we may increase some magnitude coefficient $b_i$ by $1/6$. 
Since these numbers are required to be between $0$ and $M+3d$, we will perform a preprocessing step to make sure that $b_i \leq M + 3d - 1/6$.
Notice that if $b_i = N + 3d$, 
the sum of coefficients of $q_i$ appearing in 
$$
h( \eps^{b_1} q_1, \cdots, \eps^{b_m} q_m )
$$
should be at most $ O_C( \eps^{ M+3d } \; \snorm{L_2}{ h } ) \leq O_C( \eps^M) $.
Hence, one can create a new partial decomposition by removing these primitive polynomials and the error introduced will be at most $O_C(\eps^M)$.
After doing so, we can assume without loss of generality that $b_i < M + 3d$.

Suppose that the current partial decomposition is of size $m$. We will set $N = f(m)$, and $c$ to be some number that is sufficiently small given $N,m,C,d$.
By \Cref{lem:small-gradient-partial-combination}, we can find an integer $k$ and a vector $\vec a \in \R^m$ as specified in the lemma.
Moreover, this can be done in time $\poly(n^d) \; \poly_{m, N}(1/\eps) = \poly(n^d) \; \poly_{f,C}(1/\eps)$.
Let $S_t$ be the set of degree-$t$ harmonic polynomials within the set of primitive polynomials. 
Define the polynomial $q:\R^n \mapsto \R$ as
\begin{equation} \label{eq:linear-combination-polynomial}
q(\vec x) = \sum_{i \in S_k}
\vec a_i q_i(\vec x).    
\end{equation}
We therefore have
$$
\Pr_{\vec x \sim \normal(\vec 0, \vec I), \vec y^{(i)}\sim \normal(\vec 0, \vec I) \text{ for } 1 \leq i \leq k-1}
\lp [
\snorm{2}{
\nabla_{\vec x}
D_{\vec y^{(k-1)}}
\cdots
D_{\vec y^{(1)}}
q(\vec x)}  <\eps/2  \rp]
> \eps^{N}\, ,
$$
and that there exists $j$ such that
$j \in S_k$, 
$j  > \ell$, and $\vec a_j > \sqrt{\eps}$.
Hence, we can apply \Cref{lem:polynomial-decomposition} to $q$. This shows that we can efficiently find polynomials $\{ \alpha_i, \beta_i \}_{i=1}^{\kappa}$, where $\kappa = O_{c,N,d}(1)$, such that the following holds:
\begin{enumerate}
    \item Each polynomial $\alpha_i, \beta_i$ is of degree at most $k-1$.
    \item $\deg(\alpha_i) + \deg(\beta_i) = \deg(q) = k$.
    \item $\snorm{L_2}{ \alpha_i } \; \snorm{L_2}{\beta_i} \leq O_{c,N,d}(\eps^{-c})$.
    \item 
    There exist two error polynomials $e, r$ with $\snorm{L_2}{e} \leq O_{c,N,d}(\eps^{1-c})$, $\deg(e) \leq k$ and $\snorm{L_2}{r} \leq 
    O_{c,N,d}(\eps^{-2c})$, $\deg(r) \leq k-1$,
    such that
    \begin{align}
    \label{eq:q-low-degree-decompose}
    q(\vec x) -  \sum_{i=1}^{\kappa} \alpha_i(\vec x) \beta_i(\vec x) = e(\vec x) + r(\vec x).         
    \end{align}
\end{enumerate}
By \Cref{lem:polynomial-decomposition}, this can be done in time $\poly(n^d) \; \poly_{c,d,N}(1/\eps) = \poly(n^d) \; \poly_{f,C,d}(1/\eps)$.
Denote by $\bar e, \bar r, \bar \alpha_i, \bar \beta_i$ the normalized versions of the corresponding polynomials.
Our first step is to replace some non-original primitive polynomial $q_j$ with  
$\bar e , \bar r, \{\bar \alpha_i, \bar \beta_i\}_{i=1}^{\kappa}$. In particular, the new set of primitive polynomials is set to
$$
q_1' = q_1, \cdots, \underline{q_j' = \bar e},
q_{m}' = q_m, \underline{q_{m+1}' = \bar \alpha_1, q_{m+2}' = \bar \beta_1, \cdots, 
q_{m+ 2\kappa - 1} = \bar \alpha_{\kappa},
q_{m+2\kappa} = \bar \beta_{\kappa},
q_{m + 2 \kappa + 1}' = \bar r}.
$$
Notice that $\kappa = O_{c,N,d}(1) = O_{ f, C ,d }(1)$. This ensures the
size of the primitive polynomial will not increase by too much.
The polynomial $q_j$ to be replaced is selected using the following rules:
\begin{itemize}

    \item $j \in S_k$ and $j > \ell$.
    \item $\vec a_j > \sqrt{\eps}$.
    \item $ \vec a_j \; \eps^{b_j} $ is the largest among $S_k$.
\end{itemize}
The selection rule will be important when we analyze the complexity of the new decomposition.

We will next update the magnitude coefficient array $b_1, \cdots,b_{m+2k+1}$:
the magnitude coefficients of the lower degree polynomials $\bar \alpha_i, \bar \beta_i, \bar r$ are set to $0$, and the magnitude coefficient of $\bar e$ is set to $b_j + 1/6$.
Notice that $q_j$ is of degree $k$, 
$e$ is of degree $k$, and 
$\alpha_i, \beta_i, r$ are of degree $k-1$.
Hence, we know in the worst case the potential of the decomposition will change by
\begin{align*}
&- \omega^k \lp(  M+3d - b_i \rp)
+ \omega^{k} \lp( 
M+3d - b_i - 1/6
\rp)
+ \omega^{k-1} O_{d,C,f,M}(1) \\
&= - \omega^k/6 
+ \omega^{k-1} O_{d,C,f,M}(1).
\end{align*}
It is not hard to see that the first term is dominating since $\omega$ is larger than any finite number.
This shows that the weight of the decomposition strictly decreases.

Then we will update the composition polynomial.
Using \Cref{eq:linear-combination-polynomial}, 
we know that the term $\eps^{b_j}
q_j(\vec x)$, which is fed as an input to the original composition polynomial $h$,
can be rewritten as
$$
\vec a_j^{-1} \eps^{b_j}   \; q(\vec x) - 
\sum_{i \in S_k, i\neq j}
\vec a_j^{-1} \eps^{b_j} \vec a_i q_i(\vec x).
$$
The polynomial $q$, due to \Cref{eq:q-low-degree-decompose}, can be expressed in $\bar \alpha_i, \bar \beta_i, \bar e, \bar r$, which gives us
\begin{align*}
&\sum_{i=1}^{\kappa} \lp( \vec a_j^{-1} \eps^{b_j} \;
 \snorm{L_2}{ \alpha_i }
\snorm{L_2}{ \beta_i } \rp) \; 
 \bar \alpha_i(\vec x) \bar \beta_i(\vec x)
+ 
\lp( \vec a_j^{-1}  \eps^{-1/6} \snorm{L_2}{ e }
\rp) \;
\lp( \eps^{b_j+1/6} 
\bar e(\vec x) \rp)
+ 
\lp( \vec a_j^{-1}  \eps^{b_j}
\snorm{L_2}{ r } \rp)
\bar r(\vec x) \\
&- 
\sum_{i \in S_k, i\neq j}
\lp( \vec a_j^{-1} \eps^{b_j - b_i} \vec a_i \rp)
\;
\lp( \eps^{b_i}  q_i(\vec x) \rp) \;.
\end{align*}
Recall the new set of primitive polynomials is set to be
$$
q_1' = q_1, \cdots, \underline{q_j' = \bar e},
q_{m}' = q_m, \underline{q_{m+1}' = \bar \alpha_1, q_{m+2}' = \bar \beta_1, \cdots,  q_{m+ 2\kappa - 1} = \bar \alpha_{\kappa},
q_{m+2\kappa} = \bar \beta_{\kappa},
q_{m + 2 \kappa + 1}' = \bar r}.
$$
Hence, it suffices to replace every occurrence of the $j$-th variable $\vec y_j$ in the composition polynomial $h$ from last iteration with the expression
\begin{align} \label{eq:replacement}
\sum_{i=1}^{\kappa} \gamma_i \; 
 \vec y_{m + 2i - 1}
 \vec y_{m + 2i}
+ 
\theta
\vec y_{j}
+ 
\zeta
\vec y_{m + 2 \kappa + 1}
+ 
\sum_{i \in S_k, i \leq \ell }
\lambda_i
\vec y_i
+ 
\sum_{i \in S_k, i\neq j, i > \ell }
\iota_i
\vec y_i \, ,    
\end{align}
where the coefficients are defined as
\begin{align*}
    &\gamma_i := \vec a_j^{-1} \eps^{b_j} 
    \snorm{L_2}{\alpha_i} \snorm{L_2}{\beta_i} \text{ indexed by } 1 \leq i \leq \kappa \, ,  \\
    &\theta := \vec a_j^{-1} \eps^{-1/6} \snorm{L_2}{e} \, , \,
    \zeta := \vec a_j^{-1} \eps^{b_j} \snorm{L_2}{r} \, , \\
    & \lambda_i := \vec a_j^{-1} \eps^{b_j - b_i} \vec a_i
    \text{ indexed by } i \in S_k, i \neq j.
\end{align*}
It is easy to see that such a modification on the partial decomposition can be done in time $\poly(n^d) \; \poly(m + 2\kappa + 1)$.
Since we are replacing $\eps^{b_j} q_j(\vec x)$ with an equivalent expression, the error of the new decomposition stays the same.

Let $h'$ be the new composition polynomial obtained after the replacement.
It remains to argue $h'$ has its $L_2$-norm bounded by $ C' \eps^{-3d}$ for some $C' = O_{C,f,M,d}(1)$.
Due to \Cref{fact:coefficeint-norm}, we know it suffices to argue that
$$
\LC{ h' } \leq \eps^{-3d}.
$$
Consider a monomial in $h$ of the form
$$
A:= \rho \prod_{ i } \vec y_i^{\tau_i}.
$$
Each occurrence of $\vec y_j$ in $A$ will be replaced with the expression in
\Cref{eq:replacement}.
In other words, the monomial will be replaced with the following expression in $h'$.
\begin{align} \label{eq:expansion}
\rho \prod_{ i \neq j} \vec y_i^{\tau_i}
\lp( \sum_{i=1}^{\kappa} \sqrt{\gamma_i}
 \vec y_{m + 2i - 1} \; 
 \sqrt{\gamma_i}
 \vec y_{m + 2i}
+ 
\theta
\vec y_{j}
+ 
\zeta
\vec y_{m + 2 \kappa + 1}
+ 
\sum_{i \in S_k, i \leq \ell }
\lambda_i
\vec y_i
+ 
\sum_{i \in S_k, i\neq j, i > \ell }
\iota_i
\vec y_i
\rp)^{\tau_j}.
\end{align}
We first derive an upper bound for the size of the coefficients appearing in the expression above.
\begin{itemize}[leftmargin=*]
\item 
For $\gamma_i,$ we have
\begin{align*}
\abs{\gamma_i} = 
\abs{
\vec a_j^{-1} \eps^{b_j} \snorm{L_2}{\alpha_i}
\snorm{L_2}{\beta_i}}
\leq \eps^{-1/2 -2c}  < \eps^{-1} \;,
\end{align*}
since $\abs{ \vec a_j } > \sqrt{\eps}$, $b_j \geq 0$, and $\snorm{L_2}{\alpha_i},\snorm{L_2}{\beta_i} \leq \eps^{-2c}$.
\item 
For $\theta$, we have
\begin{align*}
\abs{ \theta }
= \abs{ \vec a_j^{-1} \eps^{-1/6} \snorm{L_2}{e} }
\leq \eps^{-1/2 -1/6} \; O_{c,d.N}\lp( \eps^{1-c}\rp)
\leq 1 \;,
\end{align*}
where the first inequality follows from our selection rule 
$\abs{\vec a_j} > \sqrt{\eps}$, $\snorm{L_2}{e} < O_{c,d,N}(\eps^{1-c})$, and the second inequality is true as long as $c < 0.3$ and $\eps$ is sufficiently small given $c,d,N$.
\item 
For $\zeta$, we have
\begin{align*}
    \abs{\zeta} = 
    \abs{ \vec a_{j}^{-1} \eps^{b_j} \snorm{L_2}{r} }
    \leq O_{c,d,N}\lp(\eps^{-1/2 - 2c}\rp)
    \leq \eps^{-1} \;,
\end{align*}
where the first inequality follows from that $\abs{\vec a_j} > \sqrt{\eps}$, $b_j \geq 0$, and 
$\snorm{L_2}{ r } \leq O_{c,d,N}(\eps^{-2c})$, and the second inequality is true as long as $c < 0.2$ and $\eps$ is sufficiently small given $c, d, N$.
\item 
For $\lambda_i$ and $\iota_i$, they are both defined to be
$\vec a_j^{-1} \eps^{ b_j - b_i} \vec a_i$.
However, recall that the $\lambda_i$'s are defined for $i \leq \ell$ and and the $\iota_i$'s are defined for $i > \ell$. 
For $i \leq \ell$, we always have $b_i = 0$. Combining this with the fact that $b_j \geq 0$, $\vec a_i < 1$, and $\vec a_j > \sqrt{\eps}$ gives us
$$\abs{ \lambda_i } \leq \abs{ \vec a_j^{-1} } \leq \eps^{-1/2}.
$$
For $i > \ell$, by our selection rule of $j$, we have that $\vec a_j \eps^{b_j}$ is the largest among all $i > \ell, i \in S_k$. This ensures that 
$$
\abs{\iota_i} = \abs{\vec a_j^{-1} \eps^{ b_j - b_i} \vec a_i} \leq 1.
$$
\end{itemize}
Let $B$ be some monomial among the ones obtained by expanding \Cref{eq:expansion}. 
Consider the following quantities:
\begin{itemize}[leftmargin=*]
    \item $u$: the sum of the degrees of the variables $\vec y_{i}$ for $m < i \leq m + 2 \kappa$. These are the variables that carry a leading coefficient of $\sqrt{\gamma_i}$, which is of size at most $\eps^{-1/2}$.
    \item $v$: the degree of the variable $\vec y_{m+2\kappa+1}$. This is the variable that carries a leading coefficient of $\zeta$, which is of size at most $\eps^{-1}$.
    \item $z$: the difference of the sum of the degrees of the variables $\vec y_i$ for $i \in S_k, i \leq \ell$, in monomial $A$ and in monomial $B$. These are the variables that carry a leading coefficient of $\lambda_i$, which is of size at most $\eps^{-1}$.
\end{itemize}
Except for the variables mentioned above, the other variables have leading coefficients bounded by $1$. Therefore, one can see that the coefficient of the new monomial $B$ obtained from expanding $A$ can be bounded from above by
\begin{align} \label{eq:leading-coefficient-bound}
\rho \; \eps^{ -u/2 - v - z }.    
\end{align}

For the monomials $A, B$, consider the following relevant quantities: (i) the degree of the monomial, which we denote by $U(A), V(B)$, 
(ii) the degrees of the monomials weighted by the degrees of the corresponding primitive polynomials, i.e., $V(A):=\sum_{i} \tau_i \deg(q_i) $ (and $V(B)$ is defined similarly), and
(iii) the degree of the variables corresponding to the original primitive polynomials, i.e., $ 
Z(A):= \sum_{i \leq \ell} \tau_i  $ (and $Z(B)$ is defined similarly.)
We then make the following observations:
\begin{itemize}[leftmargin=*]
    \item $U(B) = U(A) + u/2$.
    This is because the variables $\vec y_i$ for $m < i \leq m + 2\kappa$ in $B$ always appear as products of pairs in \Cref{eq:expansion} in place of $\vec y_j$ from $A$.
    Hence, if the sum of the degrees of such $\vec y_i$ is $u$ in $B$, the degree of $B$ must be $u/2$ larger than the degree of $A$.
    \item $V(B) \leq V(A) - v$.
    To see this, we note that
    (i) the degree of $ \deg(q_{m+2i-1}) +  \deg(q_{m+2i})$ for $1\leq i \leq \kappa$ is precisely $\deg(q_j)$ by \Cref{lem:polynomial-decomposition}, (ii) the degree of $q_i$ for $i \in S_k$ is the same as $q_j$ since $S_k$ is defined to be the set of polynomials with degree $k = \deg(q_j)$ and (iii) the degree of $q_{m + 2\kappa + 1}$ is at most $\deg(q_j) - 1$ again by \Cref{lem:polynomial-decomposition}. Hence, if some $\vec y_j$ is replaced with $q_{m+2\kappa+1}$, the weighted degree decreases by at least $1$ and if $\vec y_j$ is replaced with some other terms, the weighted degree remains unchanged.
    Consequently, if the degree of $\vec y_{m + 2\kappa + 1}$ is $v$ in $B$, the weighted degree of $B$ must have changed by at least $v (\deg(q_{m+2\kappa+1}) - \deg(q_j) ) \leq -v $.
    \item $Z(B) = Z(A) + z$. This follows by the definition of $z$.
\end{itemize}
Combining the above observations with statement~\eqref{eq:leading-coefficient-bound}, we can then use induction to show that the leading coefficient of $B$ is at most
$$
\eps^{  -U(B) + ( V(B) - d ) - Z(B) }.
$$
The quantities $V(B), U(B), Z(B)$, by their definitions, are all within the range $[0, d]$. Therefore, we conclude that the coefficient of any monomial $B$ that appears in $h'$
is bounded from above by $\eps^{-3d}$.
By \Cref{fact:coefficeint-norm}, this implies that the $L_2$ norm of $h'$
is at most $O_{d,m+\kappa}( \eps^{-3d} )$.
As the new decomposition indeed has bounded complexity and strictly smaller potential, this finishes the inductive step of the transfinite induction, completing the proof of \Cref{prop:super-non-singular-extension}.
\end{proof}

\section{Super Non-Singular Polynomials and (Anti-)Concentration}
\label{sec:proof-anti-concentration}
In this section, we provide the formal statement and the proof of our (anti-)concentration theorem (\Cref{thm:informal-anticoncentration}) concerning the standard Gaussian distribution conditioned on sets of the form $\vec q(\x) \in R$, where $\vec q:\R^n \mapsto \R$ is a super non-singular transformation and $R \subset \R^{m}$ is an axis-aligned rectangle.
\begin{theorem}
[Conditional (Anti-)Concentration for Super Non-Singular Transformation]
\label{thm:(anti-)concentration}
Let $d, \ell, K, C_{d, \ell, K}$ be positive integers and $\eps \in (0, 1)$.
Let $S = \{q_1, \cdots, q_{\ell}\}$ be a set of harmonic polynomials, where each $q_i:\R^n \mapsto \R$ is of degree at most $d$.
Define $\vec q:\R^n \mapsto \R^{\ell}$ to be the vector-valued polynomial such that $\vec q(\x) = ( q_1(\x), \cdots, q_{\ell}(\x) )$.
Let $R \subset \R^{\ell}$ be an axis-aligned rectangle satisfying
(i) each point in $R$ is at most $ \poly_{d, \ell} ( \log(1/\eps) )$-far from the origin, and
(ii) $\Pr_{ \vec x \sim \normal(\vec 0, \vec I) }
\lp[  \x \in R \rp] > \poly_{d, \ell}( \eps )$.
Denote by $D$ the distribution of $\normal(\vec 0, \vec I)$ conditioned on $\{ 
\x: \vec q(\x) \in R \}$.
Suppose $C_{d, \ell, K}$ is sufficiently large given $d, \ell, K$, and $\eps$ is sufficiently small given $d, \ell, K, C_{d, \ell, K}$.
Assume that $\{q_1, \cdots, q_{\ell}\}$ is $(\eps^{1/(3d^2K)}, C_{d, \ell, K})$-super non-singular.
Then for any polynomial $p:\R^n \mapsto \R$ of degree-$d$ and $\eps < t < \eps^{2/K}$, it holds
\begin{align}
\Pr_{ \vec x \sim D  }
\lp[ 
\abs{p(\vec x)} < t \; \snorm{D, L_2}{p}
\rp] \leq t^{1/(2d)} \, ,
\end{align}
and, for all $t > 0$, it holds
\begin{align}
\Pr_{ \vec x \sim D } \lp[ \abs{p(\vec x)} > t \eps^{-1/K} \snorm{D, L_2}{p}
\rp] \leq O_{d,K}\lp( t^{- K  } \rp).
\end{align}
\end{theorem}
We first outline our proof plan.
To show the above theorem, we will first use \Cref{prop:super-non-singular-extension} to extend $\{q_1, \cdots, q_{\ell}\}$ into a super-non singular decomposition of the target polynomial $p$. In particular, this gives us a super non-singular set of polynomials $\{q_1, \cdots, q_{\ell}, q_{\ell+1}, \cdots, q_{m}\}$ such that $h( q_1(\x), \cdots, q_m(\x)) \approx p(\x)$.
Define $\bar {\vec q}: \R^n \mapsto \R^m$ to be $\bar {\vec q}(\x) = (q_1(\x), \cdots, q_m(\x))$.
Since $p(\x)$ and $h( q_1(\x), \cdots, q_m(\x))$ are approximately the same, the polynomial $p$ under $\normal(\vec 0, \vec I)$ will roughly have the same (anti-)concentration properties as $h$ under $\bar {\vec q}( \normal(\vec 0, \vec I) )$.

Then it remains to show $h$ has good (anti-)concentration properties under $\bar {\vec q}( \normal(\vec 0, \vec I) )$. 
To achieve the goal, we need to first prove a structural result showing that $\bar {\vec q}( \normal(\vec 0, \vec I) )$ is close in total variation distance to some distribution comparable (see \Cref{def:comparability}) to a mixture of reasonable Gaussians (see \Cref{intro-def:reasonable-gaussian}).
Specifically, we establish the following.

\begin{proposition}[Super Non-Singular Polynomial Transformations are Reasonable]
\label{prop:Gaussian-Comparability}
Let $m,d,N, C_{m,d,N}$ be positive integers, where $C_{m,d,N}$ is sufficiently large given $m, d, N$.
Let $\delta \in (0, 1)$ be sufficiently small given the aforementioned integers. 
Let $\vec q: \R^n \mapsto \R^m$ be a $( \delta^{1/(3d)} , C_{m,d,N})$-super non-singular vector-valued polynomial of degree $d$ \footnote{Recall that we say a vector-valued polynomial is super non-singular if it consists of a super non-singular set of real-valued polynomials.}.
Then the distribution $\vec q( \normal(\vec 0, \vec I) )$ is $O( \delta^N )$-close in total variation distance to 
some distribution that is comparable to some mixture distribution
$\int \normal_{\theta} d\theta$, where each $\normal_{\theta}$
is a $\lp(\delta, \log^{ O(d) }(1/\delta  ) \rp)$-reasonable Gaussian.
\end{proposition}
Gaussian distributions are still log-concave if we condition on an axis-aligned rectangle $R$, therefore still satisfying decent (anti-)concentration properties after the conditioning.
It turns out this is also roughly true for a mixture of reasonable Gaussians (see \Cref{lem:Gaussian-Mixture-anti-concentration}) conditioned on an axis-aligned rectangle that is not too far from the origin (which is assumed in \Cref{thm:(anti-)concentration}).
With \Cref{lem:Gaussian-Mixture-anti-concentration} and \Cref{prop:Gaussian-Comparability} established, it is then not hard to conclude that the polynomial $h$ has good (anti-)concentration properties under $\bar{\vec q}(\normal(\vec 0, \vec I))$.

Sections~\ref{sec:non-singular-jacobian}, \ref{sec:approximate-linear-transform}, and \ref{sec:compare-reasonable-gaussian} are devoted to the proof of \Cref{prop:Gaussian-Comparability}.
After that, we discuss the (anti-)concentration properties of (mixture of) reasonable Gaussians (\Cref{lem:Gaussian-Mixture-anti-concentration}) in \Cref{sec:reasonable-gaussian}. 
In Sections~\ref{sec:transfer-probability}, \ref{sec:transfer-properties}, and~\ref{sec:conclude-anti-concentration}, we conclude the proof of \Cref{thm:(anti-)concentration}, where we transfer the (anti-)concentration properties from a mixture of reasonable Gaussians conditioned on $R$
to $\bar {\vec q}( \normal(\vec 0, \vec I) )$ conditioned on $R$, and then finally to $\normal(\vec 0, \vec I)$ conditioned on $\{ \x: \vec q(\x) \in R \}$.
\subsection{Non-Singularity of Jacobian}
\label{sec:non-singular-jacobian}
As we have already mentioned, the key idea to establish \Cref{prop:Gaussian-Comparability} is the observation that the polynomial transformation behaves somewhat like a non-singular linear transformation.
As our starting point, we show that the super non-singularity condition implies that the Jacobian of the polynomial transformation evaluated at $\x$ must be non-singular with high probability when $\x \sim \normal(\vec 0, \vec I)$ \footnote{As a matter of fact, this condition on the Jacobian of a polynomial transformation is equivalent to the exact definition of non-singularity used in the work of \cite{Kane12subpoly}.}.
\begin{lemma}
\label{lem:non-singular-jacobian}
Let $m,d,N,C_{m,d,N}$ be integers, where $C_{m,d,N}$ is sufficiently large given $m,d,N$.
Let $\{ q_1, \cdots, q_m | q^{(i)}: \R^n \mapsto \R\}$ be a set of $(\delta^{1/d}, C_{m,d,N})$-super non-singular polynomials.
Define $\vec q: \R^n \mapsto \R^m$ to be vector-valued polynomial whose $i$-th coordinate is given by $q_i$.
Then it holds
$$
\Pr_{\vec x \sim \normal(\vec 0, \vec I)}\lp[\sigma_{\min}\lp( \Jac_{\vec q}( \vec x ) \rp) \leq \delta\rp] < \delta^N.
$$
\end{lemma}
Our overall strategy for showing \Cref{lem:non-singular-jacobian} is by contradiction.
Suppose that
\begin{align}
\label{eq:source-contradiction}
\Pr_{\vec x \sim \normal(\vec 0, \vec I)} \lp[ 
\sigma_{\min}( \Jac_{\vec q}(\vec x) < \delta ) \rp]> \delta^N    
\end{align}
We start with the simple observation that 
\Cref{eq:source-contradiction} is essentially equivalent to
\begin{equation} \label{eq:non-singular-jacobian}
\Pr_{\vec x \sim \normal(\vec 0, \vec I)}
\lp[ 
\min_{ \vec a \in \R^m: \snorm{2}{\vec a} = 1 } 
\snorm{2}{
\nabla_{\vec x}
\lp(  \sum_{i=1}\vec a_i \; q_i(\vec x) \rp)} < \delta
\rp] > \delta^N \;.
\end{equation}
In order to obtain a contradiction against the definition of super non-singularity, there are still two issues we need to tackle:
\begin{enumerate}[leftmargin=*]
    \item The linear combination coefficients $\vec a_i$ in \Cref{eq:non-singular-jacobian} which minimize
    $\snorm{2}{  \nabla_{\x} \lp( \sum_{i=1}^m \vec a_i q_i(\x) \rp) }$ are allowed to be different for different values of $\vec x$, while the same coefficients are used across all different $\vec x$ in the definition of super non-singularity (\Cref{def:super-non-singular}).
    \item Super non-singularity poses requirements on the $k$-th order derivatives of the subset of polynomials of degree $k$, while \Cref{eq:non-singular-jacobian} poses one requirement on the first-order derivatives of the entire set of polynomials.
\end{enumerate}
To tackle the first issue, we consider a finite cover of the set of unit vectors in $\R^m$. If we further condition on the event that the gradients of the initial polynomials are themselves not too big (which follows from standard Gaussian concentration), we can conclude that at least one vector from the cover will be effective for a set of $\vec x$ with non-trivial at mass under the Gaussian distribution at the same time.
\begin{lemma}
\label{lem:small-gradient-combination}
Let $N,m$ be positive integers and $\delta \in (0,1)$ be sufficiently small given $N,m$.
Suppose that $\{q_1, \cdots, q_m\}$ is such that
\begin{align} \label{eq:singular-jacobian}
\Pr_{\vec x \sim \normal(\vec 0, \vec I)}
\lp[ 
\min_{ \vec a \in \R^m: \snorm{2}{\vec a} = 1 } 
\snorm{2}{
\nabla_{\vec x}
\lp(  \sum_{i=1}\vec a_i \; q_i(\vec x) \rp)} < \delta
\rp] > \delta^N.    
\end{align}
Furthermore, suppose that 
$\snorm{L_2}{ q_i } \leq 1$ for all $i$ in $[m]$.
Then there exists a vector $\vec a \in \R^m$ 
with $\snorm{2}{\vec a} = 1$ 
such that the polynomial
$q(\vec x) = \sum_{i=1}^m \vec a_i \; q_i(\vec x)  $ satisfies
$$
\Pr \lp[ \snorm{2}{\nabla q(\vec x)} \leq  2 \; \delta \, , \, 
\snorm{2}{\nabla q_i(\vec x)}
\leq  O_{d,m}( \log^{2d}( 1 / \delta ) )
\text{ for all } i
\rp]
\geq \delta^{N+m+1}.
$$
\end{lemma}
\begin{proof}
By \Cref{lem:gradient-norm-concentration} and union bound, we have
\begin{align} \label{eq:polynomial-bounded-l2-norm}
\Pr_{\vec x \sim \normal(\vec 0, \vec I)}
\lp [
\snorm{2}{\nabla q_i (\vec x)} \leq  
O_{d,m,N}( \log^{d/2}(1/\delta) )
\; \forall \; i
\rp]
\geq  1 - \delta^N / 2.
\end{align}
Again, by the union bound, both \Cref{eq:singular-jacobian} and \Cref{eq:polynomial-bounded-l2-norm} are true with probability at least $\delta^N / 2$ over $\x \sim \normal(\vec 0, \vec I)$.
Conditioned on that, there exists some vector $\vec a$ such that
$
\snorm{2}{ \sum_{i=1}^m \vec a_i \nabla q_i(\vec x) } \leq \delta.
$
Now consider a new vector $\vec {\tilde a}$ that is obtained by rounding each $\vec a_i$ to an integer multiple of $\delta / \log^{d/2+1}(1/\delta)$:
$$
\vec {\tilde a}_i = 
\sgn( \vec a_i ) \; \ceil{ 
\frac{  \abs{\vec a_i} \log^{d/2+1}(1/\delta)  }{ \delta }
} \; \frac{\delta}{ \log^{d/2+1}(1/\delta) }.
$$
One can verify that
$\snorm{2}{\vec {\tilde a}} \geq \snorm{2}{\vec a} = 1$ and
$ \snorm{2}{\vec a - \tilde{\vec a}} \leq \delta \; \sqrt{m} / \log^{d/2+1}(1/\delta) $.
Conditioned on the event that $\snorm{2}{\nabla q_i (\vec x)} \leq O_{d,m,N}( \log^{d/2}(1/\delta) )$ for all $i \in [m]$, 
we have
\begin{align*}
\snorm{2}{ \sum_{i=1}^m \tilde {\vec a}_i \nabla q_i(\vec x) } \leq 
\delta + 
O_{d,m,N}\lp( \log^{d/2}(1/\delta) \rp) \; \snorm{2}{ \vec a - \tilde{\vec a} }
\leq 
\eps \lp( 1 + O_{d,m,N}( \log^{-1}(1/\delta) )  \rp)
\leq
2 \delta \;,
\end{align*}
where in the first inequality we use the triangle inequality, in the second inequality we use our upper bound for $\snorm{2}{\vec a - \tilde {\vec a}}$ and the last inequality is true as long as $\delta$ is sufficiently small given $d,m,N$.
Note that there are at most $(2 \log^{d/2 + 1}(1/\delta) /\delta + 2)^m$ many vectors whose coordinates are all integer multiples of $\delta / \log^{d/2+1}(1/\delta)$ 
and lie in the range $ [- 1 - \delta / \log^{2d+1}(1/\delta), 1 + \delta / \log^{2d+1}(1/\delta)  ] $.
Therefore, there exists some $\vec a^*$ (independent of the choice of $\vec x$) such that
\begin{align*}
&\Pr_{\vec x \sim \normal(\vec 0, \vec I)}
\lp [
\snorm{2}{\nabla_{\vec x} 
\lp( \sum_{i=1}^m \vec a^*_i \; q_i(\vec x) / \snorm{2}{ \vec a^*_i }\rp) }
\leq 2 \delta \, , \,
\snorm{2}{ \nabla q_i(\vec x) } \leq O_{d,m,N}( \log^{2d}(1/\delta) ) \text{ for all } i \in [m]
\rp] \\
&\geq  
\frac{\delta^N}{2} \; (2 \log^{2d+1}(1/\delta) /\delta + 2)^{-m}
= 
\delta^{N + m} / \poly_{d, m} \lp( \log(1/\delta) \rp)
\geq
\delta^{N + m + 1} \, ,
\end{align*}
where the last inequality follows from our assumption that $\delta^{-1}$ is a sufficiently given $d,m,N$.
\end{proof}
To tackle the second issue, we need the following lemma which relates first-order derivatives of a polynomial to its higher-order derivatives.
\begin{lemma}[Proposition 11 from \cite{Kane12}]
\label{lem:iterative-directional-derivative}
Let $c,N > 0$ be real numbers and $d$ be a positive integer.
Let $\delta > 0$ be a real number that is sufficiently small given $c,N,d$.
Suppose that $q$ is a degree-$d$ polynomial such that
$$
\Pr_{ \vec x \sim \normal(\vec 0, \vec I) }\lp[ 
\snorm{2}{ \nabla q(\vec x) } < \delta
\rp] > \delta^N \;.
$$
Then we have that
$$
\Pr_{ \vec x \sim \normal(\vec 0, \vec I) }\lp[ 
\snorm{2}{\nabla_{\x} D_{\vec y} q(\x) }
< \delta^{1-c} 
\rp] > \delta^{O_{N,c,d}(1)} \;.
$$
\end{lemma}
Equipped with the above tool, we can iteratively apply it to the polynomial $\sum_{i=1}^m \vec a_i q_i$, where $\vec a$ is the global linear combination coefficient obtained through \Cref{lem:small-gradient-combination}. 
Then there are two cases to consider. 
In the first case, the sum of the coefficients $\abs{\vec a_i}$, where $q_i$ is one of the highest degree polynomials in the set is sufficiently large. In this case, it suffices to apply \Cref{lem:small-gradient-combination} $d-1$ many times and the contributions from the other low-degree polynomials in the linear combination will all vanish; 
this contradicts the definition of super non-singularity.
In the second case, we can discard the highest degree polynomials from the linear combination and recurse on the remaining polynomials.
We are now ready to present the formal proof of \Cref{lem:non-singular-jacobian}.
\begin{proof}[Proof of \Cref{lem:non-singular-jacobian}]
Let $\{q_1, \cdots, q_m\}$ be a set of polynomials with unit $L_2$ norm that does not satisfy \Cref{eq:non-singular-jacobian}. 
We will show that there exists a number $C_{d,N,m}$ that depends only on $d,N,m$
such that the set is not $(\delta^{1/d} , C_{d,N,m})$-super non-singular either.
Recall that for the set to be super non-singular, each polynomial needs to be harmonic. Hence, we assume that $\{q_1, \cdots,q_m\}$ are all harmonic since otherwise the proof is trivial.
Applying \Cref{lem:small-gradient-combination} gives us that there exists some $\vec a \in \R^m$ with $\snorm{2}{\vec a} = 1$ such that
\begin{align}
\label{eq:non-singular-combination}    
\Pr_{ \vec x \sim \normal(\vec 0, \vec I) }
\lp[ \snorm{2}{
\sum_{i=1}^m \vec a_i \nabla q_i(\vec x)
)} \leq 2 \delta \, , \,
\snorm{2}{ q_i(\vec x) } \leq O_{d,m,N}(\log^{2d}(1/\delta)) \forall i
\rp] \geq \delta^{N} \;.
\end{align}
Let $S_t$ be the set of degree-$t$ harmonic polynomials within $\{q_1, \cdots, q_m\}$.
Define $w_t$ as the weight assigned to each set $S_t$ by the vector $\vec a$, i.e., $w_t = \sqrt{\sum_{i \in S_t} \vec a_i^2}$.
Let $k$ be the largest integer such that
$w_k \geq \delta^{k/d}$. 
Such a $k$ must exist because there exists $t$ such that $w_t \geq 1/\sqrt{d} \gg \delta^{1/d}$.
Consider the polynomial
$q$ obtained by removing the contributions from $S_t$ for $t > k$.
That is, we can write:
$$
q(\vec x) = \sum_{t=1}^{k} \sum_{ i \in S_t }
\vec a_i q(\vec x) \;.
$$
We claim that this polynomial satisfies
$$
\Pr_{ \vec x \sim \normal(\vec 0, \vec I) }
\lp[ 
\snorm{2}{\nabla q(\vec x)} \leq 3\delta
\rp] \geq \delta^N \;.
$$
This is because for each $\vec x$ which satisfies the condition in \Cref{eq:non-singular-combination}, we have
\begin{align*}
\snorm{2}{q(\vec x)}
= 
\snorm{2}{  \sum_{i=1}^m \vec a_i \nabla q_i(\vec x) -  
\sum_{t=k+1}^{d} \sum_{ i \in S_t }
\vec a_i q(\vec x)
}
&\leq \snorm{2}{ \sum_{i=1}^m \vec a_i \nabla q_i(\vec x) }
+  \sum_{t=k+1}^{d} \sum_{ i \in S_t } \abs{\vec a_i} \snorm{2}{ 
 q(\vec x)} \\
& \leq 2 \delta + 
O_{d,m,N}\lp( \log^{d/2}(1/\delta) \rp) \; 
\sum_{t = k+1}^d \sum_{i \in S_t} \abs{\vec a_i} \;,
\end{align*}
where in the first inequality we use the triangle inequality, and in the second inequality we use the condition in \Cref{eq:non-singular-combination}.
Now we claim the sum $\sum_{t = k+1}^d \sum_{i \in S_t} \abs{\vec a_i}$ is bounded from above by 
$O_m \lp( \delta^{(k+1)/d} \rp)$.
This is because the inner summation is at most 
$ O_{m} \lp( \delta^{t/d} \rp) $ since the group will only be removed if $ w_t = \sqrt{ \sum_{i \in S_t} \vec a_i^2} < \delta^{t/d} $. Then we can bound from above the entire summation as a geometric sequence dominated by $k = d+1$ and conclude that it is at most $O_m \lp( \delta^{(k+1)/d} \rp)$. Thus, we further have
$$
\snorm{2}{q(\vec x)}
\leq 2 \delta + O_{d,m,N} \lp( \log^{d/2}(1/\delta) \delta^{(k+1)/d} \rp)
\leq \delta^{(k+0.75)/d} \, ,
$$
where the last inequality is true as long as $\delta$ is sufficiently small given $d,m,N$.

Now we can choose some $c$ that is sufficiently small given $k,d$ and then iteratively apply \Cref{lem:iterative-directional-derivative} $k-1$ times on $q$.
This gives us
\begin{align}
\label{eq:non-harmonic-derivative-bound}
\Pr_{ \vec x \sim \normal(\vec 0, \vec I), 
\vec y^{(i)} \sim \normal(\vec 0, \vec I) \text{ for } 1 \leq i \leq k-1
}
\lp[ \snorm{2}{\nabla_{\vec x} D_{\vec y^{(k-1)}}  
\cdots D_{\vec y^{(1)}} q(\vec x)} \leq \delta^{(k+0.5)/d}
\rp] \geq \delta^{O_{d,m,N}(1)} \;.    
\end{align}
We note that 
\begin{align}
\label{eq:vanishing-lower-order-derivative}
\nabla_{\vec x} D_{\vec y^{(k-1)}}  
\cdots D_{\vec y^{(1)}} q(\vec x)
= \nabla_{\vec x} D_{\vec y^{(k-1)}}  
\cdots D_{\vec y^{(1)}} \lp( 
\sum_{ i \in S_k } \vec a_i q_i(\vec x) 
\rp) \, ,    
\end{align}
since the contributions from the lower degree polynomials in $\{q_1, \cdots,q_m\}$ disappear after the $k$ derivative operations.
Thus, combining 
\Cref{eq:non-harmonic-derivative-bound} and
\Cref{eq:vanishing-lower-order-derivative} gives us
$$
\Pr_{ \vec x \sim \normal(\vec 0, \vec I), 
\vec y^{(i)} \sim \normal(\vec 0, \vec I) \text{ for } 1 \leq i \leq k-1
}
\lp[ \snorm{2}{\nabla_{\vec x} D_{\vec y^{(k-1)}}  
\cdots D_{\vec y^{(1)}}  \lp( 
\sum_{ i \in S_k } \vec a_i q_i(\vec x)
\rp) } \leq \delta^{ (k+0.5)/d}
\rp] \geq \delta^{O_{d,m,N}(1)} \;.
$$
The only remaining issue is that $\sqrt{\sum_{i \in S_k} \vec a_i^2} = w_k$ instead of equal to $1$.
Nonetheless, we have chosen $k$ such that $w_k \geq \delta^{k/d}$.
Hence, we can normalize $\vec a$ to obtain another vector $\bar{\vec a}$ with $\sqrt{\sum_{i \in S_k} \bar {\vec a}_i^2} = 1$ and still have that
$$
\snorm{2}{\nabla_{\vec x} D_{\vec y^{(k-1)}}  
\cdots D_{\vec y^{(1)}}  \lp( 
\sum_{ i \in S_k } \bar{\vec a}_i q_i(\vec x)
\rp) } \leq \delta^{ (k+0.5)/d - k/d}
= \delta^{1/(2d)}
$$
with probability at least $\delta^{O_{d,N,m}(1)}$ over the randomness of 
$\vec x, \vec y^{(1)}, \cdots, \vec y^{(k-1)} \sim \normal(\vec 0, \vec I)$.
This clearly violates the definition of $(\delta^{1/(2d)}, C_{d,N,m}(1))$-super non-singularity for some number $C_{d,N,m}$ that depends only on $d,N,m$.
\end{proof}

\subsection{Approximate Linear Transforms}
\label{sec:approximate-linear-transform}

When the underlying polynomials are of higher-degrees, the transformed distribution defined by them may no longer have a nice form.
However, as the Jacobian of the polynomial transformation is non-singular almost everywhere (\Cref{lem:non-singular-jacobian}), this transformation will jointly behave like some linear transformation if one restricts to ``local'' areas. In fact, in this section, we will show that the distribution of $\vec q(\normal(\vec 0, \vec I))$ is close to a mixture of $\vec p(\normal(\vec 0, \vec I))$, where each transformation $\vec p$ is ``approximately'' linear.
Notice that if each $\vec p$ were exactly linear, each transformed distribution in the mixture would indeed be some reasonable Gaussian and we would be nearly done.

Below we formalize the notion of an approximately linear transform.
One point worth noticing is that the original vector-valued polynomial $\vec q$ maps an $n$-dimensional vector to an $m$-dimensional vector where $m$ can be much smaller than $n$. 
In the special case where $\vec q$ is exactly linear, one can see that the mapping only retains a ``fraction'' of the randomness in the input variable $\vec x \sim \normal(\vec 0, \vec I)$.
In particular, the transformed distribution is equivalent to 
some linear transformation of an $m$-dimensional standard Gaussian. 
We hence formalize our definition of an approximate linear transform as an $\R^m \mapsto \R^m$ map.
\begin{definition}[Degree-$d$, $b$-bounded, $\delta$-approximate linear transform]
\label{def:approximate-linear transform}
We say that $\vec p: \R^m \mapsto \R^m$ is a degree-$d$, $b$-bounded, $\delta$-approximate-linear transform if there exist: 
\begin{itemize}
\item  a vector $\vec g \in \R^m$ with $\snorm{2}{\vec g} \leq \log^b(1/\delta)$,
\item a matrix $\vec M \in \R^{m \times m}$  with its singular values bounded by
$  \delta  \leq \sigma_i(\vec M) 
\leq \log^b(1/\delta)$, 
\item a vector-valued polynomial $\vec r: \R^m \mapsto \R^m$ of degree $d$ satisfying
$ \snorm{L_2}{\vec r} \leq \delta^{5/4}$\footnote{We remark that the choice of the constant $5/4$ 
is just to simplify the computation in some parts our proof. 
For our downstream application, we only require it to be a constant slightly larger than $1$.},
\end{itemize}
such that
$
\vec p(\vec x) 
= \vec g + \vec M \vec x + \vec r(\vec x) \;.
$
\end{definition}

The main technical lemma of this section is given below.
\begin{lemma}
\label{lem:mixture-of-linear-transform}
Let $d, m, N \in \mathbb Z^+$.
Let $\delta > 0$ be sufficiently small in $d, m, N$.
Let $\vec q: \R^n \mapsto \R^m$ be a degree-$d$ vector-valued polynomial satisfying
\begin{align} \label{eq:non-singular-assumption}
\Pr_{\vec x \sim \normal(\vec 0, \vec I)}\lp[\sigma_{\min}\lp( \Jac_{\vec q}( \vec x ) \rp) 
< \delta^{1/2}  \rp] < \delta^N \;.    
\end{align}
Then the distribution of $ \vec q( \normal(\vec 0, \vec I) ) $ is 
$O( \delta^N)$-close in total variation distance to an (infinite) mixture of 
distributions of the form
$
\vec p( \normal(\vec 0, \vec I) )
$, where 
$\vec p: \R^m \mapsto \R^m$ is a
degree-$d$, $(d+1)$-bounded,
$\delta^{3/2}$-approximate-linear transform.
\end{lemma}
We begin with the simple observation that $\vec q( \normal(\vec 0, \vec I) )$ is equivalent to the distribution of 
$ \vec q( \sqrt{1 - \delta^2} \vec x + \delta \vec z ) $, where $\vec x, \vec z$ are independently distributed as $\normal(\vec 0, \vec I)$.
Our intuition is that, due to the non-singularity condition, $\vec q$ will behave like some linear transformation locally. 
To show this formally, we will need a Taylor-style expansion of the expression
$\vec q( \sqrt{1 - \delta^2} \vec x + \delta \vec z )$. We begin with the simpler case for a scalar polynomial $q$.
\begin{lemma}
\label{eq:real-value-taylor-like expansion}
Let $q: \R^n \mapsto \R$ be a degree-$d$ harmonic polynomial with $\snorm{L_2}{q} \leq 1$, and let $\delta \in  (0, 1/4)$. 
Then there exists degree-$d$ polynomials $g: \R^n \mapsto \R$ and $e: \R^{n} \times \R^{n} \mapsto \R$ with
$ \snorm{L_2}{g} \leq 1 $, $\snorm{L_2}{e} \leq 2^{d/2} \delta^2$
such that 
$$
q( \sqrt{1 - \delta^2} \vec x + \delta \vec z )
= g(\vec x) + 
\lp(\sqrt{1 - \delta^2}\rp)^{d-1} \; 
\delta \; \nabla q(\vec x) \cdot \vec z +  e( \vec x, \vec z ) \, ,
$$
for all $\x, \vec z \in \R^n$.
\end{lemma}

\begin{remark}
{\em For the sake of clarity, we note that the expansion resembles the Taylor expansion of $q( \sqrt{1 - \delta^2}\vec x + \delta \vec z )$ viewed as a function of $\delta \vec z$ around $\vec 0$, but the two expansions are not exactly the same. In particular, if it were exactly the Taylor expansion, the coefficient in front of the linear term would be $ \nabla q( \sqrt{1 - \delta^2} \vec x)$ instead of $ \nabla q(\vec x)$. However, we do need the gradient term to be evaluated at $\vec x$ exactly to exploit the non-singularity property of $\vec q$ under the standard Gaussian distribution.
Due to this subtle difference, we need to use properties of Hermite polynomials to establish the expansion needed carefully instead of applying Taylor's Theorem in a black-box manner.}
\end{remark}
\begin{proof}
We assume that
$ q(\vec x)$ is a harmonic polynomial of degree $d$ with Hermite decomposition
$ q(\vec x) = \sum_{ \vec s: \abs{\vec s} = d }  \vec T_{\vec s}  H_{\vec s}(\vec x)$.
Applying \Cref{fact:hermite-addition-formula} then gives
\begin{align*}
q( \sqrt{1 - \delta^2} \vec x + \delta \vec z  )
&= \sum_{ \vec s: \abs{\vec s} = d  }
\vec T_{\vec s}
\sum_{ \vec k \leq \vec s }
\sqrt{ \binom{\vec s}{\vec k} }
\sqrt{1 - \delta^2}^{ \abs{\vec s - \vec k} }
\delta^{\abs{\vec k}}
H_{\vec s - \vec k}(  \vec x)
\; H_{\vec k}(\vec z) \\
&= 
\sum_{j=0}^d
\sqrt{1 - \delta^2}^{ d-j }
\delta^{j}
\sum_{ \vec s: \abs{\vec s} = d }
\vec T_{\vec s}
\sum_{ \vec k \leq \vec s, \abs{\vec k} = j }
\sqrt{ \binom{\vec s}{\vec k} }
H_{\vec s - \vec k}(  \vec x)
\; H_{\vec k}(\vec z) \;.
\end{align*}
Notice that 
$H_{\vec s - \vec k}(  \vec x)
\; H_{\vec k}(\vec z)$
and $H_{\vec s' - \vec k'}(  \vec x)
\; H_{\vec k'}(\vec z)$
will be orthogonal if either $\vec s \neq \vec s'$ or $\vec k \neq \vec k'$.
Hence, this is an orthogonal decomposition of the polynomial $q( \sqrt{1 - \delta^2} \vec x + \delta \vec z  )$. 
We will analyze the summation for the terms $j > 1$, $j = 1$, and $j = 0$ separately.
Define 
\begin{align}
\label{eq:error-e-def}
e(\vec x, \vec z)
&= 
\sum_{j=2}^d
\sqrt{1 - \delta^2}^{d - j}
\delta^j
\sum_{ \vec s: \snorm{1}{\vec s} = d }
\vec T_{\vec s}
\sum_{ \vec k \leq \vec s, \snorm{1}{\vec k} = j }
\sqrt{ \binom{\vec s}{\vec k} }
H_{\vec s - \vec k}(  \vec x)
\; H_{\vec k}(\vec z)\, ,  \\
m(\vec x, \vec z)
&= 
\sqrt{1 - \delta^2}^{d - 1}
\delta^1
\sum_{ \vec s: \snorm{1}{\vec s} = d }
\vec T_{\vec s}
\sum_{ \vec k \leq \vec s, \snorm{1}{\vec k} = 1 }
\sqrt{ \binom{\vec s}{\vec k} }
H_{\vec s - \vec k}(  \vec x)
\; H_{\vec k}(\vec z)\, ,  \nonumber \\
g(\vec x)
&= 
\sqrt{1 - \delta^2}^{d}
\sum_{ \vec s: \snorm{1}{\vec s} = d }
\vec T_{\vec s}
H_{\vec s}(  \vec x) \, , \nonumber
\end{align}
which correspond to the terms with $j > 1$, $j = 1$, $j = 0$ respectively.

For $e$, we show that it is a polynomial with $L_2$-norm at most $2^{d/2} \delta^2$.
Exploiting the fact that \Cref{eq:error-e-def} is an orthogonal decomposition of 
$e(\vec x, \vec z)$, 
we have the squared $L_2$-norm of $e$ is given by
\begin{align*}
\snorm{L_2}{e}^2
&= \sum_{j=2}^d
\sum_{ \vec s: \snorm{1}{\vec s} = d }
\sum_{ \vec k \leq \vec s, \snorm{1}{\vec k} = j }
\lp( 
\sqrt{1 - \delta^2}^{d - j}
\delta^{j} 
\vec T_{\vec s}  \sqrt{ \binom{\vec s}{\vec k} } \rp)^2 \\
&= 
\sum_{ \vec s: \snorm{1}{\vec s} = d }
\vec T_{\vec s}^2 \;
\sum_{j=2}^d
(1 - \delta^2)^{d - j}
\delta^{2j}
\sum_{ \vec k \leq \vec s, \abs{\vec k} = j }
{ \binom{\vec s}{\vec k} } \\
&\leq 
\sum_{ \vec s: \snorm{1}{\vec s} = d }
\vec T_{\vec s}^2 \;  
\delta^4
\sum_{ \vec k \leq \vec s, \snorm{1}{\vec k} \geq 2}
{ \binom{\vec s}{\vec k} } \, ,
\end{align*}
where in the inequality we use the elementary fact
$ (1 - \delta^2)^{ d - j } \delta^{2j} \leq \delta^4$
for $j \geq 2$ and $0 < \delta < 1$. 
Now for the summation 
$\sum_{ \vec k \leq \vec s, \abs{\vec k} \geq 2}
{ \binom{\vec s}{\vec k} }^2$, 
we claim that it can be bounded above by $2^d$.
In particular, it is bounded above by
\begin{align*}
\sum_{ \vec k \leq \vec s}
{ \binom{\vec s}{\vec k} }^2
= 
\prod_{i=1}^n \sum_{ \vec k_i = 0 }^{ \vec s_i } 
{ \binom{\vec s_i}{\vec k_i} }
\leq \prod_{i=1}^n 2^{ \vec s_i } \, ,
\end{align*}
where in the last inequality we use the Binomial theorem. Since we also have the restriction that $\abs{\vec s} = d$, it follows that the expression is bounded above by $2^d$.
Since $\sum_{\vec s: \abs{\vec s} = d} \vec T_{\vec s}^2 = \snorm{L_2}{q}^2 \leq 1$, it follows that
$\snorm{L_2}{e} \leq 2^{d/2} \delta^2$.

For the $j = 1$ term, we can simplify the expression as
\begin{align*}
m(\vec x, \vec z)=    
\sqrt{1 - \delta^2}^{d - 1}
\delta
\sum_{ \vec s: \abs{\vec s} = d }
\vec T_{\vec s}
\sum_{i: \vec s_i > 0}
\sqrt{\vec s_i}
H_{ \vec s - \mathbf{ \mathcal I }(i)   }(  \vec x)
\; \vec z_i \, ,
\end{align*}
where we use $ \mathcal I (i)$ to denote the indicator vector that is $1$ on the $i$-th coordinate and $0$ everywhere else. 
Notice that $ \sqrt{\vec s_i} H_{\vec s - \mathcal I (i)} (\vec x) $ is exactly the partial derivative
$ \frac{\partial}{\partial \vec x_i}
H_{\vec s}(\vec x)$.
Accordingly, for a fixed coordinate $i$, we have
$$
\sum_{ \vec s: \abs{\vec s} = d }
\vec T_{\vec s}
\sqrt{\vec s_i}
H_{\vec s - \mathcal I (i)}(  \vec x)
= \frac{\partial}{\partial \vec x_i}
q(\vec x) \;.
$$
Summing over all coordinates $i$ then gives
$$
m(\vec x, \vec z) = 
\sqrt{1 - \delta^2}^{d-1} \; \delta \;
\nabla q(\vec x) \cdot \vec z \;.
$$
Finally, for $j = 0$, it is easy to see that
$$
g(\vec x) = \sqrt{1 - \delta^2}^d q(\vec x) \;.
$$
Hence, $\snorm{L_2}{g} < \snorm{L_2}{q} \leq 1$.
\end{proof}

As an immediate corollary, we obtain a generalized version of the above lemma for vector-valued polynomials.
\begin{corollary}
\label{cor:taylor-like-expansion}
Let $ \{ q_1, \cdots, q_m \}$ be a set of real-valued harmonic polynomials of degree at most $d$. 
Define $\vec q: \R^n \mapsto \R^m$ to be the vector-valued polynomial such that $\vec q(\x) = ( q_1(\x), \cdots, q_m(\x) )$ 
denote the vector-valued polynomial 
$\vec q_i(\vec x) = q^{(i)}(\vec x)$.
Assume that $\snorm{L_2}{\vec q} \leq 1$.
Then it holds
$$
\vec q(\sqrt{1 - \delta^2} \vec x + \delta \vec z)
= \vec g(\vec x) + 
\delta \;  \vec D^{(\vec q)} \; \Jac_{\vec q}(\vec x) \; \vec z + \vec e(\vec x, \vec z) \;,
$$
where 
$\vec D^{\vec q} \in \R^{m \times m}$ is a diagonal matrix such that $\vec D^{\vec q}_{i,i} = \lp(\sqrt{1 - \delta^2}\rp)^{ \deg(q_i) - 1}$ for $i \in [m]$, 
$\snorm{L_2}{\vec g_i} \leq 1 $, 
and $\snorm{L_2}{\vec e_i} \leq 2^{d/2} \; \delta^2$.
\end{corollary}
\begin{proof}
The result follows by applying \Cref{eq:real-value-taylor-like expansion} to each polynomial $\vec q_i(\vec x)$.
\end{proof}

To focus on ``local'' behaviors of the transformed distribution $\vec q( \sqrt{1 - \delta^2} \vec x + \delta \vec z )$, we can condition on $\vec x$ being some fixed value $\vec x^\ast$ and view $\vec q\lp( \sqrt{1 - \delta^2} \vec x^\ast + \delta \vec z \rp)$ as a function of $\vec z$. 
Then using \Cref{cor:taylor-like-expansion}, it is not hard to check that the linear component dominates over the higher order terms in $\vec e(\vec x^\ast, \vec z)$ while viewed as a function of $\vec z$, as long as the value $\vec x^\ast$ ensures that $\Jac_{\vec q}(\vec x)$ is non-singular and satisfies some other mild technical conditions.
Since $\vec q$ consists of a non-singular set of polynomials, we know that $\vec x \sim \normal(\vec 0, \vec I)$ indeed satisfies the conditions with high probability.
The detailed argument is given below.
\begin{proof}[Proof of \Cref{lem:mixture-of-linear-transform}]
Instead of analyzing the distribution of $\vec q(\vec x)$ 
for $\vec x \sim \normal(\vec 0, \vec I)$ 
directly, we consider the distribution of 
$\vec q( \sqrt{1 - \delta^2} \vec x + \delta \vec z )$ 
for $\vec x, \vec z$ being two \iid random variables from $\normal(\vec 0, \vec I)$.
It is easy to see that the two distributions are the same since the distribution of $\sqrt{1 - \delta^2} \vec x + \delta \vec z$ is still $\normal(\vec 0, \vec I)$.

Using \Cref{cor:taylor-like-expansion}, 
we have
$$
\vec q(\sqrt{1 - \delta^2} \; \vec x + \delta \; \vec z)
= \vec g(\vec x) + 
\sqrt{1-\delta^2}^{d-1} \; 
\delta \; \vec D^{\vec q} \; \Jac_{\vec q}(\vec x) \; \vec z
+ \vec e(\vec x, \vec z).
$$
We will first condition on $\vec x$ being some fixed value $\vec x^*$ that satisfies the following:
\begin{enumerate}
\item 
$\snorm{2}{\vec g(\vec x^*)} \leq 
\sqrt{m} \; N^{d} \log^{d}(1/\delta)$. 
\item $\snorm{F}{ \vec J_{\vec q}(\vec x^*) } \leq 
\sqrt{m} \; N^{d}\log^{d}(1/\delta)$.
\item The singular values of $\vec J_{\vec q}(\vec x^*) $ are lower bounded by $\delta^{1/2}$. 
Notice that this immediately implies that $\sigma_{\min} \lp( \vec D^{\vec q} \; \vec J_{\vec q}(\vec x^*) \rp) \geq \Omega( \delta^{1/2} )$
\item $\vec e(\vec x^\ast, \vec z)$ as a function of $\vec z$ has small $L_2$-norm, i.e., 
$$ 
\sqrt{ \E_{\vec z \sim \normal(\vec 0, \vec I)} \lp[ 
\snorm{2}{ \vec e(\vec x^\ast, \vec z) }^2
\rp] } \leq \delta^{ 1.9 }.
$$
\end{enumerate}
Recall that $\vec D^{\vec q} \in \R^{m \times m}$ is a diagonal matrix, where $\vec D^{\vec q}_{i,i} = \lp( \sqrt{1 - \delta^2} \rp)^{ \deg(q_i)-1 }$ for $i \in [m]$ as specified in \Cref{cor:taylor-like-expansion}.
When $\delta$ is sufficiently small given $d \geq \deg(q_i)$, we then have $ (1 - o(1)) \vec I \preceq \vec D^{\vec q} \preceq \vec I $. Denote $\vec J_{\vec x^*} := \vec D^{\vec q} \Jac_{\vec q}(\x^\ast)$. 
We thus have $\snorm{F}{ \vec J_{\vec x^*}} \leq \sqrt{m} N^d \log^d(1/\delta)$ and 
$\sigma_{\min}\lp( \vec J_{\x^*} \rp) \geq \delta^{1/2}/2$.

Notice that we condition on $\vec x$ being some \emph{fixed} value $\vec x^\ast$ that satisfies the above restrictions rather than conditioning on $\vec x$ satisfying the conditions.
After such a conditioning, the randomness comes from $\vec z$ solely.
Now consider the row space of 
$\vec J_{\x^*}$.
We will next decompose $\vec z$ in terms of its projection in the row space and in the orthogonal complement of the row space.
Since $\vec J_{\x^*}$ is non-singular, the row space is $m$-dimensional. 
Let $\vec P_{\vec x^\ast} \in  \R^{n \times m}$ be a matrix obtained by concatenating a set of orthonormal vectors from the row space of  $\vec J_{\x^*}$, and
$\vec P_{\vec x^\ast}^{\perp} \in \R^{n \times (n-m)}$ be a matrix obtained by concatenating a set of orthonormal vectors from the orthogonal complement of the row space of $\Jac_{\vec q}(\vec x^\ast)$.
Then the distribution of $\vec z \sim \normal(\vec 0 , \vec I_n)$ is the same as 
$\vec P_{\vec x^\ast} \; \vec u + \vec P^{\perp}_{\vec x^\ast} \; \vec v$
for $\vec u \sim \normal(\vec 0, \vec I_m)$ and $\vec v \sim \normal(\vec 0, \vec I_{n-m})$.
Therefore, the distribution after the conditioning on $\vec x = \vec x^\ast$ is equivalent to
$$
\vec g(\vec x^\ast) + \delta \; \vec J_{\x^*} \; 
\lp( \vec P_{\vec x^\ast} \vec u + \vec P_{\vec x^\ast}^{\perp} \vec v  \rp)
+ \vec e( \vec x^\ast, \vec P_{\vec x^\ast} \vec u + \vec P_{\vec x^\ast}^{\perp} \vec v ) \, ,
$$
where $\vec u \sim \normal(\vec 0, \vec I_{m})$ and $\vec v \sim \normal(\vec 0, \vec I_{n-m})$.
Notice that we always have
$\vec J_{\x^*} \vec P_{\vec x^\ast}^{\perp} \vec v = 0  $ by the definition of $\vec P_{\vec x^\ast}^{\perp}$.
We claim the $L_2$-norm of $\vec e( \vec x^\ast, \vec P_{\vec x^\ast} \vec u + \vec P_{\vec x^\ast}^{\perp} \vec v )$ as a function of $\vec u, \vec v$ is the same as the $L_2$-norm of $\vec e(\vec x^\ast, \vec z)$ as a function of $\vec z$.
This is because
\begin{align*}
\E_{\vec u \sim \normal(\vec 0, \vec I_m),
\vec v \sim \normal(\vec 0, \vec I_{n-m})}
\lp[ 
\vec e^2(\vec x^\ast, 
\vec P_{\vec x^\ast} \vec u
+ \vec P_{\vec x^\ast}^{\perp} \vec v
)
\rp]
&= 
\E_{\vec u \sim \normal(\vec 0, \vec I_m),
\vec v \sim \normal(\vec 0, \vec I_{n-m})}
\lp[ 
\vec e^2\lp(\vec x^\ast, 
\lp[ \vec P_{\vec x^\ast} \vec P_{\vec x^\ast}^{\perp} \rp] 
\begin{bmatrix}
    \vec u\\
    \vec v
\end{bmatrix}
\rp)
\rp] \\
&= 
\E_{\vec z \sim \normal(\vec 0, \vec I))}
\lp[ 
\vec e^2\lp(\vec x^\ast, 
\lp[ \vec P_{\vec x^\ast} \vec P_{\vec x^\ast}^{\perp} \rp] 
\vec z
\rp)
\rp] \\
&= 
\E_{\vec z \sim \normal(\vec 0, \vec I))}
\lp[ 
\vec e^2\lp(\vec x^\ast,
\vec z
\rp)
\rp] \, ,
\end{align*}
where the last equality follows from the fact that 
$[ \vec P_{\vec x^\ast} \vec P_{\vec x^\ast}^{\perp} ]$ is a unitary matrix.
Therefore, the distribution after the conditioning on $\vec x = \vec x^\ast$ is equivalent to
$$
\vec g(\vec x^\ast) + \delta \; \vec J_{\x^*} \; 
\vec P_{\vec x^\ast} \vec u 
+ \vec e'_{\vec x^\ast}( \vec u, \vec v )
$$
for some degree-$d$ polynomial $\vec e'_{\vec x^\ast}$ with $L_2$-norm bounded above by $ \delta^{ 1.9 } $.

We next condition on $\vec v = \vec v^\ast$ such that
$ \vec e'_{\vec x^\ast}( \vec u, \vec v^\ast )  $ as a function of $\vec u$ has $L_2$-norm bounded by $\delta^{1.875}$, i.e., 
\begin{align}
\label{eq:v-criteria}
\sqrt{\E_{ \vec u \sim \normal(\vec 0, I_m) }
\lp[ \vec e'_{\vec x^\ast}( \vec u, \vec v^\ast ) \rp]}
\leq \delta^{1.875}.
\end{align}

We will denote the new function after fixing $\vec v = \vec v^\ast$ as
$ \vec  r_{\vec x^\ast, \vec v^\ast}$.
The distribution after the conditioning is equivalent to
$$
\vec g(\vec x^\ast) + \delta \; \vec J_{\x^*} \; 
\vec P_{\vec x^\ast} \vec u 
+ \vec r_{\vec x^\ast, \vec v^\ast}( \vec u) \, ,
$$
for $\vec u \sim \normal(\vec 0, I_m)$.




Suppose that $\vec J_{\x^*}$ has the (compact) singular value decomposition
$$
 \vec J_{\x^*} = \vec U_{\vec x^\ast} \Sigma_{\vec x^\ast}
\vec V_{\vec x^\ast} \, ,
$$
where $\vec U_{\vec x^\ast} \in \R^{m \times m}$, $\Sigma_{\vec x^\ast} \in \R^{m \times m}$ and $\vec V_{\vec x^\ast} \in \R^{m \times n}$.
Since both $\vec V_{\vec x^\ast}$ and $\vec P_{\vec x^\ast}$ are semi-unitary, one can verify that $
\vec V_{\vec x^\ast} \vec P(\vec x^\ast)$ is some unitary matrix of dimension $\R^{m \times m}$.
Hence, $\vec J_{\x^*} \vec P_{\vec x^\ast}$ is some $\vec M_{\vec x^\ast} \in \R^{m \times m}$ whose singular values are bounded below by $\sqrt{\delta}/2$ (since its singular values are exactly those of $\vec J_{\x^*}$).
Furthermore, since 
$\snorm{F}{ \vec J_{\x^*} }$ is at most 
$\sqrt{m} N^{2d} \log^{2d}(1/\delta)$, it follows that the singular values of $\vec M_{\vec x^\ast}$ are also bounded above by $\sqrt{m} N^{2d} \log^{2d}(1/\delta)$.
Hence, we conclude that the distribution after the conditioning of $\vec x = \vec x^\ast, \vec v = \vec v^\ast$ is that of
\begin{align} \label{eq:qbar-residue-form}
\vec g(\vec x^\ast) + \delta \; \vec M_{\vec x^\ast} \vec u 
+ \vec r_{\vec x^\ast, \vec v^\ast}( \vec u) \, ,
\end{align}
where $\vec g(\vec x^\ast) \in \R^m$, $\vec M_{\vec x^\ast} \in \R^{m \times m}$, and $\vec r_{\vec x^\ast, \vec v^\ast}: \R^m \mapsto \R^m$. 

We know that $\snorm{2}{ \vec g(\vec x^\ast) } \leq 
O_{d,m,N}( \log^{d}(1/\delta) )$ by our choice of $\vec x^\ast$. This is indeed bounded above by $\log^{d+1}(1/\delta)$ when $\delta$ is sufficiently small given $d,m,N$.
We have just shown that the singular values of  $\vec M_{\vec x^\ast}$ are between $\delta$ and $O_{d,m,N}\lp(\log^{d}(1/\delta)\rp) < \log^{d+1}(1 / \delta)$. Hence, the singular values of $\delta \vec M_{\vec x^\ast}$ should be between $\delta^{3/2}$ and $\delta \log^{d+1}(1/\delta)$.
For $\vec r_{\vec x^\ast, \vec v^\ast}$, our choice of $\vec x^\ast$ and $\vec v^\ast$ guarantees that its $L_2$-norm is at most $\delta^{1.875} = \lp(\delta^{3/2}\rp)^{5/4}$.
This verifies that the transformation in \Cref{eq:qbar-residue-form} is indeed a degree-$d$, $(d+1)$-bounded, $\delta^{3/2}$-approximate linear transform.

It remains to argue that $\x  \sim \normal(\vec 0, \vec I)$
satisfies the criteria of $\vec x^\ast$ (specified at the beginning of the proof) with probability at least $1 - O(\delta^N)$, and for each valid $\vec x^\ast$,
$\vec v \sim \normal(\vec 0, \vec I_{n-m})$ satisfies the criteria of $\vec v^\ast$ (\Cref{eq:v-criteria}) with probability at least $1 - O(\delta^N)$.
If so, it follows that the overall distribution is $O(\delta^N)$ close in total variation distance to a mixture of distributions of the desired form. 

\paragraph{Concentration of valid $\vec x^\ast$. }
We will examine the criteria for $\vec x^\ast$ one by one and argue that each criterion fails with a small probability for $\vec x \sim \normal(\vec 0, \vec I)$.
We start with the condition $ \snorm{2}{\vec g(\vec x^\ast)} \leq \sqrt{m} N^d \log^d(1/\delta)$.
For each of the coordinates, we have
$\snorm{L_2}{ \vec g_i } \leq 1$.
This implies that
$$
\snorm{L_2}{ \vec g }^2
= \sum_{i=1}^m \snorm{L_2}{ \vec g_i }^2
\leq m.
$$
Then using \Cref{lem:vector-norm-concentraion}, we have
$$
\snorm{2}{ \vec g(\vec x)}
\leq  O_d(1) \; N^{d/2} \; \log^{d/2}(1/\delta) \; 
\snorm{2}{L_2}{\vec g}
\leq \sqrt{m} \; N^{d} \; \log^{d}(1/\delta) \, ,
$$
with probability at least $1 - \delta^N$.

For $\snorm{2}{ \vec J_{q}(\vec x^\ast) }$, since $\snorm{L_2}{ \vec q_i } \leq 1$ for each $i$, we can use \Cref{lem:Jacobian-Frob-concentration} and conclude that
$$
\snorm{2}{\Jac_{\vec q}(\vec x)}
\leq 
\snorm{F}{\Jac_{\vec q}(\vec x)}
\leq N^{d} \log^{d}( 1/\delta) \sqrt{m} \, ,
$$
with probability at least $1 - \delta^N$.


Next, the singular values of $\Jac_{\vec q}(\vec x^\ast)$ are bounded from below by $\sqrt{\delta}$ with probability at least $1 - \delta^N$ by our assumption that $\vec q$ satisfies \Cref{eq:non-singular-assumption}.

Finally, 
we wish to show that
$$
\sqrt{
\E_{\vec z \sim \normal(\vec 0, \vec I)}
\lp[ 
\snorm{2}{ \vec e(\vec x, \vec z) }^2
\rp]
}
\leq \delta^{1.9} \, ,
$$
with probability at least $1 - \delta^N$ over the randomness of 
$\vec x \sim \normal(\vec 0, \vec I)$.
Our starting point is that
$\snorm{L_2}{ \vec e_i } \leq 2^{d/2} \; \delta^2$.
It follows that
$$
\sqrt{\E_{\vec x, \vec z}
\lp[ \snorm{2}{\vec e(\vec x, \vec z)}^2 \rp]}
= 
\sqrt{\sum_{i=1}^m 
\snorm{L_2}{ \vec e_i }^2}
\leq \sqrt{m} \; 2^{d/2} \; \delta^2.
$$
Define the restricted function
$\vec e_{\vec x}(\vec z) = \vec e(\vec x, \vec z)$.
Using \Cref{lem:restricted-polynomial-norm}, we have
$$
\sqrt{
\E_{\vec z \sim \normal(\vec 0, \vec I) }
\lp[ 
\snorm{2}{\vec e_{\vec x}(\vec z)}^2
\rp]
}
\leq N^{d} \; \log^{d}(1/\delta) \; \sqrt{m} \; 2^{d/2} \; \delta^2 \, ,
$$
with probability at least $1 - \delta^N$ over the randomness of $\vec x \sim \normal(\vec 0, \vec I)$.
One can check that the right-hand side is indeed bounded by $ \delta^{1.9}$ when $\delta$ is sufficiently small given $N,d,m$.

\paragraph{Concentration of $\vec v^\ast$. }
The proof is similar to the argument of the last criterion of $\vec x^\ast$.
We already have that $
\snorm{L_2}
{\vec e'_{\vec x^\ast}(\vec u, \vec v)}
\leq \delta^{1.9}.
$
Define the restricted function 
$\vec r_{\vec x^\ast, \vec v}(\vec u) = \vec e'_{\vec x^\ast}(\vec u, \vec v)$.
By \Cref{lem:restricted-polynomial-norm}, it holds
$
\snorm{L_2}{ \vec r_{\vec x^\ast, \vec v} }
\leq  \poly_{N,d,m}( \log(1/\delta) ) \; \delta^{1.9}
\leq \delta^{1.875}
$
with probability at least $1 - \delta^N$ over the randomness of $\vec v \sim \normal(\vec 0, \vec I_{n-m})$.
This concludes the proof of \Cref{lem:mixture-of-linear-transform}.
\end{proof}

\subsection{Comparing to Reasonable Gaussians}
\label{sec:compare-reasonable-gaussian}
Note that the approximate linear transforms constructed via
\Cref{lem:mixture-of-linear-transform} are not exactly linear due to an extra polynomial error factor. 
Consequently, applying these approximate linear transforms on Gaussian inputs will not give distributions that are exactly Gaussian.
 However, as we require the $L_2$-norm of the super-linear polynomial component to be significantly smaller than the minimum singular value of the linear component, the resulting distribution will be close in total variation distance to some distribution \emph{comparable} (see \Cref{def:comparability} for the notion of distribution comparability) to a reasonable Gaussian distribution (see \Cref{intro-def:reasonable-gaussian} for the definition of a reasonable Gaussian). After that, we will combine 
 the structural result of this section (\Cref{lem:approximate-linear-sandwich}) with
 the tools developed in Sections~\ref{sec:non-singular-jacobian} and \ref{sec:approximate-linear-transform} to conclude the proof of \Cref{prop:Gaussian-Comparability}.


Consider an approximate linear transform $\vec p: \R^n \mapsto \R^m$
of the form
\begin{align}
\label{eq:approximate-linear-transform-ref}
\vec p(\vec w)    
= \vec g + \vec M \vec w + \vec r(\vec w) \, ,
\end{align}
where $\vec g$ is the offset vector, $\vec M$ is the linear transformation matrix and $\vec r$ is the residual polynomial, all satisfying the conditions of \Cref{def:approximate-linear transform} with appropriate parameters.
The main goal of this subsection is to show that 
$\vec p(\normal(\vec 0, \vec I))$ will be close to some distribution whose probability density function is within constant factors of the probability density function of some Gaussian distribution. 
In particular, we will show it is close to some distribution that is comparable to $\normal(\vec g, \vec M \vec M^T)$ in total variation distance. Moreover, by the assumptions that $\vec g, \vec M$ satisfy the conditions of \Cref{def:approximate-linear transform}, it is not hard to see that $\normal(\vec g, \vec M \vec M^T)$ is a reasonable Gaussian with appropriate parameters.
The formal statement, which is the main result of this subsection, is given below.
\begin{lemma}
\label{lem:approximate-linear-sandwich}
Let $\vec p: \R^m \mapsto \R^m$ be a degree-$d$, $b$-bounded, $\delta$-approximate linear transform. 
Furthermore, assume that $\delta$ is sufficiently small given $d,m,b$.
Then the distribution $\vec p( \normal(\vec 0, \vec I) )$ is $\delta^{ \Omega( \log(1/\delta) ) }$-close in total variation distance to some distribution that is comparable to some $(\delta, \log^{2b}(1/\delta))$-reasonable Gaussian.
\end{lemma}
The rest of this subsection is devoted to the proof of \Cref{lem:approximate-linear-sandwich}.

Analyzing the probability density function of $\vec p( \normal(\vec 0, \vec I) )$ at points far away from the origin is quite challenging due to the influence of the polynomial component. 
To circumvent this issue, we will exploit the fact that $\vec p$ is invertible within some origin-centered ball $B$ of radius bounded by $\poly_{d,b} \lp( \log(1/\delta) \rp)$.
That way, using the change of variable formula of probability density functions, we are at least able to compute explicitly the probability density function of $\vec p( \normal(\vec 0, \vec I) \mid B )$, where $\normal(\vec 0, \vec I) \mid B$ represents the normal distribution truncated to the set $B$.
Then if we can show that $\vec p( \normal(\vec 0, \vec I) \mid B )$ is comparable to $\normal(\vec g, \vec M \vec M^T )$, \Cref{lem:approximate-linear-sandwich} will then follow from the fact that $\vec p( \normal(\vec 0, \vec I) \mid B )$ and $\vec p( \normal(\vec 0, \vec I))$ are close in total variation distance.
In particular, since $\normal(\vec 0, \vec I) \mid B$ and $\normal(\vec 0, \vec I)$ are close in total variation distance when the radius of $B$ is sufficiently large, it should not be hard to conclude that the original transformed Gaussian $\vec p(\normal(\vec 0, \vec I))$ 
is close to being comparable to $\normal(\vec g, \vec M \vec M^T )$ as well. 
In the rest of this subsection, for convenience, we will use the notation
\begin{align}
\label{eq:M-bar-ref}
    \mathcal M:= \vec p( \normal(\vec 0, \vec I) ) \, , \, 
    \mathcal {\bar M}:= \vec p(\normal(\vec 0, \vec I) \mid B) \, ,
\end{align}
where $B$ is an origin-centered ball of radius $\ell:= \log(1/\delta)$.

We start by showing that $\vec p$ is locally invertible. Moreover, as one would expect, its inverse will be close to that of its linear part.
\begin{lemma}
\label{lem:p-invertibility}
Let $\vec p: \R^m \mapsto \R^m$ be the degree-$d$, $b$-bounded, $\delta$-approximate linear transform in \Cref{eq:approximate-linear-transform-ref}. 
Furthermore, assume that $\delta$ is sufficiently small given $d,m,b$.
Then $\vec p$ is invertible within some origin-centered ball $B$ with radius $\log(1/\delta)$. Moreover, if we denote by $\vec p^{-1}$ the inverse of $\vec p$ restricted to the ball $B$ 
\footnote{As $\vec p$ is not necessarily invertible globally, from this point on, we always mean the inverse of $\vec p$ restricted to $B$ if we write $\vec p^{-1}$.}, we have
\begin{align}
\snorm{2}{ \vec p^{-1}(\vec y) - \vec M^{-1}(\vec y - \vec g)  } \leq \delta^{1/8}.
\end{align}
\end{lemma}
\begin{proof}
Let $\ell = \log(1/\delta)$, and $B(\ell)$ be the origin-centered ball centered with radius $\ell$.
By definition, we can write the transformation as
\begin{align*}
\vec p(\vec w) = \vec g + \vec M \vec w + \vec r(\vec w) \, ,
\end{align*}
where $\vec g \in \R^m, \vec M \in \R^{m \times m}, \vec r: \R^m \mapsto \R^m$ satisfy the conditions specified in \Cref{def:approximate-linear transform}.
To show that $\vec p$ is invertible on the restricted domain 
$B( \ell  )$, it suffices to show that for any $\vec u \neq \vec v \in B(\ell)$, 
$\vec p(\vec u) \neq \vec p(\vec v)$.
Using the definition of $\vec p$, we have that
$$
\vec p(\vec u) - \vec p(\vec v)
= 
\vec M \lp(\vec u - \vec v\rp)  + 
\vec r(\vec u) - \vec r(\vec v).
$$
On the one hand, 
since the singular values of $\vec M$ is bounded below by $\delta$, 
we have that
$$
\snorm{2}{ \vec M (\vec u -\vec v) }
\geq  \delta \snorm{2}{\vec u - \vec v}.
$$
On the other hand, we know that 
$ \snorm{L_2}{\vec r} \leq \delta^{5/4}$ and
$\vec u, \vec v \in B_m(\ell)$.
Therefore, by \Cref{lem:bounded-lipchitz}, we have that
$$
\snorm{2}{\vec r(\vec u) - \vec r(\vec v)}
\leq O_{d,m}(1) \; \ell^d \; \delta^{5/4} \; \snorm{2}{\vec u - \vec v}.
$$
Hence, it holds that
\begin{align} \label{eq:dev-lower-bound}
\snorm{2}{ \vec p(\vec u) - \vec p(\vec v) }
\geq \frac{\delta}{2} \snorm{2}{ \vec u - \vec v } \, ,    
\end{align}
when $\delta$ is sufficiently small given $m,b,d$.
This then implies that $\vec p(\vec u) \neq \vec p(\vec v)$ for $\vec u \neq \vec v$, and hence concludes the argument for the invertibility of $\vec p$ within $B$.

We next show that for $\vec y$ within the image of $B_m(\ell)$ under $\vec p$, 
it holds
\begin{align*} \label{eq:deviation-from-linear}
\snorm{2}{\vec p^{-1}(\vec y) - \vec M^{-1} (\vec y - \vec g) }
\leq \delta^{1/8}.    
\end{align*}
By definition, there exists $\vec w \in B_m(\ell)$ such that
$$
\vec y =  \vec M \vec w + \vec g + \vec r(\vec w).
$$
Then the left hand side of \Cref{eq:deviation-from-linear} can be simplified as
$$
\snorm{2}{\vec w - \vec M^{-1} ( \vec M \vec w + \vec g + \vec r(\vec w) - \vec g) }
=
\snorm{2}{\vec M^{-1} \vec r(\vec w) }.
$$
Since the smallest singular value of $\vec M$ is bounded below by $\delta$, it holds $\snorm{2}{ \vec M^{-1} } \leq \delta^{-1}$.
On the other hand, by \Cref{lem:bounded-vector-evaluation}, we have that $
\snorm{2}{\vec r(\vec w)}
\leq O_{m,d}(1) \; \ell^d \; \delta^{5/4}.
$
It then follows that 
\begin{align*}
\snorm{2}{\vec M^{-1} \vec r(\vec w) } \leq 
O_{m,d}(1) \; \ell^d \; \delta^{1/4} \leq
\delta^{1/8} \, ,
\end{align*}
when $\delta$ is sufficiently small compared to $m, d, \ell$.
This completes the proof of \Cref{lem:p-invertibility}.
\end{proof}
After knowing that $\vec p$ is locally invertible over some ball $B$ of bounded radius, we can apply the change of variables formula to deduce the probability density function of the transformed distribution $\bar{\mathcal M} = \vec p\lp(\normal(\vec 0, \vec I) \mid B \rp)$. 
That is,
for all $\vec y \in \vec p(B)$, we have
$$
     \mathcal {\bar M}( \vec y )
     = 
     \frac{1}{ \lp(\sqrt{2 \pi}\rp)^{m} } \;
     \frac{1}{  
     \Pr_{ \vec x \sim \normal(\vec 0, \vec I)  }\lp[\vec x \in B \rp]} \; 
     \exp \lp( - \snorm{2}{\vec p^{-1}(\vec y)}^2  \rp) \;
    \frac{1}{ 
     \det \lp(  \Jac_{\vec p} ( \vec p^{-1} (\vec y) ) \rp)
     }.
$$
We would like to compare $\mathcal {\bar M}( \vec y )$ to the probability density function of $\normal(\vec g, \vec M \vec M^T)$:
$$
     \normal(\vec y ; \vec g, \vec M \vec M^T) 
     = 
     \frac{1}{ \lp(\sqrt{2 \pi}\rp)^m } \;
     \exp \lp( - (\vec y - \vec g)^T \lp( \vec M \vec M^T \rp)^{-1} (\vec y - \vec g)   \rp)
     \frac{1}{ \det(M) } \, ,
$$
where we denote by $\normal(\vec y ; \vec g, \vec M \vec M^T) $ the probability density function of $\normal(\vec g, \vec M \vec M^T)$ evaluated at $\vec y$.
Specifically, we would like to show that $\mathcal {\bar M}(\vec y)$ is pointwise within a constant factor of $\normal(\vec y ; \vec g, \vec M \vec M^T)$ for all $\vec y \in \vec p( B )$.
It is not hard to see that $\Pr_{ \vec x \sim \normal(\vec 0, \vec I) } \lp[ \vec x \in B\rp]$ is close to $1$, since the radius of $B$ is sufficiently large.
Moreover, the expressions within the exponential function are also close to each other due to the approximate linearity of $\vec p^{-1}$ shown in \Cref{lem:p-invertibility}.
Hence, it remains to show that the determinant of the Jacobian of $\vec p$ is close to that of $M$.
In fact, we will show a stronger property: the singular values of the two matrices are close up to a multiplicative factor of at most $1 \pm \poly(\delta)$.
\begin{lemma}
\label{lem:Jacobian-bound}
Let $\vec p: \R^m \mapsto \R^m$ be the degree-$d$, $b$-bounded, $\delta$-approximate linear transform in \Cref{eq:approximate-linear-transform-ref}, $\ell = \log(1/\delta)$, and $B$ be an origin-centered ball of radius $\ell$.
Furthermore, assume that $\delta$ is sufficiently small given $d,m,b$.
Then for all $\vec x \in B$ it holds
$$
( 1 - O( \delta^{0.1} )) \sigma_i(M) \leq \sigma_i( \Jac_{\vec p}(\vec x) )
\leq 
(1 +O(\delta^{0.1}) ) \sigma_i( M ).
$$
Consequently, if $\delta$ is sufficiently small given $m$, it holds
\begin{align}
0.5 \det (M) \leq \det \lp( \Jac_{\vec p}(\vec x) \rp) \leq 2 \det (M).
\end{align}
\end{lemma}
\begin{proof}
By \Cref{eq:approximate-linear-def}, we have
$$
\Jac_{\vec p}
\lp( \vec w \rp)
= \vec M  +  \Jac_{\vec r}( \vec w ) \, ,
$$
where $M$ is a matrix whose singular values are bounded below by $\delta$, and
$ \vec r: \R^m \mapsto \R^m$ satisfies
$\snorm{L_2}{\vec r} \leq \delta^{5/4}$.
By \Cref{lem:bounded-Jacobian-evaluation}, we have
$$
\snorm{F}{ \Jac_{\vec r}(\vec w) }
\leq O_{d,m}(1) \; 
\ell^d \;  \delta^{5/4}.
$$
It follows that all the singular values of $\Jac_{\vec r}(\vec w)$ are also bounded above by $O_{d,m}(1) \; 
\ell^d \;  \delta^{5/4}$.
Therefore, 
using Weyl's inequality (\Cref{fact:weyl}) for singular values, we have
$$
\abs{\sigma_i( \vec M_{\vec x} + \Jac_{\vec r}( \vec w ))
- 
\sigma_i(\vec M_{\vec x})}
\leq \sigma_{\max}( \Jac_{\vec r}( \vec w ) )
\leq O_{m,d}(1) \; 
\ell^d \;  \delta^{5/4} \leq \delta^{1.1} \, ,
$$
where the last inequality follows from our choice of $\delta$ to be sufficiently small.
As $\sigma_i( \vec M ) \geq \delta$, 
we have
$$
\frac{ \sigma_i( \vec M + \Jac_{\vec r}( \vec w )) }
{ \sigma_i( \vec M )} 
= (1 \pm  O( \delta^{0.1} ) ).
$$
It then follows that
$$
 \det(\Jac_{\vec p}(\vec w)  ) =  \lp( 1 \pm o(1) \rp) \det( M ).
$$    
This completes the proof of \Cref{lem:Jacobian-bound}.
\end{proof}
With \Cref{lem:p-invertibility} and \Cref{lem:Jacobian-bound}, we can then bound the ratio between the probability density functions of $\normal(\vec g, \vec M \vec M^T)$ and $\mathcal {\bar M}$.
\begin{lemma}
\label{lem:pdf-ratio-bound}
Let $\vec p: \R^m \mapsto \R^m$ be the degree-$d$, $b$-bounded, $\delta$-approximate linear transform in \Cref{eq:approximate-linear-transform-ref}, $\ell = \log(1/\delta)$, $B$ be an origin-centered ball of radius $\ell$, and $\bar {\mathcal M}$ be the distribution defined in \Cref{eq:M-bar-ref}.
Furthermore, assume that $\delta$ is sufficiently small given $d,m,b$.
Then for all $\vec y$ within the support of $\bar{\mathcal M}$, 
it holds
$$
(1 - O( \delta^{0.1} ) )\normal(\vec y; \vec g, \vec M \vec M^T)
\leq \bar{\mathcal M}(\vec y)
\leq 
(1 + O( \delta^{0.1} ) )\normal(\vec y; \vec g, \vec M \vec M^T).
$$
\end{lemma}
\begin{proof}
Recall that $\mathcal {\bar M} = \vec p( \normal(\vec 0, \vec I) \mid B )$.
Therefore, 
the support of $\mathcal {\bar M}$ is exactly $\vec p(B)$.
By \Cref{lem:p-invertibility}, $\vec p$ is invertible over $B$.
Hence, we can apply the change of variable formula of probability density function. 
In particular, this gives us that for all $\vec y \in \vec p(B)$,  we have that
\begin{align*}
     \mathcal {\bar M}( \vec y )
     = 
     \frac{1}{ \sqrt{2 \pi} } \;
     \frac{1}{  \Pr_{ \vec x \sim \normal(\vec 0, \vec I) } \lp[ \vec x \in B \rp] } \; 
     \frac{1}{ 
     \det \lp(  \Jac_{\vec p} ( \vec p^{-1} (\vec y) ) \rp)
     }
     \exp \lp( - \snorm{2}{\vec p^{-1}(\vec y)}^2  \rp).
\end{align*}
The term $\Pr_{ \vec x \sim \normal(\vec 0, \vec I) } \lp[ \vec x \in B \rp]$ is at least  $1 -  \delta^{\log(1/\delta)}$ by our choice of $\ell$.  The Jacobian term is of order 
$\lp(1 \pm o(1) \rp) \det^{-1}(M)$ by \Cref{lem:Jacobian-bound}. 
It remains to analyze the expression inside the exponential function.
By \Cref{lem:p-invertibility}, the inverse function is approximately linear. Hence, we can write 
\begin{align*}
\snorm{2}{\vec p^{-1}(\vec y)}^2
&= \lp( M^{-1}\lp(\vec y - \vec g\rp) + \vec \xi_{\vec y} \rp)^T
\lp( M^{-1}\lp(\vec y - \vec g\rp) + \vec \xi_{\vec y} \rp) \\
&=    
\lp(\vec y - \vec g\rp)^T 
\lp(\vec M \vec M^T\rp)^{-1}
\lp(\vec y - \vec g\rp)^T 
+ 2 \; \vec \xi_{\vec y}^T \; M^{-1}  \;\lp(\vec y - \vec g\rp)
+ \snorm{2}{ \xi_{\vec y} }^2 \, ,
\end{align*}
where the term $\xi_{\vec y}$ is some error term depending on $\vec y$
with $\snorm{2}{\xi_{\vec y}} \leq \delta^{1/8}$.
Except for the first quadratic term that is standard in the probability density function of a normal distribution, we need to argue the two remaining terms are both small. 
For the constant term, since $\snorm{2}{ \xi_{\vec y} } \leq \delta^{1/8}$, it immediately follows that the term is of order $o(1)$.

For the middle linear term, we need to take advantage of the fact that $\vec y$ is a vector from the set $\vec p(B)$.
By definition, we know that there exists $\vec w \in B$ such that
$$
\vec y = \vec g + \vec M \vec w + \vec r(\vec w) \, ,
$$
which implies that
\begin{align*}
\snorm{2}{\vec M^{-1}  \;\lp(\vec y - \vec g\rp)}
&= \snorm{2}{ \vec w + \vec M^{-1} \vec r(\vec w) } \\
&\leq \snorm{2}{\vec w}   + \snorm{2}{ \vec M^{-1} \vec r(\vec w) }
\leq \ell +  \snorm{2}{ \vec M^{-1} \vec r(\vec w) } \, ,
\end{align*}
where we first apply the triangle inequality and in the second inequality we bound $\snorm{2}{\vec w}$ by $\ell$ since $\vec w \in B$, which is a ball of radius $\ell$.
Recall that $\snorm{L_2}{\vec r} \leq \delta^{5/4}$ by our assumption.
Hence, using \Cref{lem:bounded-vector-evaluation}, we have that
$$
\snorm{2}{ \vec r(\vec w) }
\leq O_{m,d}(1) \; \ell^d \; \delta^{5/4}
= o(\delta).
$$
By the assumption that $\vec p$ is a 
degree-$d$, $b$-bounded, $\delta$-approximate linear transform, we have that $\sigma_{\max}\lp( \vec M^{-1} \rp) = 
1 / \sigma_{\min}\lp( \vec M \rp)
\leq \delta^{-1}$.
Hence, it follows that
$$ 
\snorm{2}{\vec M^{-1}  \;\lp(\vec y - \vec g\rp) } 
\leq  \ell +
\snorm{2}{ \vec M^{-1} \vec r(\vec w) }
\leq
\ell + o(1).
$$
This further implies that
$2 \; \vec \xi_{\vec y}^T \; M^{-1}  \;\lp(\vec y - \vec g\rp) \leq O(\ell \; \delta^{1/8}) = o(1)$.
Hence, overall, we have
\begin{align*}
\exp\lp( \snorm{2}{ \vec p^{-1}(\vec y) } \rp)
&=  
\exp \lp( \lp(\vec y - \vec g\rp)^T 
\lp(\vec M \vec M^T\rp)^{-1}
\lp(\vec y - \vec g\rp)^T \pm o(1) \rp) \\
&= \lp( 1 \pm o(1) \rp)\; 
\exp \lp( \lp(\vec y - \vec g\rp)^T 
\lp(\vec M \vec M^T\rp)^{-1}
\lp(\vec y - \vec g\rp)^T \rp).
\end{align*}
Combining the above gives that for all $\vec y \in \vec p( B )$,  we have
\begin{align*}
     \mathcal {\bar M}( \vec y )
    &= (1 \pm o(1)) \;  \normal\lp( \vec y; \vec g, \vec M \vec M^T  \rp).
\end{align*}    
This concludes the proof of \Cref{lem:pdf-ratio-bound}.
\end{proof}

At this point, it is tempting to conclude that the distribution $\bar {\mathcal M}$ is comparable to $\normal(\vec g, \vec M\vec M^T)$.
However, we emphasize that the above lemma applies only for points \emph{within the support of $\bar{\mathcal M}$}. 
Unfortunately, the supports of the two distributions are not aligned.
Yet, recall that our final goal is only to show closeness to some distribution comparable to the Gaussian distribution $\normal(\vec g, \vec M \vec M^T)$ in total variation distance.
Hence, the final missing piece is to show that the support of $\bar{\mathcal M}$ (which is the image of $B$) under $\vec p$) is not too ``light''.
More accurately, we need to show that the points outside the support of $\bar{\mathcal M}$ have negligible mass under $\normal(\vec g, \vec M \vec M^T)$.
To show this, we require some basic understanding of what the set $\vec p(B)$ looks like. 
By our assumption, $\vec p$ is approximately a linear map. If $\vec p$ were exactly a linear map, $\vec p(B)$ would then be an ellipsoid.
The fact that $\vec p$ is only approximately linear makes the exact characterization of $\vec p(B)$ difficult.
Nonetheless, we manage to show below that $\vec p(B)$ still \emph{includes} an ellipsoid. 
Specifically, it includes the ellipsoid $ \vec M B' + \vec g  $, where $B'$ is a ball of slightly smaller radius than $B$.
\begin{lemma}
\label{lem:support-include-ellipsoid}
Let $\vec p: \R^m \mapsto \R^m$ be the degree-$d$, $b$-bounded, $\delta$-approximate linear transform in \Cref{eq:approximate-linear-transform-ref}, and $B$ be an origin-centered ball of radius $\ell = \log(1/\delta)$. 
Furthermore, assume that $\delta$ is sufficiently small given $d,m,b$.
Then the image of $B$ under $\vec p$ includes the ellipsoid $\vec M B' + \vec g$, where $B'$ is an origin-centered ball of radius $0.9 \ell$.
\end{lemma}
The proof of \Cref{lem:support-include-ellipsoid} is similar to that of Lemma 1.3 in Chapter XIV of \cite{lang2012real}.
There, the author shows that if a function $\vec f$ is close of being the identity mapping, then the image of a ball under $\vec f$ will include a ball of only a slightly smaller radius. 
The proof strategy is the following. 
Suppose that one wants to show that a point $\vec y$ within this ball of a smaller radius belongs to the image of some ball $B$ under the function.
One can then use $\vec y$, the function, and its derivatives to construct a contraction mapping from $B$ to itself and demonstrate the fixed point of the mapping is exactly $\vec y$.
In our setting, our function $\vec p$ is only close to being a non-singular linear transformation. 
Nonetheless, it turns out that one can use Newton–Raphson's iteration to construct a contraction mapping and mimic the proof strategy of Lemma 1.3 in \cite{lang2012real}.
\begin{proof}[Proof of \Cref{lem:support-include-ellipsoid}]
We will show the following. For all $\vec y \in \vec g + \vec M B'$, we can find $\vec x \in B $ such that
$ \vec p(\vec x) = \vec y $.
Consider the mapping $T_{\vec y}: B \mapsto \R^n$ defined using a slightly modified Newton-Raphson iteration:
$$
T_{\vec y}(\vec x) = \vec x - 
\vec M^{-1}
\lp( \vec p(\vec x) - \vec y\rp) \, ,
$$
where $\vec M$ is the non-singular matrix in the linear term in \Cref{eq:approximate-linear-def} \footnote{One can see that $\Jac_{\vec q}(\vec x)$ is very close to $\vec M$. Hence, this is almost the standard Newton–Raphson's iteration.}.
We will show that the map is a contraction on $B$.
In particular, for all $\vec x, \vec x' \in B$, we claim that
\begin{align*}
\snorm{2}{ T_{\vec y}(\vec x) - T_{\vec y}(\vec x') }    
\leq o(1) \; \snorm{2}{ \vec x - \vec x' }.
\end{align*}
Using the definition of $T_{\vec y}$, we can write
\begin{align} \label{eq:map-expansion}
\snorm{2}{ T_{\vec y}(\vec x) - T_{\vec y}(\vec x') }    
=
\snorm{2}{ 
\vec M^{-1}
\lp( 
\vec M \lp( \vec x - \vec x' \rp)
- \lp( \vec p(\vec x) - \vec p(\vec x') \rp)
\rp)
}.
\end{align}
By the intermediate value theorem, there exists some
$\vec {\tilde x} = t \vec x + (1- t) \vec x'$ for $t \in (0, 1)$ such that
$$
\vec p(\vec x) - \vec p(\vec x')
= \Jac_{\vec p}(\vec {\tilde x}) \lp( \vec x - \vec x' \rp).
$$
We can then further rewrite \Cref{eq:map-expansion} as 
$$
\snorm{2}{ T_{\vec y}(\vec x) - T_{\vec y}(\vec x') }  
=
\snorm{2}{
\vec M^{-1} 
\lp(\vec M - \Jac_{\vec p}(\vec {\tilde x})\rp) \lp( \vec x - \vec x' \rp)
}.
$$
On the other hand, since $\vec {\tilde x} \in B$, 
we have
\begin{align*}
\snorm{F}{ \vec M - \Jac_{\vec p}(\vec {\tilde x}) }    
= 
\snorm{F}{ \Jac_{\vec r}(\vec {\tilde x}) }    
\leq O_{d,m}(1) \; \ell^{d} \; \snorm{L_2}{\vec r} 
< o(\delta) \, ,
\end{align*}
where in the equality we use the defining equation of $\vec p$ (\Cref{eq:approximate-linear-def}), 
in the first inequality we use \Cref{lem:bounded-Jacobian-evaluation}, and the second inequality is true since $\snorm{L_2}{\vec r} \leq \delta^{5/4}$ and $\delta$ is sufficiently small given $d,m,b$.
Since the operator norm of matrices is sub-multiplicative, we then further have 
$$
\snorm{2}{ T_{\vec y}(\vec x) - T_{\vec y}(\vec x') }  
\leq 
\sigma_{\max} (M^{-1}) \; \sigma_{\max} \lp( \vec M - \Jac_{\vec p}(\vec {\tilde x}) \rp)
\snorm{2}{ \vec x - \vec x' }
< o(1) \; \snorm{2}{ \vec x - \vec x' } \, ,
$$
since the singular value of $M$ is bounded below by $\delta$ and the singular value of $\vec M - \Jac_{\vec p}(\vec {\tilde x})$ is bounded above by $o(\delta)$.
It then follows that $T_{\vec y}$ is actually a contraction mapping from $B$ to itself.
By the contraction mapping theorem, there is a fixed point of $T_{\vec y}$.
Lastly, since $M^{-1}$ is non-singular, it is easy to see that the fixed point $\vec x^*$ must satisfy $f(\vec x^*) = \vec y$.
This concludes our proof.
\end{proof}
Then we can easily bound the mass of the points outside the support of $\bar{\mathcal M}$ under $\normal(\vec g, \vec M \vec M^T)$ via standard Gaussian concentration argument.
\begin{corollary}
\label{cor:outside-support-mass}
Let $\vec p: \R^m \mapsto \R^m$ be the degree-$d$, $b$-bounded, $\delta$-approximate linear transform in \Cref{eq:approximate-linear-transform-ref} and $\bar {\mathcal M}$ be the distribution defined in \Cref{eq:M-bar-ref}.
Furthermore, assume that $\delta$ is sufficiently small given $d,m,b$.
Then it holds
$
\Pr_{ \vec x \sim \normal(\vec g, \vec M \vec M^T) } \lp[ \vec x \not \in \text{support}(\bar{\mathcal M}) \rp] \leq  \delta^{ \Omega(\log(1/\delta)) }.
$
\end{corollary}
\begin{proof}
Let $B'$ be an origin-centered ball of radius $0.9 \ell$.
By \Cref{lem:support-include-ellipsoid}, 
it suffices to argue that
$$
\Pr_{ \vec x \sim \normal(\vec g, \vec M \vec M^T) }
\lp[ 
\vec x \in \vec M B' + \vec g
\rp] \geq 1- \delta^{ \Omega( \log(1/\delta) ) }.
$$
Notice that, by applying the transformation $\x \mapsto  \vec M^{-1}\lp(\x - \vec g\rp)$, the left hand side is equal to
$
\Pr_{ \vec x \sim \normal(\vec 0, \vec I) }
\lp[  \vec x \in B' \rp].
$
If $m = 1$, deriving a lower bound of this probability boils down to standard Gaussian tail bounds. 
For $m > 2$, we show that this situation is not much different when $\delta$ is sufficiently small given $m$.
\begin{claim}
\label{clm:low-dim-gaussian-tail}
It holds 
$
\Pr_{ \vec x \sim \normal(\vec 0, \vec I_m) }
\lp[  \snorm{2}{\vec x} \geq v \rp]
\leq \exp\lp( - \Omega(v^2)  \rp)
$
when $v$ is sufficiently large given $m$.
\end{claim}
\begin{proof}
The above is equivalent to $\snorm{2}{ \vec x }^2 \leq v^2$ and $\snorm{2}{ \vec x }^2$ is exactly the $\chi^2$ distribution with mean $m$.
By the tail bound of the $\chi^2$ distribution, we hence have
$$
\Pr_{ \vec x \sim \normal(\vec 0, \vec I) }
\lp[  
\sum_{i=1}^m \vec x_i^2 \geq \frac{v^2}{m} \; m 
\rp]
\leq \lp( \frac{v^2}{m} \; \exp \lp(1 - \frac{v^2}{m} \rp) \rp)^{m/2}
\leq \exp \lp( - \Omega\lp( \ell^2 \rp)  \rp) \, ,
$$
where the last inequality is true as long as $v$ is sufficiently large given $m$.
\end{proof}
Applying \Cref{clm:low-dim-gaussian-tail} with $v = 0.9 \ell$ then concludes the proof of \Cref{cor:outside-support-mass}.
\end{proof}

Now, we are ready to conclude the proof of \Cref{lem:approximate-linear-sandwich}.

\begin{proof}[Proof of \Cref{lem:approximate-linear-sandwich}]
By definition, we can write the transformation as
\begin{align} \label{eq:approximate-linear-def}
\vec p(\vec w) = \vec g + \vec M \vec w + \vec r(\vec w)\, ,
\end{align}
where $\vec g \in \R^m, \vec M \in \R^{m \times m}, \vec r: \R^m \mapsto \R^m$ satisfy the conditions specified in \Cref{def:approximate-linear transform}.
It follows from the definition that $\normal(\vec g, \vec M \vec M^T)$ is a $(\delta, \log^{2b}(1/\delta))$-reasonable Gaussian.
Our goal is to show that the transformed distribution of $\vec p$ applied to $\normal(\vec 0, \vec I)$ is close to some distribution
comparable to $\normal( \vec g, \vec M \vec M^T )$.

It is challenging to analyze the behavior of the transformed distribution $\mathcal M = \vec p(\normal(\vec 0, \vec I))$
in regions that are far away from the origin.
On the other hand, we know that $\vec p$ is locally invertible within some origin-centered ball with bounded radius by \Cref{lem:p-invertibility}.
For this purpose, we will restrict our attention to some origin-centered ball $B$ of radius $\ell := \log(1/\delta)$.
This allows us to consider the distribution $\bar{\mathcal M} = \vec p( \normal(\vec 0, \vec I) \mid B )$ instead.
Using the data-processing inequality for total variation distance, we know that the total variation distance between 
$\vec p( \normal(\vec 0, \vec I) \mid B )$ and $\vec p( \normal(\vec 0, \vec I) )$ can be bounded above by
\begin{align} \label{eq:truncated-transform-distance}
    \dtv \lp( \vec p( \normal(\vec 0, \vec I) \mid B ),  \vec p( \normal(\vec 0, \vec I) )\rp)
    \leq \dtv \lp(\normal(\vec 0, \vec I) \mid B, \normal(\vec 0, \vec I)\rp)
    \leq \delta^{\Omega\lp( \log(1/\delta) \rp)} \, ,
\end{align}
where in the second inequality we again use \Cref{clm:low-dim-gaussian-tail} with $v = \ell$.

By \Cref{lem:pdf-ratio-bound}, we have
\begin{align*}
     \mathcal {\bar M}( \vec y )
    &= (1 \pm o(1)) \;  \normal\lp( \vec y; \vec g, \vec M \vec M^T  \rp).
\end{align*}
Consider the positive measure $\mathcal D: \R^m \mapsto \R$ defined as
\begin{align*}
    \mathcal D(\vec y) = 
    \begin{cases}
\mathcal {\bar M}(\vec y) &\text{ if  $\vec y  \in \vec p(B)$} \, ,\\
\normal\lp( \vec y; \vec g, \vec M \vec M^T  \rp) &\text{ otherwise.}
\end{cases} 
\end{align*}
On the one hand, it is easy to see that
\begin{align} \label{eq:density-ratio-bound}
1 - o(1) \leq \frac{\mathcal D(\vec y)}{ \normal\lp( \vec y; \vec g, \vec M \vec M^T  \rp)  } \leq  1 + o(1)     
\end{align}
for all $\vec y \in \R^m$.
On the other hand,
$\mathcal D$ and $\mathcal {\bar M}$ only differ on the complement of the support of $\mathcal {\bar M}$. By \Cref{cor:outside-support-mass}, the mass of the points outside the support of $\mathcal {\bar M}$ is at most $\delta^{ \Omega( \log(1/\delta) ) }$ under $\normal(\vec g, \vec M \vec M^T)$.
Hence, if we denote by $\mathcal K$ the support of $\mathcal {\bar M}$, we have
\begin{align} \label{eq:L1-distance}
\int_{ \vec y }
\abs{\mathcal D(\vec y) - \mathcal {\bar M}(\vec y)}
&\leq 
\Pr_{ \vec y \in \normal\lp( \vec g, \vec M \vec M^T  \rp)  }\lp[ 
\vec y \not \in \mathcal K \rp] 
\leq\delta^{\Omega( \log(1/\delta) )}.
\end{align}
Denote by $\mathcal {\hat D}$ the distribution obtained by normalizing the measure $\mathcal D$.
One can check Inequalities~\eqref{eq:density-ratio-bound} and \eqref{eq:L1-distance} still hold if we replace $\mathcal D$ with $\mathcal {\hat D}$. 
In summary, this gives us a distribution $\mathcal {\hat D}$ such that
$$
0.5 \leq \frac{\mathcal {\hat D}(\vec y)}{ \normal\lp( \vec y; \vec g, \vec M \vec M^T  \rp)  } \leq  2     
$$
and that
$$
\dtv \lp( 
\mathcal {\hat D} , \mathcal M
\rp)
\leq 
\dtv  \lp( 
\mathcal {\hat D} , \mathcal {\bar M}
\rp)
+ 
\dtv \lp( \mathcal {\bar M} , \mathcal {M} \rp)
\leq \delta^{ \Omega( \log(1/\delta) ) }.
$$
It then follows that the original transformed distribution $\mathcal M$ is $\delta^{ \Omega( \log(1/\delta) ) }$-close in total variation distance to $\mathcal {\hat D}$, which is comparable to $\normal(\vec g, \vec M \vec M^T )$.
This concludes the proof of \Cref{lem:approximate-linear-sandwich}
\end{proof}

With the tools developed in the past three sections, we are finally able to conclude the proof of \Cref{prop:Gaussian-Comparability}.
\begin{proof}[Proof of \Cref{prop:Gaussian-Comparability}]
By the assumption in \Cref{prop:Gaussian-Comparability}, $\vec q$ is a $\lp( \delta^{1/(3d)} , C_{d,m,N}\rp)$-super non-singular polynomial transformation of degree at most $d$.
By \Cref{lem:non-singular-jacobian}, we know that $\Jac_{\vec q}( \normal(\vec 0, \vec I) )$ is non-singular with high probability. In particular, we have
$$
\Pr_{ \vec x \sim \normal(\vec 0, \vec I) }
\lp[ \sigma_{\min} \lp( \Jac_{\vec q}(\vec x) \rp) < \delta^{1/3} \rp] < \delta^N.
$$
By \Cref{lem:mixture-of-linear-transform}, we hence have that $\vec q( \normal(\vec 0, \vec I) )$ is $O(\delta^N)$-close in total variation distance to some mixture distribution
$\int_{ \theta }  \vec p_{\theta}(\normal(\vec 0, \vec I)) d \theta $, where each $\vec p_{\theta}$ is a degree-$d$, $(d+1)$-bounded, $ \lp( \delta^{1/3} \rp)^{2 \; \frac{3}{2}} = \delta$-approximate linear transform.
By \Cref{lem:approximate-linear-sandwich}, each $\vec p_{\theta}( \normal(\vec 0, \vec I) )$ is $  \delta^{ \Omega( \log(1/\delta) ) }$-close in total variation distance to some distribution $\mathcal M_{\theta}$ that is comparable to
a $\lp(\delta, \log^{ 2(d+1) }(1/\delta  ) \rp)$-reasonable Gaussian distribution $\normal_{\theta}$.
Hence, when $\log(1/\delta)$ is a sufficiently large multiple of $N$, it follows that
$\vec q(\normal(\vec 0, \vec I))$ is $O( \delta^N ) + \delta^{\Omega(\log(1/\delta))} = O(\delta^N)$-close in total variation distance to $\int_{\theta} \mathcal M_{\theta} d\theta$, which is comparable to $\int_{\theta} \normal_{\theta} d\theta$.
This concludes the proof of \Cref{prop:Gaussian-Comparability}.
\end{proof}
\subsection{Reasonable Gaussians}
\label{sec:reasonable-gaussian}
Using \Cref{prop:Gaussian-Comparability}, we have reduced the proof of \Cref{thm:(anti-)concentration} to show (anti-)concentration properties of mixture of reasonable Gaussians. This goal will be accomplished in the current subsection. The formal statement of the main result of this subsection is given below.
\begin{lemma}[(Anti-)Concentration of Mixture of Reasonable Gaussians]
\label{lem:Gaussian-Mixture-anti-concentration}
Let $d,\ell,m$ be positive integers where $\ell < m$, $\delta \in (0, 1)$, and $\kappa > 1$.
Let $\int \normal_{\theta} d \theta$ be a mixture of $(\delta, \kappa)$-reasonable Gaussians in $\R^m$ and $h:\R^m \mapsto \R$ be a degree-$d$ polynomial. 
Let $I_1, \cdots,I_{\ell} $ be a set of intervals such that each interval is at most $\kappa$-far from the origin.
Let $R$ be the axis-aligned rectangle $\{ \vec x \in \R^m: \vec x_i \in I_i \text{ for all } i \in [\ell] \}$ \footnote{Notice that the last $m - \ell$ coordinates of $R$ are left unrestricted.}.
Then for any $t > 0$, we have the anti-concentration bound
\begin{align*}
    \Pr_{\vec x \sim  \int \normal_{\theta} d \theta \mid R }[ 
    \abs{h(\vec x)} < t \snorm{\int \normal_{\theta} d \theta \mid R, L_2}{h}  ] < O_{d,m}(1) \; \lp(  t \delta^{-2d} \kappa^{6d} \rp)^{1/d} \, ,
\end{align*}
and the concentration bound
\begin{align*}
    \Pr_{ \vec x \sim \int \normal_{\theta} d \theta \mid R }
    \lp[ \abs{h(\vec x)} > t  \;
    \snorm{ \int \normal_{\theta} d \theta \mid R, L_2  }{h} \rp]
    \leq \exp \lp( -\Omega_{d,m}\lp(   \delta^{2d} \kappa^{-6d}  t \rp)^{1/d} \rp).    
\end{align*}
\end{lemma}
We now give the proof overview of \Cref{lem:Gaussian-Mixture-anti-concentration}.
Gaussian distributions conditioned on the convex set $R: = \{ \vec x \in \R^m: \vec x_i \in I_i \text{ for all } i \in [\ell]\}$ are log-concave distributions. 
Hence, Carbery-Wright Theorem (\Cref{lem:CW-concentration}) and the polynomial concentration lemma (\Cref{lem:polynomial-concentration}) directly implies that each component within the mixture will satisfy good (anti-)concentration properties. 
The main issue is that the ``scale'' of each distribution is different. 
More specifically, the $L_2$-norms of the target polynomial $h$ under different Gaussian distributions $\normal_{\theta}$ conditioned on $R$ may be different.
Hence, the main step towards establishing anti-concentration of the conditional Gaussian mixture is to argue that the $L_2$-norms of the target polynomial under different reasonable conditional Gaussians are within a factor of
$\poly_{d,m,b}(\delta)$ from each other.

As a warm up, we show that for low-dimensional reasonable Gaussians (without any conditioning), the $L_2$-norms of $h$ under these distributions will be bounded by a factor of $\poly_{d,m,b}(\delta)$ from each other.
This is due to the fact that the $L_2$-norm of a polynomial under a reasonable Gaussian is the same as the $L_2$-norm of the polynomial composed with some linear transformation under the standard Gaussian. 
As we have shown in \Cref{fact:coefficeint-norm}, 
the $L_2$-norm of a low-dimensional and low-degree polynomial under the standard Gaussian is roughly proportional to its maximum coefficient.
Therefore, as long as the linear transformation does not change the maximum coefficient of the polynomial by too much, the bound will follow.
Specifically, we show the following:
\begin{lemma}
\label{lem:reasonable-norm}
Let  $\normal(\vec \mu, \vec \Sigma)$ be a $(\delta, \delta^{-1})$-reasonable Gaussian and $h: \R^m \mapsto \R$ be a degree-$d$ polynomial.
Then it holds
$$
\Theta_{d,m}( \delta^d ) \snorm{L_2}{h}
\leq 
\snorm{\normal(\vec \mu, \vec \Sigma), L_2 }{h}
\leq 
 \Theta_{d,m}( \delta^{-d} )
\snorm{L_2}{h}.
$$
\end{lemma}
\begin{proof}
Define the degree-$d$ polynomial $h': \R^m \mapsto \R$ as
$$
h'(\vec x) = h( \vec \Sigma^{1/2} \vec x + \vec \mu ).
$$
By the definition of a reasonable Gaussian, we have that
$$
 \delta \vec I \preceq \vec \Sigma \preceq \delta^{-1}  \vec I \, , \,
 \snorm{2}{\vec \mu} \leq \delta^{-1}.
$$
This further implies that
$$
\Theta_{d,m} \lp( \delta^{d} \rp)
  \LC{h}
\leq 
\LC{h'}
\leq 
 \Theta_{d,m} \lp( \delta^{-d} \rp)
\LC{h}.
$$
On the other hand, we have that
$
\snorm{ \normal(\vec \mu, \vec \Sigma),  L_2}{h} = 
\snorm{ L_2 }{h'}
$.
Combining this with the fact that $ \snorm{ L_2 }{h'} = \Theta_{d,m}( \LC{h'})$, and $ \snorm{ L_2 }{h} = \Theta_{d,m}( \LC{h})$ (\Cref{fact:coefficeint-norm}) then yields the bound in the lemma.
\end{proof}
Next we try to analyze the effect of conditioning on a reasonable Gaussian.
Though the conditioning considered in \Cref{lem:Gaussian-Mixture-anti-concentration} involves intervals, we will consider the more extreme case where a subset of coordinates are conditioned to be equal to some specific values. 
After such extreme conditioning, the resulting distribution becomes a lower-dimensional Gaussian, and hence no longer reasonable. However, it is not hard to show that if we isolate out the coordinates used for conditioning, the marginal distribution on the remaining coordinates remains a reasonable Gaussian.
\begin{lemma}
\label{lem:coordinate-restriction}
Let $d,\ell,m$ be positive integers, where $\ell < m$.
Let $\normal(\vec \mu, \vec \Sigma)$ be a $(\delta, \kappa)$-reasonable Gaussian in $\R^m$. 
Let $\vec a \in \R^{\ell}$ be a vector such that $\vec a_i < \kappa$.
Then $\normal(\vec \mu, \vec \Sigma)$ conditioned on the set $\{ \vec x \in \R^m: \vec x_i = \vec a_i \text{ for all } i \in [\ell] \}$ is a $(\delta, \delta^{-d} \kappa^d )$-reasonable Gaussian.
\end{lemma}
\begin{proof}
We first leverage the block decomposition of $\vec \mu$ and $\vec \Sigma$:
\begin{align}
\vec \mu = 
\begin{bmatrix}
    \vec \mu_1 \\
    \vec \mu_2
\end{bmatrix}   \, , \,
\vec \Sigma = \begin{bmatrix}
    \vec \Sigma_{1,1} &\vec \Sigma_{1,2} \\ 
    \vec \Sigma_{2,1} &\vec \Sigma_{2,2}
\end{bmatrix} \, ,
\end{align}
where $\mu_1 \in \R^k$ is the mean of the $k$ coordinates being conditioned, $\mu_2 \in \R^{m-k}$ is the mean of the remaining coordinates, and $\vec \Sigma_{1,1} \in \R^{k \times k}, \vec \Sigma_{1,2} \in \R^{k \times (m-k)}, \vec \Sigma_{2,1} \in \R^{k \times (m-k)}, \vec \Sigma_{2,2} \in \R^{(m-k) \times (m-k)}$ are the corresponding sub-covariance matrices.
Since $\sigma_{\min} \lp( \vec \Sigma \rp) > \delta$, the same holds for $\vec \Sigma_{2,2}$.
Notice that the marginal distribution $\normal(\vec \mu, \vec \Sigma)$ under the conditioning is some distribution whose 
last $m-k$ coordinates form
another $(m-k)$-dimensional Gaussian distribution 
$\normal( \vec {\tilde \mu}, \vec {\tilde \Sigma} )$
with parameters
\begin{align*}
&\vec {\tilde \mu} = \vec \mu_{1} + \vec \Sigma_{1,2} \; \vec \Sigma_{2,2}^{-1}  \; (\vec a - \vec \mu_2) \, , \\
&\vec {\tilde \Sigma}
= \vec \Sigma_{1,1} - 
\vec \Sigma_{1,2} 
\vec \Sigma_{2,2}^{-1} \vec \Sigma_{2,1}.
\end{align*}
We proceed to bound the parameters.
Since $\snorm{2}{\vec \mu} \leq \kappa $ and
$ \delta \vec I \preceq   \vec \Sigma  \preceq \kappa \vec I$, 
the entries in $\vec \mu_1$, $\vec \mu_2$, $\vec \Sigma_{1,1}$, $\vec \Sigma_{1,2}$, $\vec \Sigma_{2,1}$ are all bounded above by $\kappa$ and the entries in $\vec \Sigma_{2,2}^{-1}$ are bounded above by $\delta^{-1}$.
It then follows that $\snorm{2}{ \tilde \mu } \leq 
O_{m}( \delta^{-1} \; \kappa^2 )$.
For $\vec {\tilde \Sigma}$, we remark that its expression is exactly the Schur complement of $\vec \Sigma_{2,2}$ in $\vec \Sigma$.
Therefore, we can employ the following fact.
\begin{fact}
Let $\vec \Sigma \in \R^{m \times m}$.
Suppose \begin{align}
\vec \Sigma = \begin{bmatrix}
    \vec \Sigma_{1,1} &\vec \Sigma_{1,2} \\ 
    \vec \Sigma_{2,1} &\vec \Sigma_{2,2}
\end{bmatrix} \, ,
\end{align}
 where $\vec \Sigma_{1,1} \in \R^{k \times k}, \vec \Sigma_{1,2} \in \R^{k \times (m-k)}, \vec \Sigma_{2,1} \in \R^{k \times (m-k)}, \vec \Sigma_{2,2} \in \R^{(m-k) \times (m-k)}$.
 Denote by $\vec S$ the Schur complement $$
 \vec S
 = \vec \Sigma_{1,1} - 
\vec \Sigma_{1,2} 
\vec \Sigma_{2,2}^{-1} \vec \Sigma_{2,1}.
 $$
 If $\vec \Sigma$ is positive definite, then it holds
 $$
 \lambda_t( \vec \Sigma )
 \geq \lambda_t(\vec S)
 \geq \lambda_{t + k}(\vec \Sigma).
 $$
\end{fact}
This then gives us
$$
 \delta \vec I \preceq \vec {\tilde \Sigma}
 \preceq \kappa \vec I.
$$    
We can thus conclude that the marginal of the conditional distribution is a $(\delta, O_m(\delta^{-1} \kappa^2) )$-reasonable Gaussian.
\end{proof}
As we have mentioned in the proof overview of \Cref{lem:Gaussian-Mixture-anti-concentration}, the main step is to show the $L_2$ norms of $h$ under the mixture of conditional Gaussians are roughly at the same scale.
We next combine \Cref{lem:reasonable-norm} and \Cref{lem:coordinate-restriction} to establish the above intuition formally.
Specifically, we show that:
\begin{lemma}
\label{lem:mixture-gaussian-norm-bound}
Let $\int \normal_{\theta} d \theta$  be a mixture of $(\delta, \kappa)$-reasonable Gaussians and $h: \R^m \mapsto \R$ be a polynomial of degree $d$.
Let $I_1, \cdots, I_{\ell}$ be a set of intervals that are at most $\kappa$ far from the origin and $R$ be the set of vectors defined as $R:= \{ \vec x \in \R^m: \vec x_i \in I_i \text{ for all } i \in [\ell] \}$.
Then for any reasonable Gaussian $\normal_{\theta^{*}}$ within the mixture, it holds 
$$
\Theta_{d,m}( \delta^{2d} \kappa^{-6d} ) 
\snorm{ \normal_{\theta^{*}} | R, L_2 }{ h }
\leq  \snorm{ \int \normal_{\theta} d \theta | R, L_2}{h}
\leq \Theta_{d,m}( \delta^{2d} \kappa^{-6d} )
\snorm{ \normal_{\theta^{*}} | R, L_2 }{ h }.
$$
\end{lemma}
\begin{proof}
From the definition of the mixture, we know that $\snorm{ \int \normal_{\theta} d \theta | R, L_2 }{h}^2$ is a linear combination of $\snorm{ \normal_{\theta} | R, L_2 }{h}^2 d\theta$, where $\normal_{\theta}$ is a reasonable Gaussian in the mixture.
For a vector $\vec a \in \R^{\ell}$, define the set $R(\vec a)$
as $\{ \vec x \in \R^m : \vec x_i = \vec a_i \text{ for all } i \in [\ell] \}$.
Then for a reasonable Gaussian $\normal_{\theta}$, we have that
$
\snorm{ \normal_{\theta } | R, L_2 }{ h }^2
$
is a linear combination of
$ \snorm{ \normal_{\theta } |  R(\vec a) , L_2 }{ h }^2  $, where $\vec a \in I_1, \cdots, I_{\ell}$.
Hence, it suffices to show that for any two $(\delta, \kappa)$-reasonable Gaussians $\normal_{\theta}, \normal_{\theta'}$
and any two vectors $\vec a, \vec a' \in I_1, \cdots, I_{\ell}$, it holds
\begin{align}
\label{eq:reasonable-ratio-bound}
\Theta_{d,m}( \delta^{2d} \kappa^{-6d} ) 
\snorm{ \normal_{\theta } | R(\vec a), L_2 }{ h }
\leq  \snorm{\normal_{\theta' } | R(\vec a'), L_2 }{ h }
\leq \Theta_{d,m}( \delta^{-2d} \kappa^{6d} )
\snorm{\normal_{\theta } | R(\vec a), L_2 }{ h }.    
\end{align}

For a vector $\vec a$, we can define the restricted polynomial function 
$h_{\vec a}: \R^{m - \ell} \mapsto \R$ as $h_{\vec a}(\vec x):= h(\vec a, \vec x)$.
We will also denote by $\normal_{\theta, \vec a}$ the 
$(m-\ell)$-dimensional marginal distribution of the last $m - \ell$ coordinates of $\normal_{\theta} \mid R(\vec a)$.
Then it is not hard to see that
\begin{align}
\label{eq:reasonable-ratio-bound-step1}
\snorm{ \normal_{\theta } | R(\vec a), L_2 }{h}
= \snorm{ \normal_{ \theta, \vec a }, L_2 }{h_{\vec a} }.    
\end{align}
By \Cref{lem:coordinate-restriction}, $\normal_{ \theta, \vec a }$ is a $(\delta, O_m( \delta^{-1} \kappa^2 ) )$-reasonable Gaussian.
Hence, applying \Cref{lem:reasonable-norm}, we have
\begin{align}
\label{eq:reasonable-ratio-bound-step2}    
\Theta_{d,m}( \delta^{d} \kappa^{-2d} ) \;
\snorm{ L_2 }{ h_{\vec a} }
\leq 
\snorm{ \normal_{ \theta, \vec a }, L_2 }{ h_{\vec a} }
\leq
\Theta_{d,m}( \delta^{-d} \kappa^{2d} )
\snorm{ L_2 }{ h_{\vec a} }.
\end{align}
Since the size of each coordinate of $\vec a$ is bounded above $\kappa$, we have
\begin{align}
\label{eq:reasonable-ratio-bound-step3}
\Theta_{d,m}( \kappa^{-d}) \snorm{L_2}{h} \leq  \snorm{L_2}{ h_{\vec a} }
\leq
\Theta_{d,m}( \kappa^d) \snorm{L_2}{h}.    
\end{align}
Combining \Cref{eq:reasonable-ratio-bound-step1}, \Cref{eq:reasonable-ratio-bound-step2}, and \Cref{eq:reasonable-ratio-bound-step3} then gives us
$$
\Theta_{d,m}( \delta^{d} \kappa^{-3d}) \snorm{L_2}{h}
\leq
\snorm{ \normal_{\theta } | R(\vec a), L_2 }{h}
\leq 
\Theta_{d,m}( \delta^{-d} \kappa^{3d}) \snorm{L_2}{h}.
$$
\Cref{eq:reasonable-ratio-bound} hence follows, completing the proof of \Cref{lem:mixture-gaussian-norm-bound}.
\end{proof}

We are ready to conclude the proof of our main lemma in this subsection.
\begin{proof}[Proof of \Cref{lem:Gaussian-Mixture-anti-concentration}]
Define $R$ to be the set of vectors $\{ \vec x \in \R^m : \vec x_i \in I_i \text{ for all } i \in [\ell] \}$.
We start with the anti-concentration bound.
For any $t > 0$ and each reasonable Gaussian $\normal_{\theta}$ in the mixture, applying the Carbery Wright Theorem (\Cref{lem:CW-concentration}) gives us that
\begin{align*}
    \Pr_{ \vec x \sim \normal_{\theta} | R }
    \lp[ \abs{h(\vec x)} < t \snorm{\normal_{\theta} | R, L_2  }{h} \rp]
    \leq O(d  t^{1/d}).
\end{align*}
Applying \Cref{lem:mixture-gaussian-norm-bound} gives us
\begin{align*}
    \Pr_{ \vec x \sim \normal_{\theta} | R }
    \lp[ \abs{h(\vec x)} < 
    \Theta_{d,m} \lp( \delta^{2d} \kappa^{-6d} \rp)
    t \snorm{ \int \normal_{\theta} d \theta | R, L_2  }{h} \rp]
    \leq O(t^{1/d}).
\end{align*}
By reparametrizing the variable $t$ in the equation, we know it is equivalent to
\begin{align*}
    \Pr_{ \vec x \sim \normal_{\theta} | R }
    \lp[ \abs{h(\vec x)} < 
    t \snorm{ \int \normal_{\theta} d \theta | R, L_2  }{h} \rp]
    \leq O_{d,m}\lp( \lp(  t \delta^{-2d} \kappa^{6d} \rp)^{1/d} \rp).
\end{align*}
Since this holds for each Gaussian $\normal_{\theta}$ in the mixture, the anti-concentration of the entire mixture hence follows.

Now we turn our attention to the concentration bound.
For any $t > 0$ and each reasonable Gaussian $\normal_{\theta}$ in the mixture, applying \Cref{lem:polynomial-concentration} gives us
\begin{align*}
    \Pr_{ \vec x \sim \normal_{\theta} | R }
    \lp[ \abs{h(\vec x)} > t \snorm{\normal_{\theta} | R, L_2  }{h} \rp]
    \leq \exp ( -\Omega(t)^{1/d} ).
\end{align*}
Applying \Cref{lem:mixture-gaussian-norm-bound} gives us
\begin{align*}
    \Pr_{ \vec x \sim \normal_{\theta} | R }
    \lp[ \abs{h(\vec x)} > t \; 
    \Theta_{d,m} \lp( \delta^{-2d} \kappa^{6d} \rp) \;
    \snorm{ \int \normal_{\theta} d \theta | R, L_2  }{h} \rp]
    \leq \exp ( -\Omega(t)^{1/d} ).
\end{align*}
By reparametrizing the variable $t$ in the equation, we know it is equivalent to
\begin{align*}
    \Pr_{ \vec x \sim \normal_{\theta} | R }
    \lp[ \abs{h(\vec x)} > t  \;
    \snorm{ \int \normal_{\theta} d \theta | R, L_2  }{h} \rp]
    \leq \exp \lp( -\Omega_{d,m}\lp(   \delta^{2d} \kappa^{-6d}  t \rp)^{1/d} \rp).
\end{align*}
This completes the proof of \Cref{lem:mixture-gaussian-norm-bound}.
\end{proof}

\subsection{Proof of \texorpdfstring{\Cref{thm:(anti-)concentration}}: Global Definitions}
\label{sec:global-definition}
The current subsection and the remaining three subsections \Cref{sec:transfer-probability}, \Cref{sec:transfer-properties}, and \Cref{sec:conclude-anti-concentration} are all devoted to conclude the proof of \Cref{thm:(anti-)concentration}. 
For that reason, throughout the rest of the section, we are going to repeatedly use some variables and mathematical objects. 
To avoid restating their definitions in different places, we will define these variables and objects globally at the beginning of this subsection. 

\paragraph{Global Parameters} $\eps, \delta$ will be used to denote real numbers in $(0, 1)$ and 
$d, \ell, K, M, N, C_{K, d, \ell}$ will be used to denote positive integers. 
Their sizes are required to satisfy the following conditions. 
$M,N$ will be numbers that depend on $K, d, \ell$ only and they are required to be sufficiently large given $K, d, \ell$. 
$C_{K, d, \ell}$ will be some sufficiently large function of $K, d, \ell, M, N$ (which is simply some other sufficiently large function of $K, d, \ell$, since $M, N$ depends only on $K, d, \ell$).
The parameters $\delta, \eps$ satisfy the equality $\delta^{d^2 K} = \eps$, and 
$\eps$ is required to be sufficiently small given all the aforementioned integers (which is simply some sufficiently small function of $d, \ell, K$).

\paragraph{Global Mathematical Objects} The mathematical objects used will be as follows:
\begin{itemize}
    \item $\{q_1, \cdots, q_{\ell}\}$ are a $\lp(\eps^{1/(3d^2K)} = \delta^{1/3}, C_{K, d, \ell} \rp)$-super non-singular set of polynomials.
    \item $\{I_i\}_{i=1}^{\ell}$ is a set of intervals such that each $I_i$ is at most $\poly_d(\log(1/\eps))$-far from the origin.
    \item $\mathcal P$ is defined to be the set $\mathcal P:= \{\vec x: q_i(\vec x) \in I_i \forall i \in [\ell]\}$ and $R$ is defined to be the axis-aligned rectangle
    $ R:= I_1 \times I_2 \cdots \times I_{\ell} \times \R^{m - \ell} $.
    \item  $D$ is the distribution $\normal(\vec 0, \vec I)$ conditioned on $\mathcal P$.
    \item $p:\R^n \mapsto \R$ is the given target polynomial of degree $d$. We are interested in showing concentration and anti-concentration properties of $p(D)$.
    Since these properties are invariant up to scaling, we assume without loss of generality that $\snorm{L_2}{p} = 1$.
    \item The vector-valued polynomial $\vec q: \R^n \mapsto \R^m$, the composition polynomial $h: \R^m \mapsto \R$ and the error polynomial $e: \R^n \mapsto \R$
    are constructed as follows.
    Since we assume that the set of polynomials $\{q_1, \cdots, q_{\ell}\}$ is $\lp(\eps^{1/(3d^2K)}, C_{K, d, \ell}\rp)$-super non-singular and
    $C_{K, d, \ell}$ is sufficiently large given $d, \ell, M, N$, we can use \Cref{prop:super-non-singular-extension} to show that for some $m=O_{d, \ell, M, N}(1) = O_{K, d, \ell}(1)$ there exists a $(\delta, N)$-super non-singular set of polynomials 
$\{q_1, \cdots, q_{\ell}, q_{\ell+1}, \cdots, q_m \}$, a composition polynomial $h: \R^m \mapsto \R$,  and an error polynomial $e: \R^n \mapsto \R$ with $\snorm{L_2}{e} \leq \delta^M$
such that
\begin{align}\label{eq:global-decomposition}
p(\vec x) = h( q_1(\vec x), \cdots, q_m(\vec x) ) + e(\vec x).    
\end{align}
We define $\vec q:\R^n \mapsto \R^m$ as the vector-valued polynomial whose $i$-th coordinate is $q_i$.
    \item  $\mathcal M$ and $\int_{\theta} \normal_{\theta} d \theta$ are distributions in $\R^m$. They are constructed as follows.
    Using \Cref{prop:Gaussian-Comparability}, we know that $\vec q(\normal(\vec 0, \vec I))$ is $O(\delta^N)$-close in total variation distance to some distribution $\mathcal M$ that is comparable to a mixture of $(\delta, \poly_{d,m,N} ( \log(1/\delta) ) )$-reasonable Gaussians, i.e., $\int_{\theta} \normal_{\theta} d \theta$.
\end{itemize}
\subsection{Proof of \texorpdfstring{\Cref{thm:(anti-)concentration}}: Transferring Probabilities}
\label{sec:transfer-probability}
By \Cref{lem:Gaussian-Mixture-anti-concentration}, the composition polynomial $h$ should have decent (anti-)concentration properties under $\int_{\theta} \normal_{\theta} d \theta | R$.
To conclude the proof of \Cref{thm:(anti-)concentration}, we will need to ``transfer'' the (anti-)concentration properties back to the original polynomial $p$ under $D = \normal(\vec 0, \vec I) \mid \mathcal P$.
In this subsection, our goal is to show that the probability of $p$ being small/large under $D$ is not too much larger than the probability of $h$ being small/large under $\int_{\theta} \normal_{\theta} d \theta | R$. The formal statement is given below.
\begin{lemma}
\label{lem:transfer-probability-bound}
For any threshold $t > \delta^{M/2}$, it holds
\begin{align*} 
\Pr_{\vec x \sim D}[ \abs{p(\vec x)} > 2t ]
\leq O(1) \; \Pr_{\vec x \sim \int_{\theta} \normal_{\theta} d \theta}[  
\abs{h(\vec x)} > t
\mid R ] + \exp\lp( - \Omega\lp( t \delta^{-M} \rp)^{1/d} \rp) + O(\delta^{N/2}).
\end{align*}
\begin{align*} 
\Pr_{\vec x \sim D}[ \abs{p(\vec x)} < t ]
\leq O(1) \; \Pr_{\vec x \sim \int_{\theta} \normal_{\theta} d \theta}[  
\abs{h(\vec x)} < 2 t \mid R ] + \exp\lp( - \Omega\lp( t \delta^{-M} \rp)^{1/d} \rp) + O(\delta^{N/2}).
\end{align*}
\end{lemma}
\begin{proof}
The proof of the above lemma is divided into two steps.
\begin{enumerate}
    \item We bound the probability of $p$ being small/large under $D$ by the probability of $h$ being small/large under $\vec q(\normal(\vec 0, \vec I)) \mid R$.
    \item We bound the probability of $h$ being small/large under $\vec q(\normal(\vec 0, \vec I)) \mid R$  by the probability of $h$ being small/large under $\int_{\theta} \normal_{\theta} d \theta \mid R$.
\end{enumerate}
We start with the first step. The key observation is that if we define $\tilde p(\vec x) = p(\vec x) - e(\vec x) =  h(\vec q(\vec x))$, the distribution of $\tilde p( D )$ is actually exactly the same as $ h(  \vec q(\normal(\vec 0, \vec I)) \mid R ) $.
Therefore, it suffices to control the influence of the error polynomial $e$ on the probability. 
To show that, we will exploit the fact that the error polynomial $e$ has its $L_2$-norm bounded by $\delta^M$, which is tiny when $M$ is sufficiently large, and has good concentration under $\normal(\vec 0, \vec I)$. As a side remark, this is precisely the source of the additive term $\exp\lp( - \Omega\lp( t \delta^{-M} \rp)^{1/d} \rp)$ in \Cref{lem:transfer-probability-bound}.
\begin{lemma}[Transfer (Anti-)Concentration Probabilities]
\label{lem:first-transfer}
For any threshold $t > \delta^{M/2}$, it holds
\begin{align*}
\Pr_{\vec x \sim D}[ \abs{p(\vec x)} < t ]
\leq \Pr_{\vec y \sim \vec q(\normal(\vec 0, \vec I)) }[  
h(\vec y) < 2t \mid R] + \exp\lp( - \Omega\lp( t \delta^{-M} \rp)^{1/d} \rp).
\end{align*}   
\begin{align*}
\Pr_{\vec x \sim D}[ \abs{p(\vec x)} > 2t ]
\leq \Pr_{\vec y \sim \vec q(\normal(\vec 0, \vec I)) }[  
h(\vec y) > t \mid R] + \exp\lp( - \Omega\lp( t \delta^{-M} \rp)^{1/d} \rp).
\end{align*} 
\end{lemma}
\begin{proof}
We start by showing a concentration bound of $e$ under the conditional distribution $D$. In particular, for $t> \delta^{M/2} $, we show that
\begin{equation} \label{eq:error-concentration}
\Pr_{ \vec x \sim D } \lp[ \abs{e(x)} > t  \rp] < \exp\lp( - \Omega\lp( t \delta^{-M} \rp)^{1/d} \rp).    
\end{equation}
We know that $e$ has good concentration under the full Gaussian distribution $\normal(\vec 0, \vec I)$.
By \Cref{lem:polynomial-concentration}, with some slight reparameterization, we have
$$
\Pr_{ \vec x \sim \normal(\vec 0, \vec I) } \lp[ \abs{e(x)} > t  \rp] < \exp\lp( - \Omega\lp( t \delta^{-M} \rp)^{1/d} \rp).
$$
Since we assume that the conditioning set has mass at least $\poly_{d, \ell}(\eps)$, the tail bound gets magnified by at most a factor of $\poly_{d, \ell}(1/\eps)$ after the conditioning. This gives us
$$
\Pr_{ \vec x \sim D } \lp[ \abs{e(x)} > t  \rp] < \poly_{d, \ell}(1/\eps) \; \exp\lp( - \Omega\lp( t \delta^{-M} \rp)^{1/d} \rp) 
< 
\exp\lp( - \Omega\lp( t \delta^{-M} \rp)^{1/d}  + O_{d, \ell}( \log(1/\eps) )\rp) .
$$
Notice that when $t > \delta^{M/2}$, the expression in the exponential function is
still $\Omega\lp( t \delta^{-M} \rp)^{1/d}$ when $M$ is sufficiently large given $d, \ell, K$. This concludes the proof of Inequality~\eqref{eq:error-concentration}.

We now proceed to show the first inequality in \Cref{lem:first-transfer}.
By our decomposition, we have
\begin{align*}
\Pr_{\vec x \sim D}[ \abs{p(\vec x)} < t ]    
= \Pr_{ \vec x \sim D } [ \abs{ f( \vec q(\vec x) ) ) + e(\vec x)} < t].
\end{align*}
We claim that the event $A:=\abs{ f( \vec q(\vec x) ) ) + e(\vec x)} < t$
implies the event $\abs{ f( \vec q(\vec x) ) } < 2t$ \emph{or} $\abs{e(\vec x)} > t$.
To show this, for the sake of contradiction, we assume that $\abs{ f( \vec q(\vec x) ) } > 2t$ \emph{and} 
$\abs{e(\vec x)} < t$. Then by the triangle inequality, we would have $\abs{ f( \vec q(\vec x) ) ) + e(\vec x)} > t$, which contradicts with event $A$.
With the implication in mind, we can apply the union bound:
\begin{align} 
\Pr_{\vec x \sim D}[ \abs{p(\vec x)} < t ]    
&\leq 
\Pr_{\vec x \sim D}[ \abs{ h(\vec q(\vec x)) } < 2t ] + \Pr_{\vec x \sim D}[ \abs{e(\vec x)} > t ] \nonumber \\
&=
\Pr_{\vec x \sim \vec q(\normal(\vec 0, \vec I)}[ \abs{ h(\vec x) } < 2t \mid R ] + \Pr_{\vec x \sim D}[ \abs{e(\vec x)} > t ].
\end{align}
Bounding the second term $\Pr_{\vec x \sim D}[ \abs{e(\vec x)} > t ]$ via \Cref{eq:error-concentration} then concludes the proof of the first inequality.
The argument for the second inequality is symmetric to the first one, and is therefore omitted.
This concludes the proof of \Cref{lem:first-transfer}.
\end{proof}


\end{proof}
For the second step, we need to exploit the fact that $\vec q(\normal(\vec 0, \vec I))$ is close in total variation distance to some distribution $\mathcal M$, which is comparable to $\int_{\theta} \normal_{\theta} d \theta$.
One technical issue is that we actually want to bound the conditional probabilities of the relevant distributions on the set $R$, which will magnify the distribution distances between them. 
This is where we use our assumption that the mass of $R$ is not too light under $\vec q(\normal(\vec 0, \vec I))$ in the statement of our theorem, i.e., 
$\Pr_{\vec x \sim \normal(\vec 0, \vec I)} \lp[ \vec q(\vec x) \in R \rp] 
= \Pr_{\vec x \sim \normal(\vec 0, \vec I)} \lp[ q_i(\vec x) \in I_i \forall i \in [\ell] \rp] 
> \poly_{d, \ell}(\eps)$.
With this assumption in mind, it is then easy to see that the total variation distance is increased to at most $\delta^{N/2}$ if we take $N$ to be sufficiently large given $d, \ell$.
\begin{claim}
\label{clm:conditional-tv-closeness}
Suppose that $N$ is sufficiently large given $d, \ell, K$.
The surrogate distribution $\vec q(\normal(\vec 0, \vec I)) \mid R$ is $\delta^{N/2}$-close in total variation distance to $\mathcal M \mid R$.
\end{claim}
\begin{proof}
We know that $\vec q(\normal(\vec 0, \vec I))$ is $O(\delta^N)$-close in total variation distance to $\mathcal M$ (see \Cref{sec:global-definition} for their definitions).
Hence, their probability masses over $R$ differ by at most $O(\delta^N)$.
By the assumption of \Cref{thm:(anti-)concentration}, we know that the mass of $R$ under $\vec q(\normal(\vec 0, \vec I))$ is at least $\poly_{d, \ell}(\eps) = \poly_{d, \ell, K}(\delta)$ (by the definition of $\delta$ such that $\delta^{d^2 K} = \eps$).
Therefore, when $N$ is sufficiently large given $d, \ell, K$, it follows that $\vec q(\normal(\vec 0, \vec I)) \mid R$ is $\delta^{N/2}$-close in total variation distance to $\mathcal M \mid R$.
\end{proof}
Then we know that the probability of any set under $\mathcal M \mid R$ is within a constant factor of $\int_{\theta} \normal_{\theta} d \theta \mid R$, since $\mathcal M$ and $\int_{\theta} \normal_{\theta} d \theta$ are comparable to each other.
\begin{claim}
\label{clm:second-transfer}
For any set $S \subset \R^m$, it holds
\begin{align*}
\Pr_{\vec x \sim   \vec q(\normal(\vec 0, \vec I))}[ \vec x \in S  \mid   R]
\leq O(1) \; \Pr_{\vec x \sim \int_{\theta} \normal_{\theta} d \theta }[  
\vec x \in S \mid R] + O( \delta^{N/2} ).
\end{align*}    
\end{claim}
From here, \Cref{lem:transfer-probability-bound} readily follows from \Cref{lem:first-transfer} and \Cref{clm:second-transfer}.

\subsection{Proof of \texorpdfstring{\Cref{thm:(anti-)concentration}}: Relating the Norms}
\label{sec:transfer-properties}
From \Cref{lem:transfer-l2-norm-bound}, we know that the probability of $p$ being smaller/larger than a threshold $t$ under $D$ can be appropriately bounded by the probability of $h$ being smaller/larger than the same threshold $t$ (up to a constant factor) under $\int_{\theta} \normal_{\theta} d\theta$ (plus some additive terms that can be ignored if we take $M,N$ to be sufficiently large).
However, notice that in the standard formulation of (anti-)concentration inequalities, the size of the polynomial should be normalized by its $L_2$-norm under the corresponding distribution.
Therefore, before we can conclude the proof of \Cref{thm:(anti-)concentration}, we need to show that the $L_2$ norm of $h$ under $\int_{\theta}  \normal_{\theta} d\theta | R$ is on the same scale as that of $p$ under $D$.
\begin{lemma}
\label{lem:transfer-l2-norm-bound}
It holds
$$
\snorm{D, L_2}{p} = \Theta(1) \; \snorm{ \int_{\theta} \normal_{\theta} d \theta \mid R , L_2}{ h }.
$$
\end{lemma}
Similar to the proof of \Cref{lem:transfer-probability-bound}, we will divide our proof into steps: 
\begin{align}
\label{eq:step1}
    &\snorm{D, L_2}{p} = \Theta(1) \snorm{ \vec q(\normal(\vec 0, \vec I)) \mid R, L_2 }{h} \, , \\
\label{eq:step2}    
    &\snorm{ \vec q(\normal(\vec 0, \vec I)) \mid R, L_2 }{h} = \Theta(1) \snorm{ \mathcal M \mid R, L_2 }{h} \, , \\
\label{eq:step3}        
    &\snorm{ \mathcal M \mid R, L_2 }{h} = \Theta(1) \snorm{ \int_{\theta} \normal_{\theta} d \theta \mid R, L_2 }{h}.
\end{align}
For the first step, we start with the simple observation that
$\snorm{D, L_2}{p}$ and $\snorm{ \vec q(\normal(\vec 0, \vec I)) \mid R, L_2 }{h}$ differ by at most $\snorm{D, L_2}{e}$, which is a tiny term since $\snorm{D}{e} \leq \delta^M$.
This can be easily shown using the triangle inequality and the definition of these distributions/polynomials. 
\begin{claim}
\label{clm:step1-additive}
It holds $ \abs{\snorm{D, L_2}{p} - \snorm{ \vec q(\normal(\vec 0, \vec I)) \mid R , L_2 }{h}} \leq \delta^{M/2}$.
\end{claim}
\begin{proof}
By the triangle inequality, we have
$$
\abs{\snorm{D, L_2}{p} - \snorm{ \vec q(\normal(\vec 0, \vec I)) \mid R , L_2 }{h}} \leq 
\snorm{D, L_2}{e}.
$$
Recall that $\snorm{L_2}{e} \leq \delta^M$ by the guarantees of \Cref{prop:super-non-singular-extension}, $D := \normal(\vec 0, \vec I) \mid \mathcal P$, and the mass of $\mathcal P$ under $\normal(\vec 0, \vec I)$ is at least $\poly_{d, \ell}(\eps)$ by our assumption.
Moreover, $\eps, \delta$ satisfy the equality $ \delta^{d^2K} = \eps $.
Hence, when $M$ is sufficiently large given $d, \ell, K$, it follows that 
$ \snorm{D, L_2}{e} \leq  \poly_{d, \ell}(1/\eps) \delta^M \leq \delta^{M/2}$.
\end{proof}
Eventually, our goal is to argue that the $L_2$-norms of interest are within a constant factor of each other.
This additive bound in \Cref{clm:step1-additive} will be sufficient for this purpose given that the conditional $L_2$ norm, i.e., $\snorm{\vec q(\normal(\vec 0, \vec I)) \mid R, L_2}{h}$, $\snorm{D, L_2}{p}$, is not much smaller than the corresponding unconditional versions, i.e., $\snorm{ \vec q(\normal(\vec 0, \vec I)) , L_2 }{h}$, $\snorm{L_2}{p}$.
It turns out that we can take advantage of the fact that these polynomials, i.e., $p, h$, are decently anti-concentrated 
to argue that the conditioning cannot trivialize their $L_2$ norms.
\begin{lemma}
\label{lem:conditional-l2-norm}
It holds 
\begin{align*}
&\snorm{\vec q(\normal(\vec 0, \vec I)) \mid R, L_2}{h} \geq 
\poly_{d, \ell}(\eps) \snorm{ \vec q(\normal(\vec 0, \vec I)) , L_2 }{h} \, ,\\    
&\snorm{D, L_2}{p} \geq 
\poly_{d, \ell}(\eps) \snorm{L_2}{p} = \poly_{d, \ell}(\eps).
\end{align*}
\end{lemma}
\begin{proof}
It holds
$$
\snorm{\vec q(\normal(\vec 0, \vec I)) \mid R, L_2}{h}^2
\geq \E_{ \vec x \sim  \normal(\vec 0, \vec I) }
\lp[ 
(h \circ \vec q)^2(\vec x) \; \mathbbm 1 \{ \vec q(\vec x) \in R \}
\rp] \, ,
$$
where we denote by $h \circ \vec q$ the composition of $h$ and $\vec q$, i.e., 
$h \circ \vec q(\vec x) = h( \vec q_1(\vec x), \cdots, \vec q_m(\vec x) )$.
By the guarantees of \Cref{prop:super-non-singular-extension}, $h \circ \vec q$ is a degree at most $d$ polynomial.
Define 
$\kappa := \Pr_{ \vec x \sim \normal(\vec 0, \vec I) } \lp[ \vec q(\vec x) \in R \rp] $.
By the Carbery Wright Theorem (\Cref{lem:CW-concentration}), we have that
$$
\Pr_{ \vec x \sim \normal(\vec 0, \vec I) }\lp[ 
\abs{ (h \circ \vec q)(\vec x) } <  (\kappa / C)^d  \snorm{L_2}{h \circ \vec q}  
\rp] > \kappa / 2 \, ,
$$
for some sufficiently large universal constant $C$.
By the union bound, we hence have
$$
\Pr_{ \vec x \sim \normal(\vec 0, \vec I) }
\lp[  \abs{\lp(h \circ \vec q \rp)(\vec x)} > (\kappa / C)^d  \snorm{L_2}{h \circ \vec q}, \vec q(\vec x) \in R \rp] > \kappa/2.
$$
Hence, we can derive the lower bound
\begin{align*}
\snorm{\vec q(\normal(\vec 0, \vec I)) \mid R, L_2}{h}^2
&\geq
\kappa^{-1} \; \E_{ \vec x \sim  \normal(\vec 0, \vec I) }
\lp[ 
(h \circ \vec q)^2(\vec x) \; \mathbbm 1 \{ \vec q(\vec x) \in R \} \;
\mathbbm 1 \{ \abs{\lp(h \circ \vec q \rp)(\vec x)} > (\kappa / C)^d  \snorm{L_2}{h \circ \vec q} \}
\rp] \\
&\geq   
\kappa^{-1} \; 
\frac{\kappa}{2} \; 
(\kappa / C)^d  \snorm{L_2}{h \circ \vec q}
\geq \poly_{d, \ell}(\eps) \snorm{ \vec q(\normal(\vec 0, \vec I)), L_2 }{h}^2 \, ,
\end{align*}
where in the last inequality we use that $\kappa \geq \poly_{d, \ell}(\eps)$
and that $\snorm{L_2}{ h \circ \vec q } = \snorm{\vec q(\normal(\vec 0, \vec I)), L_2}{h}$.
Using almost identical arguments, one can show the second inequality in \Cref{lem:conditional-l2-norm} as well.
This concludes the proof of \Cref{lem:conditional-l2-norm}.
\end{proof}
\begin{proof}[Proof of Step(1) (\Cref{eq:step1})]
From the definitions of $h, \vec q, p, e$ and the triangle inequality, we have
$$
\abs{ \snorm{\vec q(\normal(\vec 0, \vec I)), L_2}{h} 
- \snorm{ L_2}{p}  } \leq \snorm{L_2}{e} \leq \delta^M.
$$
Since we assume that $\snorm{L_2}{p} = 1$, it follows that $\snorm{\vec q(\normal(\vec 0, \vec I)), L_2}{h}  = \Omega(1)$.
Then \Cref{lem:conditional-l2-norm} gives us $\snorm{\vec q(\normal(\vec 0, \vec I)) \mid R, L_2}{h}  = \poly_{d, \ell}(\eps) \gg \delta^M$, when $M$ is sufficiently large given $d, \ell, K$.
Combining this with \Cref{clm:step1-additive} then concludes the proof of Step (1) (\Cref{eq:step1}).    
\end{proof}

We now turn to Step (2) (\Cref{eq:step2}). If it were the case that $\vec q(\normal(\vec 0, \vec I)) = \mathcal M$ , Step (2) would be trivial.
The situation becomes more complicated when $\vec q(\normal(\vec 0, \vec I))$ is only close in total variation distance to  $\mathcal M$.
In order to relate the $L_2$ norms of $h$ under $\vec q(\normal(\vec 0, \vec I))$ and under $\mathcal M$, we need to exploit the fact that $h$ satisfies good concentration under $\vec q(\normal(\vec 0, \vec I))$ and under $\mathcal M$. 
\begin{claim}
\label{clm:polynomial-concentration}
For any threshold $t > 0$, it holds
\begin{align*}
&\Pr_{ \vec x \sim  \vec q(\normal(\vec 0, \vec I))}
\lp[ \abs{ h(\vec x) } > t \snorm{\vec q(\normal(\vec 0, \vec I)), L_2}{h} \rp] < O \lp(  2^{-t^{2/d}} \rp) \, ,  \\
&\Pr_{ \vec x \sim  \mathcal M}
\lp[ \abs{ h(\vec x) } > t \snorm{\mathcal M, L_2}{h} \rp] < O \lp( 
 \exp\lp( - \lp(t \eps^{1/K} \rp)^{1/d} \rp) \rp).  
\end{align*}
\end{claim}
\begin{proof}
We note that the distribution of $h(\vec x)$, where $\vec x \sim \vec q(\normal(\vec 0, \vec I))$, is the same as the distribution of 
$h \circ \vec q(\vec x)$, where $\vec x \sim \normal(\vec 0, \vec I)$. By the guarantees of \Cref{prop:super-non-singular-extension}, we know that $h \circ \vec q$ is a degree at most $d$ polynomial. 
Hence, the first inequality simply follows from the concentration of polynomials under the standard Gaussian (\Cref{lem:polynomial-concentration}).

For the second inequality, we will exploit the fact that $\mathcal M$ is comparable to $\int_{\theta} \normal_{\theta} d\theta$.
In particular, we have
\begin{align*}
\Pr_{ \vec x \sim \mathcal M }  \lp[ \abs{h(\vec x)} > t \snorm{\mathcal M, L_2}{h} \rp]
\leq 2 \; 
\Pr_{ \vec x \sim \int_{\theta} \normal_{\theta} d\theta}  \lp[ \abs{h(\vec x)} > t/2 \snorm{\int_{\theta} \normal_{\theta} d\theta, L_2}{h} \rp].
\end{align*}
Recall that each $\normal_{\theta}$ is a $\big(\delta, \poly_{d,m,N}(\log(1/\delta))\big)$-reasonable Gaussian. Therefore, 
using the concentration property of the reasonable Gaussian mixture
(\Cref{lem:Gaussian-Mixture-anti-concentration}) \footnote{We apply the lemma with the conditioning set $S$ being the entire space.}, we further have
\begin{align*}
\Pr_{ \vec x \sim \int_{\theta} \normal_{\theta} d\theta}  \lp[ \abs{h(\vec x)} > t/2 \snorm{\int_{\theta} \normal_{\theta} d\theta, L_2}{h} \rp]
\leq 
O(1) \; 
\exp \lp( -\Omega_{d,m}\lp(   \delta^{2d} \; \poly_{d,m,N}(\log^{-1}(1/\delta)) \;  t\rp)^{1/d} \rp).
\end{align*}
When $\delta$ is sufficiently small given $d,m,N$, we thus have
$\Omega_{d,m} \lp( \delta^{2d} \poly_{d,m,N}(\log^{-1}(1/\delta)) \rp) 
\geq \delta^{2d+1} > \eps^{1/K}.$
This concludes the proof of the second inequality.
\end{proof}
When a polynomial satisfies good concentration under a distribution, we can take advantage of this concentration property to control the contribution of a small fraction of the points from the distribution to the $L_2$ norm of the polynomial under the distribution.
\begin{claim}[$L_2$-Norm Stability] \label{clm:l2-norm-stability}
Let $\alpha \in (0, 1)$.
Let $\mathcal E, \mathcal L$ be two distributions satisfying 
$ \alpha \mathcal E \leq \vec q(\normal(\vec 0, \vec I)), \alpha \mathcal L \leq \mathcal M$
\footnote{Recall the notation $D \leq D'$ for two distributions $D, D'$ means that the probability density function of $D$ is pointwise bounded above by that of $D'$.}.
Then it holds
\begin{align*}
&\alpha \snorm{\mathcal L, L_2}{ h }^2 < O_d(\alpha)  \; \eps^{-2/K} \; \log^{2d}(1/\alpha) \; \snorm{\mathcal M, L_2}{h}^2 \, , \\
&\alpha \snorm{\mathcal E, L_2}{ h }^2 < O_d(\alpha)\; \log^{d}(1/\alpha) \; \snorm{\vec q(\normal(\vec 0, \vec I)), L_2}{h}^2.
\end{align*}
\end{claim}
\begin{proof}
Using \Cref{clm:polynomial-concentration}, we know that for any positive integer $i$, it holds
\begin{align} \label{eq:integer-step-concentration}
\Pr_{ \vec x \sim \mathcal M } \lp[ 
\abs{h(\vec x)} > \eps^{-1/K} \; i^d  \; \log^d(1/\alpha)
\snorm{\mathcal M, L_2}{h}
\rp] <  C \; \alpha^i \, ,
\end{align}
for some universal constant $C$.
We will use $F_i$ to denote the set 
$$
F_i:= \{ \vec x \in \R^n \text{ such that }
\eps^{-1/K} i^d  \; \log^d(1/\alpha) \; \snorm{\mathcal M, L_2}{h}
\leq \abs{h(\vec x)} \leq 
\eps^{-1/K} (i+1)^d  \; \log^d(1/\alpha)  \; \snorm{\mathcal M, L_2}{h}\}.
$$
In particular, each $F_i$ is constructed such that
$
\Pr_{ \vec x \sim \mathcal M } [\vec x \in F_i] \leq C \alpha^i.
$
Then we have
\begin{align*}
\alpha \E_{\vec x \sim \mathcal E}    
\lp[ h^2(\vec x)  \rp]
&= 
\sum_{i=0}^{\infty}
\alpha \E_{\vec x \sim \mathcal E}    
\lp[ h^2(\vec x) \; \mathbbm 1 \{ \vec x \in F_i \}  \rp] \\
&\leq 
\alpha \; \eps^{-2/K} \; \log^{2d}(1/\alpha) \; \snorm{\mathcal M, L_2}{h}^2
+ \sum_{i=1}^{\infty}
\lp( \alpha \Pr_{ \vec x \sim \mathcal E }
 \lp[  \vec x \in F_{i} \rp]  \rp)
 \eps^{-2/K} \; (i+1)^{2d} \; \log^{2d}(1/\alpha) \; \snorm{\mathcal M, L_2}{h}^2
 \\
&\leq 
\alpha \; \eps^{-2/K} \; \log^{2d}(1/\alpha) \; \snorm{\mathcal M, L_2}{h}^2
+ 
\sum_{i=1}^{\infty}
\Pr_{ \vec x \sim \mathcal M }\lp[ \vec x \in F_{i} \rp]
 \eps^{-2/K} \; (i+1)^{2d} \; \log^{2d}(1/\alpha) \; \snorm{\mathcal M, L_2}{h}^2 \\
&\leq 
\alpha \; \eps^{-2/K} \; \log^{2d}(1/\alpha) \; \snorm{\mathcal M, L_2}{h}^2
+ 
C \; \eps^{-2/K}  \log^{2d}(1/\alpha) \; \lp( \sum_{i=1}^{\infty}
\alpha^i
  \; (i+1)^{2d} \; \rp)  
\snorm{\mathcal M, L_2}{h}^2 \, ,
\end{align*}
where in the first inequality we bound $h^2(\vec x) \; \mathbbm 1 \{ \vec x \in F_i \}$ 
from above by $(i+1)^{2d} \; \log^{2d}(1/\alpha) \; \snorm{\mathcal M, L_2}{ h }^2$ using the definition of $F_i$, 
in the second inequality
we bound $\alpha \; \Pr_{ \vec x \sim \mathcal E }
 \lp[  \vec x \in F_i \rp]$ from above by
 $ \Pr_{\vec x \sim \mathcal M} \lp[  \vec x \in F_i \rp]$ using the fact that
 $\alpha \; \mathcal E \leq \mathcal M$, and in the last inequality we use \Cref{eq:integer-step-concentration}.
 Now observe that the series
 $ \sum_{i=1}^{\infty} \alpha^i (i+1)^{2d}$ is converging and of order $O_d(\alpha)$
 Hence, the entire expression is bounded above by
 $O_d(\alpha) \; \eps^{-2/K} \; \log^{2d}(1/\alpha) \; \snorm{\mathcal M, L_2}{h}^2$.
 This concludes the proof of the first inequality in \Cref{clm:l2-norm-stability}. 
 The second inequality can be shown in an almost identical way and we therefore omit its proof.
 This concludes the proof of \Cref{clm:l2-norm-stability}.
\end{proof}
The fact that $\vec q(\normal(\vec 0, \vec I)) | R$ and $\mathcal M | R$ are close in total variation distance allows us to view them as ``noisy versions'' of each other. Combining this with \Cref{clm:l2-norm-stability} allows us to relate the $L_2$-norms of $h$ under them.
\begin{proof}[Proof of Step(2)]
We first argue the ``unconditional version'' of the lemma. In particular, we claim that
\begin{align}
\label{eq:intermediate-multiplicative-bound}    
\snorm{\vec q(\normal(\vec 0, \vec I)) , L_2}{h} = \Theta(1) \snorm{\mathcal M , L_2}{h}.
\end{align}
Since $\vec q(\normal(\vec 0, \vec I))$ and $\mathcal M$ are $O(\delta^N)$-close in total variation distance, we have the equality
$$
\vec q(\normal(\vec 0, \vec I)) - \alpha \mathcal E
= \mathcal M - \alpha \mathcal L \, ,
$$
where $\alpha = O(\delta^N)$, $\alpha \mathcal E \leq \vec q(\normal(\vec 0, \vec I))$, and
$ \alpha \mathcal L \leq \mathcal M $.
Using the definition of $L_2$-norm, we thus have
$$
\snorm{\vec q(\normal(\vec 0, \vec I)), L_2  }{ h }^2
 - \alpha \snorm{ \mathcal E, L_2 }{h}^2
= \snorm{\mathcal M, L_2  }{ h }^2
- \alpha \snorm{ \mathcal L, L_2 }{h}^2.
$$
Recall that $\alpha = O(\delta^N)$.
Hence, when $N$ is sufficiently large 
and $\eps$ is sufficiently small given $d, K$, applying \Cref{clm:l2-norm-stability} gives us
\begin{align} \label{eq:E-L-bound}
\alpha \snorm{ \mathcal E, L_2 }{h}^2 \leq \delta^{N/2} \snorm{\vec q(\normal(\vec 0, \vec I)), L_2  }{ h }^2 \, , \,
\alpha \snorm{ \mathcal L, L_2 }{h}^2 \leq \delta^{N/2} \snorm{\mathcal M, L_2  }{ h }^2.    
\end{align}
We therefore have
$$
\abs{ \snorm{\vec q(\normal(\vec 0, \vec I)) , L_2}{h}^2 -  
\snorm{\mathcal M , L_2}{h}^2
} \leq \delta^{N/2} \lp(  \snorm{\vec q(\normal(\vec 0, \vec I)), L_2}{h}^2 +
\snorm{\mathcal M, L_2}{h}^2 \rp).    
$$
Since $\delta^{N/2} = o(1)$, \Cref{eq:intermediate-multiplicative-bound} hence follows.

Next we turn to the conditional $L_2$-norms. 
Notice that even if we restrict to the set $R$, we still have the equality
\begin{align*}
&\E_{ \vec x \sim \vec q(\normal(\vec 0, \vec I)) }
\lp[ 
h^2(\vec x) \mathbbm 1\{\vec x \in R \}
\rp]
-
\alpha \E_{ \vec x \sim \mathcal E }
\lp[ 
h^2(\vec x) \mathbbm 1\{\vec x \in R \}
\rp] \\
&= 
\alpha \E_{ \vec x \sim \mathcal M }
\lp[ 
h^2(\vec x) \mathbbm 1\{\vec x \in R \}
\rp]
-
\alpha \E_{ \vec x \sim \mathcal L }
\lp[ 
h^2(\vec x) \mathbbm 1\{\vec x \in R \}
\rp].    
\end{align*}
It is easy to see that $\E_{ \vec x \sim \mathcal E }
\lp[ 
h^2(\vec x) \mathbbm 1\{\vec x \in R \}
\rp]$ is at most $\snorm{ \mathcal E , L_2 }{h}^2$ (and similarly for $\E_{ \vec x \sim \mathcal L }\lp[ h^2(\vec x) \mathbbm 1\{\vec x \in R \}\rp]$).
Combining this with \Cref{eq:E-L-bound} gives us
\begin{align*} 
\abs{ \E_{ \vec x \sim \vec q(\normal(\vec 0, \vec I)) }
\lp[ 
h^2(\vec x) \mathbbm 1\{\vec x \in R \}
\rp]
-
\E_{ \vec x \sim \mathcal M }
\lp[ 
h^2(\vec x) \mathbbm 1\{\vec x \in R \}
\rp] 
}
\leq 
\delta^{N/2} \lp(  \snorm{\vec q(\normal(\vec 0, \vec I)), L_2}{h}^2 +
\snorm{\mathcal M, L_2}{h}^2 \rp).    
\end{align*}
\Cref{eq:intermediate-multiplicative-bound} implies that $\snorm{\vec q(\normal(\vec 0, \vec I)), L_2}{h}^2, \snorm{\mathcal M, L_2}{h}^2$ are within a constant factor of each other.
Hence, we can further simplify the right hand side of the inequality above, which gives us
\begin{align*} 
\abs{ \E_{ \vec x \sim \vec q(\normal(\vec 0, \vec I)) }
\lp[ 
h^2(\vec x) \mathbbm 1\{\vec x \in R \}
\rp]
-
\E_{ \vec x \sim \mathcal E }
\lp[ 
h^2(\vec x) \mathbbm 1\{\vec x \in R \}
\rp] 
}
\leq 
O\lp( \delta^{N/2} \rp) \; \snorm{\vec q(\normal(\vec 0, \vec I)), L_2}{h}^2.
\end{align*}
Define $\kappa := \Pr_{ \vec x \sim \vec q(\normal(\vec 0, \vec I)) }\lp[ \vec x \in R\rp]$.
By our assumption, we have $\kappa \geq \poly_{d, \ell}(\eps)$.
We can normalize the above inequality with $\kappa$, which leads to
\begin{align} \label{eq:false-normalization}
\abs{ \E_{ \vec x \sim \vec q(\normal(\vec 0, \vec I)) \mid R }
\lp[ 
h^2(\vec x)
\rp]
-
\E_{ \vec x \sim \mathcal M }
\lp[ 
h^2(\vec x) \mathbbm 1\{\vec x \in R \}
\rp] / \kappa
}
\leq 
\delta^{N/4}  \snorm{\vec q(\normal(\vec 0, \vec I)), L_2}{h}^2.
\end{align}
We are almost done here except that we have the wrong normalization factor for $\snorm{\mathcal M \mid R, L_2}{h}^2$. Fortunately, the current normalization factor $\kappa$ is not too different from $\Pr_{\vec x \sim \mathcal M}{ \vec x \in R }$.
Due to the fact that $\vec q(\normal(\vec 0, \vec I))$ and $\mathcal M$ are close in total variation distance, we must have $ \abs{ \Pr_{ \vec x \sim \mathcal M }\lp[ \vec x \in R\rp] - \kappa } \leq O(\delta^N) $.
This gives us
\begin{align} \label{eq:correct-normalization}
\abs{
\E_{ \vec x \sim \mathcal M }
\lp[ 
h^2(\vec x) \mathbbm 1\{\vec x \in R \}
\rp] / \kappa
-  \E_{ \vec x \sim \mathcal M \mid R }
\lp[ 
h^2(\vec x)
\rp]
}
&\leq \delta^{N/2} \E_{ \vec x \sim \mathcal M }
\lp[ 
h^2(\vec x) \mathbbm 1\{\vec x \in R \}
\rp] \nonumber \\
& \leq
\delta^{N/2} \snorm{\mathcal M, L_2}{ h }^2
\leq O \lp( \delta^{N/2} \rp) \snorm{\vec q(\normal(\vec 0, \vec I)), L_2}{ h }^2 \, ,
\end{align}
where in the last inequality we again use \Cref{eq:intermediate-multiplicative-bound}.
Combining \Cref{eq:correct-normalization}, \Cref{eq:false-normalization} and the triangle inequality then gives us
$$
\abs{ \snorm{\vec q(\normal(\vec 0, \vec I)) \mid R, L_2}{h}^2 -  
\snorm{\mathcal M \mid R, L_2}{h}^2
} \leq O \lp( \delta^{N/4} \rp) \snorm{\vec q(\normal(\vec 0, \vec I)), L_2}{h}^2.
$$
Lastly, by \Cref{lem:conditional-l2-norm}, $\snorm{\vec q(\normal(\vec 0, \vec I)), L_2}{h}$ is at most $\poly_{d, \ell}(1/\eps) \; \snorm{\vec q(\normal(\vec 0, \vec I)) \mid R, L_2}{h}$.
When $N$ is sufficiently small, $\delta^{N/4} \; \poly_{d, \ell}(1/\eps) = o(1)$. This then allows to conclude that the two terms on the left hand side of the above inequality are within a constant factor of each other.
\end{proof}

Finally, we show Step (3) (\Cref{eq:step3}), which easily follows from the definition of distribution comparability.
\begin{proof}[Proof of Step (3) (\Cref{eq:step3})]
By the definition of distribution comparability, there exists a constant $C$ such that
$$
1/C \int_{\theta} \lp(\normal_{\theta} \mid R\rp)( \vec x ) \; d \theta \leq \mathcal M \mid R(\vec x) \leq C \int_{\theta} \lp(\normal_{\theta} \mid R\rp)( \vec x ) \; d \theta
$$
for all $\vec x$.
We hence have
$$
\snorm{ \mathcal M | R, L_2 }{h}^2
= \int_{\vec x} \lp( \mathcal M \mid R \rp)(\vec x) h^2(\vec x) d \vec x
\leq C \; \int_{\vec x} 
\int_{\theta} \lp(\normal_{\theta} \mid R\rp)(\vec x) \; d\theta \;
 h^2(\vec x) d \vec x
= C \snorm{\int_{\theta} \normal_{\theta} d\theta \mid R, L_2}{h}^2.
$$
The other direction can be shown similarly.
\end{proof}
Combining the three steps (\Cref{eq:step1}, \Cref{eq:step2}, and \Cref{eq:step3}) then concludes the proof of \Cref{lem:transfer-l2-norm-bound}.
\subsection{Proof of \texorpdfstring{\Cref{thm:(anti-)concentration}}{}}
\label{sec:conclude-anti-concentration}
We are now ready to combine everything together to derive our final (anti-)concentration bounds.
We will show the anti-concentration and the concentration bounds separately.
A crucial fact that we will use is that the $L_2$-norm of $p$ under the original distribution $D$ is within constant factors of the $L_2$-norm of the surrogate polynomial $h$ under the conditional reasonable Gaussian mixture $\int_{\theta} \normal_{\theta} d\theta$, as shown in  \Cref{lem:transfer-l2-norm-bound}.
Specifically, there exists a sufficiently large universal constant $C$ such that
\begin{align} \label{eq:apply-norm-transfer}
\frac{1}{C} \snorm{ \int_{\theta} \normal_{\theta} d\theta \mid R, L_2}{f} 
\leq  \snorm{D, L_2}{p} \leq C \snorm{ \int_{\theta} \normal_{\theta} d\theta \mid R, L_2}{h}.    
\end{align}

\paragraph{Proof of Anti-concentration.}
Choose some $t$ satisfying $\eps < t < \eps^{2/K}$.
\Cref{eq:apply-norm-transfer} implies
$$
\Pr_{ \vec x \sim D } \lp[ \abs{p(\vec x)} < t \snorm{D, L_2}{p}
\rp]
\leq
\Pr_{ \vec x \sim D } \lp[ \abs{p(\vec x)} < t \; C \;  \snorm{ \int_{\theta} \normal_{\theta} d\theta \mid R , L_2}{h}
\rp].
$$
Notice that the threshold used in the right hand side of the equation can be bounded below by
 $$
 t \; C  \snorm{ \int_{\theta} \normal_{\theta} d\theta \mid R, L_2}{h} > \eps \snorm{D, L_2}{p} > \poly_{d, \ell}(\eps)
 = \poly_{d, \ell}( \delta^{d^2K} )
 > \delta^{M/2} \, ,
 $$ 
 where the first inequality uses \Cref{eq:apply-norm-transfer} and our choice of $t > \eps$, 
 the second inequality uses \Cref{lem:conditional-l2-norm}, and the last inequality is true as long as $M$ is sufficiently large given $d, \ell, K$.
 \Cref{lem:transfer-probability-bound} then allows us to relate the probability of the event under $D$ to that under the mixture $\int_{\theta} \normal_{\theta} d\theta$,  which implies that
\begin{align*}
&\Pr_{ \vec x \sim D } \lp[ \abs{p(\vec x)} < C \eps \snorm{ \int_{\theta} \normal_{\theta} d\theta \mid R, L_2}{h}
\rp] \\
&\leq 
\Pr_{ \vec y \sim \int_{\theta} \normal_{\theta} d\theta \mid R } \lp[ \abs{h(\vec y)} < 2 C t \snorm{ \int_{\theta} \normal_{\theta} d\theta \mid R, L_2}{h}
\rp] + O( \eps^{N/(2d^2K)} )
+ \exp\lp( - \Omega\lp( t \snorm{\int_{\theta} \normal_{\theta} d\theta \mid R, L_2}{h} \delta^{-M} \rp)^{1/d} \rp).
\end{align*}
We bound the three terms separately.
For the first term, since we know that $\int_{\theta} \normal_{\theta} d\theta \mid R$ is a mixture of $(\delta, \poly_{d,m,N} ( \log(1/\delta) ) )$-reasonable Gaussians conditioned on the axis-aligned rectangle $R$, where each coordinate is at most $\poly_{d,\ell}( \log(1/\delta))$-far from the origin, invoking \Cref{lem:Gaussian-Mixture-anti-concentration} gives us the bound
$$
\Pr_{ \vec x \sim \int_{\theta} \normal_{\theta} d\theta \mid R } \lp[ \abs{h(\vec x)} < 2 C t \snorm{ \int_{\theta} \normal_{\theta} d\theta \mid R, L_2}{h}
\rp]
\leq 
O_{d,m}(1) \; \lp( t \delta^{-2d} \poly_{d,m,N} ( \log(1/\delta) ) \rp)^{1/d}.
$$
Recall that $\delta^{d^2K} = \eps$ and $t < \eps^{2/K}$.
Hence, $O_{d,m}(1) \; \delta^{-2d} \ll t^{-1/2}$ when $\delta$ is sufficiently small given $d,m$.
It then follows that the first term is bounded above by $t^{1/(2d)} / 2$.
Since $\eps < t$, the second term is obviously a lower-order term when $N$ is a sufficiently large multiple of $d^2K$.
For the third term, we have already shown that 
$ t \snorm{ \int_{\theta} \normal_{\theta} d\theta \mid R, L_2}{h} \geq  \delta^{M/2} $. 
Hence, it is also obviously a lower-order term when $M$ is a sufficiently large constant multiple of $d^2K$.
Combining the bounds for the three terms then conclude the proof of anti-concentration:
$$\Pr_{ \vec x \sim D } \lp[ \abs{p(\vec x)} < t \snorm{D, L_2}{p}
\rp]
\leq
\Pr_{ \vec x \sim D } \lp[ \abs{p(\vec x)} < C t \snorm{ \int_{\theta} \normal_{\theta} d \theta, L_2}{h}
\rp] \leq t^{1/(2d)}.
$$

\paragraph{Proof of Concentration. }
Finally, we turn our attention to the concentration bound.
Notice that the relevant bound needs to hold for all $t > 0$.
A technical issue is that the concentration bound we can establish via comparing $D$ to $\int_{\theta} \normal_{\theta} d \theta$ will stop working in the extremely far tail of the distribution since we inevitably incur a cost of $O(\delta^N)$ due to the application of \Cref{lem:transfer-probability-bound}.
Fortunately, in the extreme tail region, it is enough for us to argue through the concentration of $p$ under the full Gaussian.
Specifically, we need to divide into two cases depending on the value of the threshold $t$.

\textbf{Case I: Extreme Tails.} 
In the first case, we have $t > \delta^{-N/K}$.
We have already shown that $\snorm{D, L_2}{p} \geq \poly_{d, \ell}(\eps)$.
Hence, it follows that
$$
\Pr_{ \vec x \sim \normal(\vec 0, \vec I) }\lp[ 
\abs{p(\vec x)} > t  \snorm{L_2, D}{ p }
\rp] 
\leq 
\Pr_{ \vec x \sim \normal(\vec 0, \vec I) }\lp[ 
\abs{p(\vec x)} > t \; \poly_{d, \ell}(\eps)  \snorm{L_2}{ p }
\rp].
$$
In this case, it suffices to argue through the concentration of the polynomial under the full Gaussian. 
In particular, using \Cref{lem:polynomial-concentration}, we have
\begin{align*}
\Pr_{ \vec x \sim \normal(\vec 0, \vec I) }\lp[ 
\abs{p(\vec x)} > t \; \poly_{d, \ell}(\eps)  \snorm{L_2, D}{ p }
\rp] 
\leq
\exp\lp( - \Omega( t \; \poly_{d, \ell}(\eps) )^{1/d} \rp).
\end{align*}
Using the fact that the mass of the conditioning set of $D$ under $\normal(\vec 0, \vec I)$
is at least $\poly_{d, \ell, K}(\eps)$, we further have
\begin{align*}
\Pr_{ \vec x \sim D }\lp[ 
\abs{p(\vec x)} > t   \snorm{L_2, D}{ p }
\rp] 
\leq
\exp\lp( - \Omega( t \; \poly_{d, \ell}(\eps) )^{1/d} + O_{d, \ell, K}( \log(1/\eps) ) \rp).
\end{align*}
Since we assume that $t > \delta^{-N/d}$, when $N$ is sufficiently large given $d, \ell$, and $\eps$ is sufficiently small given $d, \ell, K, N$,
it holds
$ - \Omega( t \; \poly_{d, \ell}(\eps) )^{1/d} + O_{d, \ell, K}( \log(1/\eps) ) \geq t^{1/(2d)} $.
Combining these inequalities together then gives us
$$
\Pr_{ \vec x \sim \normal(\vec 0, \vec I) }\lp[ 
\abs{p(\vec x)} > t  \snorm{L_2, D}{ p }
\rp] 
\leq \exp( - t^{1/2d}  ) < t^{-K}.
$$

\textbf{Case II: Moderate Tails.} 
In the second case, we will choose $1 < t < \delta^{-N / d}$ (as the concentration bound is trivial for $t < 1$).
In this case, the proof is similar to the one for anti-concentration.
In particular, using \Cref{eq:apply-norm-transfer}, we have
$$
\Pr_{ \vec x \sim D } \lp[ \abs{p(\vec x)} > t \eps^{-1/K} \snorm{D, L_2}{p}
\rp]
\leq
\Pr_{ \vec x \sim D } \lp[ \abs{p(\vec x)} > t \lp(\eps^{-1/K} /C\rp) \snorm{ \int_{\theta} \normal_{\theta} d \theta, L_2}{h} \rp].
$$
Since
$t \lp(\eps^{-1/K} /C\rp) \; \snorm{ \int_{\theta} \normal_{\theta} d\theta, L_2}{h} >  
\Omega \lp( \snorm{ D, L_2}{p} \rp)  > \delta^{M/2}$, we can again apply \Cref{lem:transfer-probability-bound}, which gives us
\begin{align}
&\Pr_{ \vec x \sim D } \lp[ \abs{p(\vec x)} > t \lp(\eps^{-1 / K} / C\rp) \snorm{ \int_{\theta} \normal_{\theta} d \theta, L_2}{h}
\rp]  \nonumber \\
&\leq 
\Pr_{ \vec y \sim \int_{\theta} \normal_{\theta} d \theta } \lp[ 
\abs{h(\vec y)} > \frac{t \eps^{-1/K}}{2C} \snorm{ \int_{\theta} \normal_{\theta} d \theta, L_2}{h}
\rp] + O( \delta^{N/2} ) \nonumber \\
\label{eq:concentration-error-decomposition}
&+ \exp\lp( - \Omega\lp( \frac{t \eps^{-1/K} }{2C} \snorm{\int_{\theta} \normal_{\theta} d \theta, L_2}{h} \delta^{-M} \rp)^{1/d} \rp).
\end{align}
For the first term, using the concentration bound for a mixture of reasonable Gaussians, we have that it is bounded above by
$$ 
\exp\lp( - \Omega \lp(  t \eps^{-1/K}  \frac{\delta^{2d}}{\poly_{d,\ell}( \log(1/\delta) )} \rp)^{1/d}  \rp).
$$
Notice that $\eps^{-1/K}  \frac{\delta^{2d}}{\poly_{d,\ell}( \log(1/\delta) )} \gg 1$ since $\delta^{d^2K} = \eps$ and $\delta$ is sufficiently small given $d, \ell$.
Thus, the probability can be bounded above by $\exp ( - \Omega(t)^{1/d} ) < O_{d,K} ( t^{-1/K} )$.
For the second term, since we assume that $t < \delta^{-N  / K}$, it then follows that
$O( \delta^{N / 2} ) < O\lp( t^{-1/K}  \rp)$.
For the last term, we already have that the expression inside $\Omega(\cdot)$ is at least $\eps^{M/2}$.
Hence, the term can be bounded above by $\exp \lp( \delta^{-M/(2d)} \rp)$, which is a lower order term compared to $O( \delta^{N / 2} )$.
Combining the bounds on the three terms on the right hand side of \Cref{eq:concentration-error-decomposition} then gives us
$$
\Pr_{ \vec x \sim D } \lp[ \abs{p(\vec x)} > t \eps^{-1/K} \snorm{D, L_2}{p}
\rp] \leq O_{d,K}\lp( t^{-K} \rp).
$$
This concludes the proof of \Cref{thm:(anti-)concentration}.

\section{Polynomial Set Partitioner}\label{sec:polynomial-set-cutter}
The fundamental barrier against a naive localization strategy that iteratively applies the robust margin-perceptron algorithm on the low-margin points with respect to the polynomial learned from the previous iteration is the following: the Gaussian distribution $\normal(\vec 0, \vec I)$ conditioned on the set $\{\vec x \in \R^n:\abs{p(\vec x)} \leq \eps\}$ does not necessarily satisfy Carbery-Wright type anti-concentration. 
Equipped with \Cref{prop:super-non-singular-extension} and \Cref{thm:(anti-)concentration}, we can build an efficient routine in to decompose a poorly anti-concentrated set roughly into $\poly_d(1/\eps)$ many disjoint sets such that the conditional distribution on each subset satisfies good anti-concentration (see \Cref{intro-prop:set-cutter}).


\Cref{intro-prop:set-cutter} allows us to at least kick-start our second round of perceptron learning by running it on each of the conditional distributions independently.
To make sure we can keep doing so after multiple rounds of perceptron learning, we need the routine to be capable of decomposing sets under not only the standard Gaussian distribution but also under $\normal(\vec 0, \vec I)$ conditioned on a previous set it has produced. 
Thankfully, this is made possible by the fact that the super non-singular decomposition algorithm (see \Cref{prop:super-non-singular-extension}) is ``extendible''. 
Specifically, assume that we have received a second polynomial $p'$ after conditioning on the set $\{ \vec x: (q_1(\vec x), \cdots, q_{\ell}(\vec x)) \in R \}$ for some $\ell$-dimensional axis-aligned rectangle $R$. We can then extend $\{q_1, \cdots,q_{\ell}\}$ into a larger super non-singular set $\{q_1, \cdots, q_{\ell}, q_{\ell+1}, \cdots, q_m\}$ such that we can still approximately write $p'(\vec x) \approx h'( q_1(\vec x), \cdots, q_m(\vec x) )$. 
We can similarly cover $\{\vec x \in \R^n: \abs{p'(\vec x)} \leq \eps,  (q_1(\vec x), \cdots, q_{\ell}(\vec x)) \in R \}$ with sets of the form $\{\vec x:  (q_1(\vec x), \cdots, q_{\ell}(\vec x) \in R, (q_{\ell+1}(\vec x), \cdots, q_{m}(\vec x) \in R')\}$, where $R'$ is some other $(m-\ell)$-dimensional axis-aligned rectangle.

However, after each partitioning iteration, the new sets produced will be defined by polynomials with weaker super non-singular properties.
This is because each time after we extend some super non-singualar set using \Cref{prop:super-non-singular-extension}, the resulting set produced will still be super non-singular but with slightly worse parameters.
Hence, to ensure our overall localization procedure can be used iteratively in sufficiently many rounds and that the resulting set of polynomials is still sufficiently super non-singular to establish the desired (anti-)concentration properties through \Cref{thm:(anti-)concentration}, 
we need the initial set of polynomials produced to satisfy quite strong super non-singular conditions. 
To make it easier to express the strength of super non-singularity needed in each round, we define the notion of \emph{Level-$k$ super non-singular} polynomials with respect to a tuple $(\eps, d, K, M) \in (0, 1) \times \lp(\mathbbm Z^{+} \rp)^{3}$ (these parameters encode the input to our algorithm and will be fixed in all rounds during its execution).
Intuitively, Level-$k$ super non-singularity means that the underlying set can still be extended $k$ more times and the resulting set will still be sufficiently super non-singular.
\begin{definition}[Level-$k$ Super Non-Singular Polynomial Set]
\label{def:level-k-super-non-singular}
Fix a tuple $(\eps, d, K, M) \in (0, 1) \times \lp(\mathbbm Z^{+} \rp)^{3}$.
We use the integers from the tuple to define an infinite sequence of functions
$$\{f_{d,K,M}^{(0)}, \cdots,  f_{d,K,M}^{(k)}, \cdots \}.$$
Let $f_{d,K,M}^{(0)}: \mathbb Z^+ \mapsto \mathbb Z^+$ be a function defined as $f_{d,K,M}^{(0)}(\ell) = C_{d, \ell, K}$, where $C_{d, \ell, K}$ is a number that is sufficiently large given its subscripts as required by \Cref{thm:(anti-)concentration}.
Then we define the remaining functions in the sequence recursively as
$f_{d,K,M}^{(k)}(\ell) = C_{ d, \ell, f^{(k-1)}_{d,K,M}, M } \, ,$ 
where $C_{ d, \ell, f^{(k-1)}_{d,K,M}, M }$ is a number that is sufficiently large given its subscripts as required by \Cref{prop:super-non-singular-extension}.
Let $\vec q: \R^n \mapsto \R^m$ be a degree-$d$ polynomial transformation.
We say $\vec q$ is Level-$k$ super non-singular with respect to $(\eps, d, K, M)$ if $\vec q$ is $\lp(\eps^{1 /( 3^{k+1} d^2 K )},f_{d,K,M}^{(k)}(m) \rp)$-super non-singular.
\end{definition}
Before presenting the formal statement of the main result of the section, we discuss the growth rate of the functions defined in \Cref{def:level-k-super-non-singular}.
\begin{claim}
\label{clm:growth-rate}
Fix a tuple $(\eps, d, K, M) \in (0, 1) \times \lp(\mathbbm Z^{+} \rp)^{3}$.
Let $f^{(k)}_{d,K,M}$ be the $k$-th function in the function sequence defined in \Cref{def:level-k-super-non-singular}.
Then it holds $ C_{d, \ell, f^{(k)}_{d,K,M}, M} = O_{ d, \ell, k, K, M }(1) $.
\end{claim}
\begin{proof}
It suffices to show that the number $C_{d, \ell, f^{(k)}_{d,K,M}, M}$ can be computed by a program that only takes $d, \ell, k, K, M$ as input.
We know that there exists a program that can compute the number $C_{d, \ell, f^{(k)}_{d,K,M}, M}$ given oracle access to the function $f^{(k)}_{d,K,M}$.
Hence, it suffices to show that $f^{(k)}_{d,K,M}(\ell)$ for any positive integer $\ell$ can be computed by a sub-routine that only takes $d, k, K, M, \ell$ as input.
We show this via induction on $k$.
For $k = 0$, we have that $f^{(0)}_{d,K,M}(\ell) = C_{d, \ell, K}$, which can be computed by a program that takes input $d, \ell, K$.
For $k \geq 1$, we have that
$f^{(k)}_{d,K,M}(\ell) = C_{d, \ell, f_{d,K,M}^{(k-1)},K}$.
Therefore, we know that $f^{(k)}_{d,K,M}(\ell)$ can be computed by a program given $d, \ell, K$ as input and oracle access to $f_{d,K,M}^{(k-1)}$.
By our inductive hypothesis, $f_{d,K,M}^{(k-1)}(\ell')$ for any positive integer $\ell'$ can be computed by a program that takes only the input $d,k,K,M,\ell'$.
Thus, we conclude that $f^{(k)}_{d,K,M}(\ell)$ can be computed by a program given $d, k, K, M, \ell$.
By mathematical induction, this holds for all positive integers $k$. This concludes the proof of \Cref{clm:growth-rate}.
\end{proof}

Now we are ready to state the formal guarantees of \emph{Polynomial-Set-Parititioner}.
\begin{proposition}\label{prop:set-cutter}
Let $k, d, \ell, K$ be positive integers and $\eps \in (0, 1)$. 
Define the following mathematical objects:
\begin{itemize}
    \item $p: \R^n \mapsto \R$ is a degree-$d$ polynomial with $\snorm{L_2}{p} \leq 1$.
    \item $\vec q: \R^n \mapsto \R^{\ell}$ is a vector-valued polynomial of degree at most $d$.
    \item $R$ is an axis-aligned rectangle in $\R^{\ell}$.
\end{itemize}
Assume that (i) $\eps$ is sufficiently small given $d,\ell,k,K$ (ii) $\vec q$ is Level-$k$ super non-singular with respect to the tuple $(\eps, d, K, M=3)$, and  (iii) $\Pr[ \vec q(\vec x) \in R ] \geq \poly_{d, \ell}( \eps )$.
Then for some $m = O_{d, \ell, k, K}(1)$, there exists an algorithm \emph{Polynomial-Set-Partitioner} that takes the set 
$$
\mathcal P := \{ \vec x: \abs{ p(\vec x) } < \eps \text{ and } \vec q(\vec x) \in R \}
$$ 
as input and produces at most $\eps^{-5md}$ many sets of the form 
$$
\bar {\mathcal P}_i := 
\{ \vec x: \bar {\vec q}(\vec x) \in \bar R_i \} \, ,
$$ 
where $\bar {\vec q}: \R^n \mapsto \R^m $ is a degree-$d$ vector-valued polynomial and each $\bar R_i$ is an $m$-dimensional axis-aligned rectangle.
In particular, the sets produced satisfy the following:
\begin{enumerate}
    \item The distribution $\normal(\vec 0, \vec I)$ conditioned on each $\bar {\mathcal P}_i$ satisfies strong (anti-)concentration properties. Specifically, for any polynomial $g:\R^n \mapsto \R$ of degree-$d$, for $\eps < t < \eps^{2/K}$,
it holds
\begin{align*}
\Pr_{ \vec x \sim \normal(\vec 0, \vec I) \mid  \bar {\mathcal P}_i }
\lp[ 
\abs{g(\vec x)} < t \; \snorm{ \normal(\vec 0, \vec I) \mid  \bar {\mathcal P}_i , L_2}{g}
\rp] \leq t^{1/(2d)} \, ,
\end{align*}
and for all $t > 0$, it holds
\begin{align*}
\Pr_{ \vec x \sim \normal(\vec 0, \vec I) \mid  \bar {\mathcal P}_i } \lp[ \abs{g(\vec x)} > t \eps^{-1/K} \snorm{\normal(\vec 0, \vec I) \mid  \bar {\mathcal P}_i, L_2}{g}
\rp] \leq O_{d,K}\lp( t^{- K  } \rp).
\end{align*}
\item Each set $\bar {\mathcal P}_i$ has non-trivial probability mass under $\normal(\vec 0, \vec I)$, i.e., 
$$
\Pr_{ \vec x \sim \normal(\vec 0, \vec I) }
\lp[ \vec x \in \bar {\mathcal P}_i \rp] >
\eps^{3 + 5md} \; \Pr_{\vec x \sim \normal(\vec 0, \vec I)}\lp[\vec q(\vec x) \in R \rp].
$$
    \item 
    Define $\bar {\mathcal P} := \bigcup_{i} \bar {\mathcal P}_i$.
    The points from the given set $\mathcal P$ that are not included in $\bar {\mathcal P}$ have small probability mass, i.e., 
    $$
    \Pr_{ \vec x \sim \normal(\vec 0, \vec I)  }
    \lp[ \vec x \not \in \bar {\mathcal P}, \vec x \in \mathcal P \rp] < \eps^2 
    \Pr_{ \vec x \sim \normal(\vec 0, \vec I)  }\lp[ \vec q(\vec x) \in R \rp].
    $$
    \item 
    $\bar {\mathcal P}$ is approximately contained in a slightly larger set compared to $\mathcal P$, i.e.
    $ \tilde {\mathcal P} :=  \{ \abs{p(\vec x)} < 2 \eps, \vec q(\vec x) \in R \}$. In particular, it holds
    $$
    \Pr_{ \vec x \sim \normal(\vec 0, \vec I)  }
    \lp[ \vec x \in \bar {\mathcal P}, \vec x \not \in \tilde {\mathcal P}
    \rp] < \eps^2 \;
    \Pr_{ \vec x \sim \normal(\vec 0, \vec I)  }
    \lp[ \vec q(\vec x) \in R \rp].
    $$
    \item $\bar {\vec q}$ is level-$(k-1)$ super non-singular with respect to the tuple $(\eps, d, K, M=3)$.
\end{enumerate}
Moreover, the algorithm runs in time $\poly(n^d) \; \poly_{d, \ell, k, K}(1/\eps)$.
\end{proposition}
\begin{proof}
Let $S = \{q_1, \cdots,q_{\ell}\}$ be the set of polynomials corresponding to the coordinates of $\vec q$.
By our assumption, $\vec q$ is Level-$k$ super non-singular with respect to the tuple $(\eps, d, K, M = 3)$.
Thus, by definition, the set $S$ is $(\eps^{1/(3^{k+1}d^2 K)}, f_{d,K,M=3}^{(k)}(\ell))$ super non-singular. 
Since $f_{d,K,M=3}^{(k)}(\ell) = C_{d,\ell,f^{(k - 1)}, M=3}$ by its definition, 
using \Cref{prop:super-non-singular-extension},
we can extend the set into 
another $(\eps^{1/(3^{k}d^2 K)}, f^{(k-1)}_{d,K,M=3}(m))$ super non-singular set
$\{q_1, \cdots,q_{\ell}, q_{\ell+1}, \cdots, q_m \}$, 
where $m = O_{d, \ell, f^{(k-1)}_{d, K, M=3}}(1)
= O_{d, \ell, k, K}(1)$ by \Cref{clm:growth-rate}, such that there exists a composition polynomial $h: \R^m \mapsto \R$ and an error polynomial $e: \R^n \mapsto \R$ satisfying
\begin{align}
\label{eq:cutter-decomposition}
    p(\vec x) = h(q_1(\vec x), \cdots, q_m(\vec x) ) + e(\vec x) \, , \,
    \snorm{L_2}{e} \leq \eps^3.
\end{align}
Moreover, we can find the decomposition in \Cref{eq:cutter-decomposition} in time  $\poly(n^d) \; \poly_{d, \ell, f^{(k-1)}_{d, K, M=3}, M=3}(1/\eps)$,
 which is simply $\poly(n^d) \; \poly_{d, \ell, k, K}(1/\eps)$ by \Cref{clm:growth-rate}.
 
Define $\bar{\vec q}: \R^n \mapsto \R^m$ to be the vector-valued polynomial whose $i$-th coordinate is $q_i$.
Then by definition, $\bar{\vec q}$ is Level-$(k-1)$ super non-singular with respect to the same tuple $(\eps, d, K, M=3)$. 
This guarantees Condition (5) in the proposition is satisfied.

With the decomposition and its associated polynomial transformation $\bar{\vec q}$ in our hand, we begin our search of axis-aligned rectangles in the $m$-dimensional output space of $\bar {\vec q}$.
The first step is to restrict our search space to 
a very large rectangle that is obtained by extending the original $\ell$-dimensional rectangle $R$ in all the remaining $(m-\ell)$ dimensions:
$$ 
\bar R := R \times [- \alpha_{\ell, d} \log^{d/2}(m/\eps), \alpha_{\ell, d} \log^{d/2}(m/\eps)  ]^{m - \ell} \, ,
$$
where $\alpha_{\ell, d}$ is some number that depends on $\ell, d$ that we will specify later.
Since we have $\snorm{L_2}{q_i} = 1$ for all $i \in [m]$,
by the union bound and the concentration of polynomials (\Cref{lem:polynomial-concentration}), it holds
\begin{align*}
\Pr_{ \vec x \sim \normal(\vec 0, \vec I) }\lp[ \bar{\vec q}(\vec x) \not \in \bar R \rp] < \exp\lp( - \Omega( \alpha_{\ell, d} )^{2/d} \log(1/\eps) \rp).
\end{align*}
Since we also assume that 
$\Pr_{ \vec x \sim \normal(\vec 0, \vec I) }\lp[ \vec q(\vec x) \in R \rp]
> \poly_{d, \ell}(\eps)$, it is not hard to see that if we choose $\alpha_{d, \ell}$ to be sufficiently large given $d, \ell$, we then have
\begin{align} \label{eq:giant-discard}
\Pr_{ \vec x \sim \normal(\vec 0, \vec I) }\lp[ \bar{\vec q}(\vec x) \not \in \bar R \rp] < \eps^3 \Pr_{ \vec x \sim \normal(\vec 0, \vec I) }\lp[ 
\vec q(\vec x) \in R
\rp].
\end{align}
We next partition the rectangle $\bar R$ into small rectangles $\{ \bar R_i \}$, each of side length $\eps^{4d}$.
Since the side length of the rectangle $\bar R$ is $ 2 \alpha_{d, \ell} \log^{d/2}(m / \eps) = \poly_{d, \ell, k, K}( \log(1/\eps) )$ (recall that $m = O_{d, \ell, k, K}(1)$), 
it is not hard to see that 
the number of sub-rectangles we end up with is at most
$$ 
\lp(\poly_{d, \ell, k, K}( \log(1/\eps)) \; \eps^{-4d} \rp)^m 
\leq \poly_{d, m, k, K}( \log(1/\eps) ) \; \eps^{-4dm}
\leq \eps^{-5md} \, ,
$$
where the last inequality is true if $\eps$ is sufficiently small given $d, m, k, K$.
These will be the candidate rectangles we use to define the output set $\bar {\mathcal P}_i$.

Let $\tilde \normal$ be the empirical distribution of $\samplesize = \poly_{m,d}( 1/\eps  )$ \iid samples from $\normal(\vec 0, \vec I)$.
Among the small rectangles, we pick the ones that are (i) not too light, and (ii) contain a significant fraction of points satisfying $\abs{p(\vec x) < \eps}$. More formally, the selection rule is given by
$$
\Pr_{ \vec x \sim \tilde \normal } \lp[  \bar{\vec q}(\vec x) \in {\bar R}^{(i)} \mid \vec q(\vec x) \in R\rp] > 2 \eps^{3 + 5md}
\; , \;
\Pr_{ \vec x \sim \tilde \normal } \lp[  \abs{p(\vec x)} < \eps   \mid  \bar {\vec q}(\vec x) \in {\bar R}^{(i)} \rp] > 2 \eps^3.
$$
Note that this step can be completed in time $\poly(n^d) \; \poly( \samplesize ) \; \eps^{-5md}$. The final output of the algorithm will simply be $\bar {\mathcal P}_i = \{ \x: \bar q(\x) \in \bar R_i \}$ for those $\bar R_i$ that are retained after the selection.
Hence, this shows that the total runtime is $\poly(n^d) \; \poly_{d, \ell, k, K}(1/\eps)$.

We proceed to show the sets produced satisfy Conditions (1)-(4) in \Cref{prop:set-cutter}.
Let $\mathcal K$ be the family of sub-rectangles kept and $\mathcal T$ be the family of sub-rectangles discarded after the selection procedure.
If we take $\samplesize = \poly_{m,d}( 1/\eps  )$ many samples, 
with high probability, we will have the following probability mass separation between the rectangles from $\mathcal K$ and from $\mathcal T$ under $\normal(\vec 0, \vec I)$.\\
For all the sub-rectangles $R^{(i)} \in \mathcal K$ picked, we have
\begin{align} \label{eq:picked}
\Pr_{ \vec x \sim \normal(\vec 0, \vec I) } \lp[  \bar{\vec q}(\vec x) \in {\bar R}^{(i)} \mid \vec q(\vec x) \in R\rp] >   \eps^{3 + 5md} \text{ and }
\Pr_{ \vec x \sim \normal(\vec 0, \vec I) } \lp[  \abs{p(\vec x)} < \eps   \mid  \bar {\vec q}(\vec x) \in {\bar R}^{(i)} \rp] >  \eps^3.
\end{align}
For all the sub-rectangles $R^{(i)} \in \mathcal T$ removed, we have
\begin{align} \label{eq:thrown}
\Pr_{ \vec x \sim \normal(\vec 0, \vec I) } \lp[  \bar{\vec q}(\vec x) \in {\bar R}^{(i)} \mid \vec q(\vec x) \in R\rp] <   3 \eps^{3 + 5md} \text{ or }
\Pr_{ \vec x \sim \normal(\vec 0, \vec I) } \lp[  \abs{p(\vec x)} < \eps   \mid  \bar {\vec q}(\vec x) \in {\bar R}^{(i)} \rp] <  3 \eps^3.
\end{align}
Notice that \Cref{eq:picked} establishes Condition (2) in \Cref{prop:set-cutter}.

For each $\bar R_i \in \mathcal K$, we claim that $\normal(\vec 0, \vec I)$ conditioned on $\bar {\vec q}(\vec x) \in \bar R_i$ satisfies good (anti-)concentration. 
Indeed, the mass of the set under $\normal(\vec 0, \vec I)$ is at least $\eps^{3 + 5d} \Pr_{\vec x \sim \normal(\vec 0, \vec I)}\lp[ \vec q(\vec x) \in R \rp] \geq \poly_{d, m}(1/\eps)$ by \Cref{eq:picked}, and the set $\{q_1, \cdots, q_m\}$ is at least
$\lp(\eps^{ 1/(3^{k+1} d^2 K)} , f_{d, K, M}^{(k)}(m) \rp)$-super non-singular.
As long as $k \geq 0$, we then have $\eps^{ 1/(3^{k+1} d^2 K)} \geq \eps^{ 1/(3 d^2 K)}$, and 
$ f_{d, K, M}^{(k)}(m) \geq C_{d, m, K} $.
Hence,  (anti-)concentration properties of the conditional distribution follows from \Cref{thm:(anti-)concentration}.
This establishes Condition (1) in \Cref{prop:set-cutter}.

We next verify that the union of the produced sets, i.e., $\bigcup_{ \bar R_i \in \mathcal K } \{ \vec x: \vec q(\vec x) \in R^{(i)} \} $, contains almost all the points from $\{ \vec x: \abs{p(\vec x)} < \eps \, , \, \vec q(\vec x) \in R \}$.
Throughout our algorithm, a point $\vec x$ may be missed due to three different reasons.
First, it may be removed if it does not lie in the giant rectangle where we begin our search, i.e.
$\bar {\vec q}(\vec x) \not \in \bar R$. 
\Cref{eq:giant-discard} implies that the probability mass of these points under $\normal(\vec 0, \vec I)$ is at most
$ \eps^3 \Pr_{\vec x \sim \normal(\vec 0, \vec I)} \lp[  \vec q(\vec x) \in R \rp]$.
Secondly, the point may be thrown if it lies in some small rectangle $\bar R_i \in \mathcal T$.
By \Cref{eq:thrown}, there are two subcases.
In the first subcase, this rectangle is thrown due to its probability mass being too small, i.e.,
$$
\Pr_{ \vec x \sim \normal(\vec 0, \vec I) }\lp[ \bar {\vec q}(\vec x) \in \bar R_i \rp]  < 3 \eps^{ 3 + 5 md } \; \Pr_{\vec x \sim \normal(\vec 0, \vec I)} \lp[  \vec q(\vec x) \in R \rp].
$$
In total, there are at most $\eps^{-5 m d}$ light rectangles. 
So the total probability mass of $\vec x$ discarded in this case is bounded above by 
$
3 \eps^3 \Pr_{\vec x \sim \normal(\vec 0, \vec I)} \lp[  \vec q(\vec x) \in R \rp]
$ under $\normal(\vec 0, \vec I)$.
In the second subcase, we  have
$$ \Pr_{ \vec x \sim \normal(\vec 0, \vec I) } \lp[  \abs{p(\vec x)} < \eps   | \bar{\vec q}(\vec x) \in \bar R_i \rp] < 3 \eps^3.
$$
This is equivalent to 
$$
\Pr_{ \vec x \sim \normal(\vec 0, \vec I) } \lp[  \abs{p(\vec x)} < \eps \, , \, \bar{\vec q}(\vec x) \in \bar R_i
 \rp] < 3 \eps^3 \Pr_{ \vec x \sim \normal(\vec 0, \vec I) }\lp[ 
 \bar{\vec q}(\vec x) \in \bar R_i
 \rp].
$$
By construction, the union of all rectangles $\bar R_i$ gives rise to $\bar R$. Moreover, since the first $\ell$ dimensions of $\bar R$ and $\bar {\vec q}$ are identical to those of $R$ and $\vec q$ respectively, it is not hard to see that the set $\{ \vec x: \bar{\vec q}(\vec x) \in \bar R \}$ is a subset of $\{ \vec x: \vec q(\vec x) \in R \}$.
Consequently, the total mass of the points removed due to this reason is also no more than 
$
3 \eps^3 \Pr_{\vec x \sim \normal(\vec 0, \vec I)} \lp[  \vec q(\vec x) \in R \rp]
$ under $\normal(\vec 0, \vec I)$.
In summary, the total mass discarded from the original set $\{ \vec x: \vec q(\vec x) \in R \}$ is no more than $$
O(\eps^3) \; \Pr_{\vec x \sim \normal(\vec 0, \vec I)} \lp[  \vec q(\vec x) \in R \rp] < \eps^2 \; \Pr_{\vec x \sim \normal(\vec 0, \vec I)} \lp[  \vec q(\vec x) \in R \rp]
$$ under $\normal(\vec 0, \vec I)$.
This establishes Condition (3) in \Cref{prop:set-cutter}.

Finally, we verify that the extra points 
outside the set $\{ \vec x: \abs{p(\vec x)} < 2 \eps, \vec q(\vec x) \in R \}$ included in
$\bigcup_{ \bar R_i\in \mathcal K } \{ \vec x: \bar {\vec q}(\vec x) \in \bar R_i \} $ have small probability mass.
It is clear that all the sets produced are subsets of $\{ \vec x: \vec q(\vec x) \in R \}$ by the construction of the sub-rectangles $\bar R_i$'s. 
Hence, it remains to show that the value of $p$ evaluated at these subsets is not too big.  
Recall that the polynomial $p$ can be written as
$$
p(\vec x) = h( \vec q(\vec x) ) + e(\vec x)
$$
by our decomposition.
Fix some $\bar R_i \in \mathcal K$.
By our selection rule of $\mathcal K$ (\Cref{eq:picked}), we know that there is at least some $\vec x^*$ in the set $\{ \vec x: \bar {\vec q}(\vec x) \in \bar R_i \}$ such that
$$ 
\abs{h( \vec q'(\vec x^*) ) + e(\vec x^*) } = \abs{p(\vec x^*)} < \eps.
$$
Consider another point $\vec y \in \{ \vec x: \bar{\vec q}(\vec x) \in \bar R_i \}$.
Since $R^{(i)}$ has side-length $\eps^{4d}$ by construction and its coordinates are at most 
$ \poly_{d, \ell, k, K}(\log(1/\eps))$ far from the origin, we know that
$\snorm{2}{  -\bar {\vec q}(\vec y) } \leq \sqrt{m} \eps^{4d}$ and
$\max \lp( \snorm{2}{\bar {\vec q}(\vec x^*)},  
\bar {\vec q}(\vec y)
\rp) \leq \sqrt{m} \poly_{d, \ell, k, K}(\log(1/\eps))$.
Moreover, by the guarantees of \Cref{thm:(anti-)concentration}, we have that
$\snorm{L_2}{h} < \eps^{-3d-1}$.
Notice that $h$ is a degree-$d$, dimension-$m$ polynomial. 
Hence, \Cref{lem:bounded-lipchitz} implies that
$$
\abs{ h( \bar {\vec q}(\vec x^*) ) - h( \bar {\vec q}(\vec y) ) }
\leq O_{d,m}(1) \; \poly_{d, \ell, k, K}\lp( \log(1/\eps) \rp)
\; \eps^{-3d-1} \; \eps^{4d} < \eps^2 \, ,
$$
as long as $\eps$ is sufficiently small given $d, \ell, k, K$.
It suffices to argue that $\abs{e(\vec y) - e(\vec x^*) }$ is not too large.
If it were the case that $\abs{e(\vec y) - e(\vec x^*) } < \eps^2$, we would then have
$$
\abs{p(\vec y)}
\leq \abs{p(\vec x)} + 
\abs{ h( \bar {\vec q}(\vec x^*) ) - h( \bar {\vec q}(\vec y) ) }
+ \abs{ e(\vec y) - e(\vec x^*) } \leq \eps + O(\eps^2) < 2\eps.
$$
Then the produced sets would be \emph{exactly} contained in the set $
\tilde {\mathcal P} = \{\vec x: \vec q(\vec x) \in R , \abs{p(\vec x)} < 2\eps  \}$.
Unfortunately, this is not true in general. 
However, since $\snorm{L_2}{e} < \eps^3$ by the guarantees of \Cref{prop:super-non-singular-extension}, we can show that 
the mass of the points which do not satisfy $\abs{e(\vec y) - e(\vec x^*) } < \eps^2$ is small.
In particular, by \Cref{lem:polynomial-concentration}, we know that
$$
\Pr_{ \vec x \sim \normal(\vec 0, \vec I) }
\lp[ \abs{e(\vec x)} > \alpha_{d, \ell}^{d/2} \log^{d/2}(1/\eps) \eps^3  \rp] \leq \eps^{ \Omega(\alpha_{d, \ell}) }.
$$
Recall that $\Pr_{\vec x \sim \normal(\vec 0, \vec I)}\lp[ \vec q(\vec x) \in R \rp] > \poly_{d, \ell}(\eps)$ by our assumption.
Hence, when $\alpha_{d, \ell}$ is sufficiently large given $d, \ell$, we will have 
$$
\Pr_{ \vec x \sim \normal(\vec 0, \vec I) }
\lp[ \abs{e(\vec x)} < \alpha_{d, \ell}^{d/2} \log^{d/2}(1/\eps) \eps^3
< \eps^2
\rp] > 1 - \eps^3 \Pr_{ \vec x \sim \normal(\vec 0, \vec I) } \lp[ \vec q(\vec x) \in R \rp].
$$
Since the only additional points that are included in the sets produced but excluded from $\tilde {\mathcal P}$ are those satisfying $\abs{e(\vec x)} > \eps^2$, it is not hard to conclude that Condition (4) in the proposition is satisfied.
We have verified all the properties listed in the proposition.
This concludes the proof of \Cref{prop:set-cutter}.
\end{proof}

\section{Learning Polynomial Threshold Functions using Super Non-Singular Polynomials}\label{sec:combine}

In this section, we present an efficient algorithm that learns an unknown polynomial threshold function under the Gaussian distribution in the presence of an $\opt$ fraction of adversarial corruptions, and achieves error $O_{c,d}(\opt^{1-c})$ for any desirable constant $c > 0$.
The formal statement of our theorem is as follows:
\begin{theorem}[Main Algorithmic Result, Formal Statement]\label{thm:final-thm}
Let $\eps,\opt,\eta, c\in (0, 1) ,d\in \Z_+$. Let $\D$ be the distribution of $(\x, y) \in \R^n \times \{\pm 1\}$, where $\x$ follows the standard Gaussian distribution, and $y$ is the output of some unknown degree-$d$ PTF evaluated at $\x$.
Given an $\opt$-corrupted version of a set of $n^{O(d)} \poly_{d,c}(1/\eps) \log(1/\eta)$ \iid samples from $D$,
there exists an algorithm, with computation complexity at most $n^{O(d)} \poly_{d,c}(1/\eps) \log(1/\eta)$, that computes a decision list $h:\R^n \mapsto \{\pm 1\}$ consisting of at most $\poly_{d,c}(1/\eps)$ degree-$d$ PTFs such that, with probability at least $1-\eta$, it holds
\[
\pr_{(\x, y) \sim D}[h(\x) \neq y ]
\leq A_{c,d} ~ \opt^{1-c} + \eps\,,
\]
where $A_{c,d}>0$ is an absolute constant depending only on $c,d$.
\end{theorem}
We now provide an overview of the algorithm.
The high-level framework is to implement an iterative localization strategy through the application of the Polynomial-Set-Partitioner from \Cref{prop:set-cutter} developed in \Cref{sec:polynomial-set-cutter}.
Roughly speaking, by applying the Polynomial-Set-Partitioner, we can approximately partition an unclassified area defined by polynomial transformations into subsets such that the Gaussian distribution conditioned on each subset within the partition satisfies good (anti-)concentration properties.
Once such a partition is identified, we will next run a robust margin perceptron algorithm (see \Cref{alg:perceptron}) on each of the conditional distributions.
Under good (anti)-concentration properties, 
the margin perceptron algorithm can output a PTF such that the prediction is accurate for all but $\eps^{\Omega(1/d)}$ fraction of low-margin points under $\eps$-fraction of adversarial corruptions (see \Cref{prop:perceptron}).
Consequently, combining the PTFs learned from different conditional distributions then gives a hypothesis capable of accurately classifying all but an $\eps^{\Omega(1/d)}$ fraction of data points within the region.
The process continues 
iteratively: the algorithm is given a new unclassified region and applies the perceptron update until the 
remaining unclassified mass is reduced to a negligible fraction, specifically $O_{c,d}(\opt^{1-c})$. 

Before proceeding with the proof of \Cref{thm:final-thm}, we first present our robust margin-perceptron algorithm. This routine is the main ingredient of our algorithm. Given a distribution that satisfies some good concentration and anti-concentration properties, \Cref{alg:perceptron} 
robustly estimates a weak learner that classifies almost all the points.


\begin{Ualgorithm}
	\centering
	\fbox{\parbox{6in}{
			{\bf Input:} Sample access to an at most $\eps$-corrupted oracle of the distribution $D$; access to $\x$-marginals $D_\x$, see \Cref{def:nasty-learning}, $K>0$.\\
   {\bf Parameters:} $F=\eps^{1-8/\sqrt{K}},\gamma =\eps^{4/\sqrt{K}}$ and $C>0$ a sufficiently large absolute constant. \\
{\bf Output:} A vector $\vec w$ and a region $B$ so that: $\pr_{(\x, y) \sim D}[\sgn( {\vec w} \cdot \x) \neq y \mid B ]
\leq C_{K}(\eps)^{1-O(1/\sqrt{K})}$ and $B=\{(\x,y)\mid y\neq \sign(\vec w \cdot \x),|\vec w\cdot \x|\geq \gamma\|\vec w\|_{D_\x,L_2}\}$.
\begin{enumerate}
\item $t\gets 0$.
\item Let $\vec q^{(t)}$ be an estimator of $\Ey[y\x]$ using the algorithm of \Cref{fct:estimator-robust}.
\item Let $B_t=\{(\x,y)\mid |\vec q^{(t)}\cdot \x|\geq \gamma\|\vec q^{(t)}\|_{D_\x,L_2}\}$ and $B_t'=\{(\x,y)\mid y\neq \sign(\vec q^{(t)}\cdot \x),|\vec q^{(t)}\cdot \x|\geq \gamma\|\vec q^{(t)}\|_{D_\x,L_2}\}$
\item \label{alg:while-cond}While $\pr[B_t']\geq 2F$ and $t<O(\eps^{-C})$, do
\begin{enumerate}
    \item Let $\vec p^{(t)}$ be an estimator of  $\Ey[y\1\{y\neq \sign(\vec q^{(t)}\cdot \x)\}\x\mid (\x,y)\in B_t]$ using the algorithm of \Cref{fct:estimator-robust}.
    \item $\vec q^{(t+1)}=\vec q^{(t)}+\lambda_t \vec p^{(t)}$, where $\lambda_t=(1/4)\|\vec q^{(t)}\|_{D_\x,L_2}\pr[B_t']\gamma/\|\vec p^{(t)}\|_{D_\x,L_2}^2$.
    \item $t\gets t+1$.
\end{enumerate}
\item Return $\vec q^{(t)}$.
\end{enumerate}}}
\vspace{0.2cm}
\caption{Perceptron Update} \label{alg:perceptron}
\end{Ualgorithm}

\begin{remark}
{\em   In what follows, we will overload our norm notation and for a vector $\vec w \in \R^n$ we denote by $\|\vec w\|_{D_\x, L_p}$ its distributional $L_p$ norm when it
  is treated as a linear function of $\x$, i.e.,  
  $\|\vec w\|_{D_\x, L_p} = \Big( \E_{\x \sim D_\x}[ |\vec w \cdot \x|^p \Big)^{1/p}$. We continue to denote the common norm of $\vec w$ 
  by $\|\vec w\|_p = \Big(\sum_{i=1}^n  |\vec w_i|^{p}\Big)^{1/p}$ 
  when we treat it as a vector of $\R^n$.}
\end{remark}

\begin{proposition}\label{prop:perceptron}
Let $\eps \in (0,1)$, $K \in \mathbbm Z_+$, 
$C, C_1 > 0$, and $D$ be a distribution on $\R^n \times \{\pm 1\}$. 
Assume the following statements are true.
\begin{enumerate}
    \item The samples $\x$ are labeled with respect to an unknown function $f(\x)=\sign(\wstar\cdot \x)$ for some $\wstar\in \R^n$.
    \item The covariance matrix of the marginal distribution $D_{\x}$
    satisfies $\Exx[\x^\top\x]\preccurlyeq 2\vec I$.
    \item The $\x$-marginal distribution $D_{\x}$ satisfies anti-concentration, i.e.,
    $$
    \pr_{\x \sim \D_\x}[|\vec v \cdot \x| \leq t \|\vec v\|_{D_\x,L_2}] \leq t^{C} \, ,
    $$
    for some number $C > 0$, and for all $t \in (\eps,\eps^{2/{\sqrt{K}}})$.
    \item $D_{\x}$ has concentration, i.e.,
    $$
    \pr_{\x \sim \D_\x}[|\vec v \cdot \x| > t \|\vec v\|_{D_\x,L_2}] \leq C_1  t^{-K} \, ,
    $$
    for some $C_1 > 0$ and all $t > 0$ 
\end{enumerate}
Then \Cref{alg:perceptron} given access to an $\eps$-corrupted version of $\poly(n/\eps)$ \iid 
samples from $D$, runs in sample-polynomial time, and with probability at least $2/3$ computes a weight vector $\wh {\vec w}$ such that
\[
\pr_{(\x, y) \sim D}[\sgn(\wh {\vec w} \cdot \x) \neq y \mid B ]
\leq C_1^{1/K} ~ O_{C,K,d}( \eps^{1-O(1/\sqrt{K})} )\,,
\]
where $B = \{\x : |\vec w \cdot \x | \geq \gamma \|\vec w\|_{D_\x,L_2} \}$ for $\gamma = \eps^{4/\sqrt{K}}$.
\end{proposition}

\begin{proof}
We set the parameters $\gamma=\eps^{4/\sqrt{K}}$ and $F=\eps^{1-8/\sqrt{K}}$. We denote by $D^c$ the distribution $D$ that is at most $\eps$-corrupted.
First, we show that we can robustly estimate $\Ey[y\x]$. We will use of following fact.
\begin{fact}[Proposition 2.2 of \cite{DKS18a}]\label{fct:estimator-robust}
    Fix $\eps,\xi\in(0,1)$ and let $D$ be a distribution on $\R^n\times[-1,1]$ 
    corresponding to pairs labeled according to some function $f:\R^n\mapsto[-1,1]$, i.e., $(\x,f(\x))\sim D$.  
    Assume that $D$  satisfies the following: (i) concentration, i.e.,  $\pr_{\x \sim \D_\x}[|\vec v \cdot \x| > t \|\vec v\|_{D_\x,L_2}] \leq C t^{-K} $, for some $C>0$ and for all $t>0$, (ii) bounded variance in each direction, i.e., $\Exx[\x\x^\top]\preccurlyeq\Sigma$.
    Then, there is an algorithm that given an $\eps$-corrupted set $S$ from $D$ with $|S|=O(n\poly(1/\eps,1/\xi))$ and a description of the $\x$-marginals $D_\x$, runs in $\poly(|S|)$ time and outputs $\vec q\in \R^n$, such that $\|\Exx[f(\x)\x]-\vec q\|_2\leq c \|\Sigma\|_2C^{1/K}\eps^{1-1/K}$ with probability at least $1-\xi$, where $c>0$ is an absolute constant.
\end{fact}
For the rest of the proof, let $c>0$ be the absolute constant from \Cref{fct:estimator-robust}.
Note that for our distribution and for any band $B= \{\x : |\vec w \cdot \x | \geq \gamma \|\vec w\|_{D_\x,L_2} \}$, we have that the distribution $D^c$ conditioned on $B$ is  at most $O(\eps/\pr[B])=O(\eps)$ corrupted, because $\pr[\x\in B]\geq 1-\gamma^C=1-\eps^{Cc}=\Omega(1)$.
Let $\vec q^{(0)}$ be a robust estimator (given by \Cref{fct:estimator-robust}) of $\Ey[y\x]$ with error at most $c C_1^{1/K}\eps^{1-1/K}$, i.e., $\|\Ey[y\x]-\vec q^{(0)}\|_2\leq c C_1^{1/K}\eps^{1-1/K}$ . We first show that $\vec q^{(0)}$ is well-correlated with $\wstar$.
\begin{claim}\label{clm:initial-correlation}
    It holds that $\vec q^{(0)}\cdot\wstar\geq (1/2)\eps^{2/\sqrt{K}} $.
\end{claim}
\begin{proof}
    Using that $\vec q^{(0)}$ is equal to $\Ey[f(\x)\x]$ with error at most $c C_1^{1/K}\eps^{1-1/K}$, we get that
    \[
    \vec q^{(0)}\cdot\wstar\geq \Exx[|\wstar\cdot\x|]-c C_1^{1/K}\eps^{1-1/K}\;.
    \]
    Using the anti-concentration, i.e., property (ii) of \Cref{prop:perceptron}, we get that
    $ \Exx[|\wstar\cdot\x|]\geq \rho(1-\rho^C)$. We can assume without loss of generality that $\eps$ is less than a sufficiently small constant depending on $C,K$ and $C_1^{1/K}$. Otherwise, the constant hypothesis is $O(\eps)$. Hence, it holds that $1-\eps^{C/\sqrt{K}}\geq 1/2$. It follows that for $\rho=\eps^{2/\sqrt{K}}$, we have that $\rho(1-\rho^C)\geq \eps^{2/\sqrt{K}}/2$, and therefore
    $ \vec q^{(0)}\cdot\wstar\geq \eps^{2/\sqrt{K}}/2$.
\end{proof} 
Denote as $B_t=\{|\vec q^{(t)}\cdot \x|\geq \gamma\|\vec q^{(t)}\|_{D_\x,L_2}\}$.
While the mass of the region $B_t'=\{y\neq \sign(\vec q^{(t)}\cdot \x),|\vec q^{(t)}\cdot \x|\geq \gamma\|\vec q^{(t)}\|_{D_\x,L_2}\}$ is larger than our
threshold or the maximum number of 
iterations have not exceeded our threshold (Step \eqref{alg:while-cond}), our algorithm updates the current hypothesis using the following update rule:
\[\vec q^{(t+1)}\gets \vec q^{(t)}+\lambda_t \vec p^{(t)}\;,\]
where $\vec p^{(t)}$ is a robust estimator of $\E_{(\x,y)\sim D}[f'(\x)\x\mid B_t] $, where $f'(\x)=f(\x)\1\{y\neq \sign(\vec q^{(t)}\cdot \x)\}$, obtained by the algorithm of \Cref{fct:estimator-robust}.
Let $F_t=\E_{(\x,y)\sim D}[(\x,y)\in B_t']\geq F$ (see Step \eqref{alg:while-cond}) and denote by $D_t$ the distribution $D$ conditioned on the event $B_t$. Note that the true mass of the region $B_t$ with respect to the distribution $D$ is at least $1/2$.
We claim that the distribution $D_t$ satisfies the assumptions of \Cref{fct:estimator-robust}. First, note that the corruption level of $D_t$ is at most $2\eps$, furthermore
for the distribution $D_t$, it holds that $\pr_{\x \sim (\D_t)_\x}[|\vec v \cdot \x| > t \|\vec v\|_{D_\x,L_2}] \leq 2\pr_{\x \sim \D_\x}[|\vec v \cdot \x| > t \|\vec v\|_{D_\x,L_2}]\leq  2C_1 t^{-K} $. Moreover, for any unit vector $\vec v\in \R^n$, it holds that $\E_{\x \sim (\D_t)_\x}[(\vec v\cdot\x)^2]\leq 4$. Therefore, $D_t$ satisfies the assumptions of \Cref{fct:estimator-robust} but with parameters $C=2C_1$, $\|\Sigma\|_2\leq 4$ and at most $2\eps$ corruption rate.

Using \Cref{fct:estimator-robust} on the distribution $D_t$, we have that for our robust estimator $ \vec p^{(t)}$, it holds that $\|\vec p^{(t)}-\E_{(\x,y)\sim D}[f'(\x)\x\mid (\x,y)\in B_t]\|_2\leq  8c C_1^{1/K}\eps^{1-1/K}$.
We show that by choosing appropriately the stepsize $\lambda_t$, we decrease the norm of $\|\vec q^{(t)}\|_{D_\x,L_2}$ in each iteration. We show the following:
\begin{claim}\label{clm:decrease}
    Assume that $\lambda_t= (1/2)\gamma F_t/ \|\vec p^{(t)}\|_{D_\x,L_2}^2 $. Then it holds that $\|\vec q^{(t+1)}\|_{D_\x,L_2}\leq \|\vec q^{(t)}\|_{D_\x,L_2}(1-(1/16)F_t^{2/K} \gamma^2/C_1^{2/K})^{1/2}$.
\end{claim}
\begin{proof}
Note that by definition, we have that
\begin{align}
     \vec q^{(t)}\cdot \vec p^{(t)}&\leq  \Ey[f'(\x)   \vec q^{(t)}\cdot \x\mid B_t]+8c C_1^{1/K}\eps^{1-1/K}\|\vec q^{(t)}\|_{D_\x,L_2}\nonumber
     \\&=-\Ey[   |\vec q^{(t)}\cdot \x|\1\{y\neq\sign(\vec q^{(t)}\cdot \x)\}\mid(\x,y)\in B_t]+8c C_1^{1/K}\eps^{1-1/K}\|\vec q^{(t)}\|_{D_\x,L_2}\nonumber
     \\&\leq -(\gamma F_t- 8c C_1^{1/K}\eps^{1-1/K})\| \vec q^{(t)}\|_{D_\x,L_2}\leq -(\gamma F_t/2)\| \vec q^{(t)}\|_{D_\x,L_2} \;,\label{eq:inner-corr}
\end{align}
where we used that $\gamma F_t\geq\eps^{1-2/\sqrt{K}}\geq 8c C_1^{1/K}\eps^{1-1/K}$ for all $F_t\geq F$.
Next, we calculate the norm of $\vec p^{(t)}$. We have that
\begin{align*}
    \|\vec p^{(t)}\|_{D_\x,L_2}&\leq \sup_{\vec v\in \R^n,\|\vec v\|_{D_\x,L_2}=1}\E_{(\x,y)\sim D_t}[|   \vec v\cdot \x|\1\{y\neq \sign(\vec q^{(t)}\cdot \x)\} ]\leq 2C_1^{1/K}F_t^{1-1/(K-1)}\;,
\end{align*}
where in the last inequality we used the concentration of the tails.
Therefore, we have that
\begin{align*}
    \|\vec q^{(t+1)}\|_{D_\x,L_2}^2&=\|\vec q^{(t)}\|_{D_\x,L_2}^2+2\lambda_t\vec q^{(t)}\cdot \vec p^{(t)} + \lambda_t^2\|\vec p^{(t)}\|_{D_\x,L_2}^2 
    \\&\leq \|\vec q^{(t)}\|_{D_\x,L_2}^2-\lambda_t\gamma F_t\| \vec q^{(t)}\|_{D_\x,L_2}+ 4C_1^{2/K}F_t^{-2/K}\lambda_t^2
    \\&\leq \|\vec q^{(t)}\|_{D_\x,L_2}^2-\lambda_t( F_t\gamma/2)\| \vec q^{(t)}\|_{D_\x,L_2}
    \\&=\|\vec q^{(t)}\|_{D_\x,L_2}^2(1-(1/16)F_t^{2/K} \gamma^2/C_1^{2/K})\;,
\end{align*}
where we used that $\lambda_t= (1/2)\gamma F_t/ \|\vec p^{(t)}\|_{D_\x,L_2}^2 $.
\end{proof}
Finally, we show that $\vec p^{(t)}$ does not decrease the correlation with $\wstar$ by a lot. We show the following

\begin{claim}
    It holds that $\wstar\cdot \vec p^{(t)}\geq - 8c C_1^{1/K}\eps^{1-1/K}$.
\end{claim}
\begin{proof}
Note that $\vec p^{(t)}$ is a robust estimator of $\E[f'(\x) \x\mid |\vec q^{(t)}\cdot \x|\geq \gamma\|\vec q^{(t)}\|_{D_\x,L_2}]$ with error $8c C_1^{1/K}\eps^{1-1/K}$ and it holds that $\E[f'(\x) \wstar\cdot\x\mid |\vec q^{(t)}\cdot \x|\geq \gamma\|\vec q^{(t)}\|_{D_\x,L_2}]\geq 0$, therefore we have that 
\begin{align*}
         \wstar\cdot \vec p^{(t)}\geq -8c C_1^{1/K}\eps^{1-1/K}\;.
\end{align*}
\end{proof}
From \Cref{clm:decrease}, we have that after $t=\Theta(\log(1/\eps))C_1^{2/K}/(\gamma^2F^{2/K})$ iterations, we have that
\begin{align*}
    \vec q^{(t)}\cdot \wstar&\leq \|\vec q^{(t)}\|_{D_\x,L_2}\leq    \|\vec q^{(0)}\|_{D_\x,L_2}(1-\eps^{C_3/(2K)+2/K})^{t/2}\leq  \eps^{1/2}\;.
\end{align*}
On the other hand, we have that 
\begin{align*}
    \vec q^{(t)}\cdot \wstar&\geq \vec q^{(t-1)}\cdot \wstar+\lambda_t\vec p^{(t-1)}\cdot \wstar
    \\&\geq \vec q^{(0)}\cdot \wstar - 8c C_1^{3/K}\eps^{1-1/K}t \gamma /F^{1-2/(K-1)}
    \\&\geq  (1/2)\eps^{2/\sqrt{K}}-C'C_1^{3/K}\eps^{(1-1/K)(8/\sqrt{K})}\log(1/\eps)
    \\&\geq (1/2)\eps^{2/\sqrt{K}}-C' C_1^{3/K}\eps^{3/\sqrt{K}}\geq (1/4)\eps^{2/\sqrt{K}}\;,
\end{align*}
where $C'>0$ is an absolute constant and we used that $\gamma=\eps^{4/\sqrt{K}}$ and $F=\eps^{1-8/\sqrt{K}}$, therefore we get that
$(1/4)\eps^{2/\sqrt{K}}\leq \eps^{1/2}$, which is a contradiction. Therefore, after at most $O(\log(1/\eps)/(\eps^{c_3}))$ iterations for some absolute constant $c_3>0$, we have that $\E[\1\{y\neq \sign(\vec q^{(t)}\cdot \x),|\vec q^{(t)}\cdot \x|\geq \gamma\|\vec q^{(t)}\|_{D_\x,L_2}\}]\leq F=\eps^{1-8/\sqrt{K}}$. To conclude the proof, note that by the anti-concentration assumption, we have that 
$\E[\1\{|\vec q^{(t)}\cdot \x|\geq \gamma\|\vec q^{(t)}\|_{D_\x,L_2}\}]\geq 1-\gamma^C=1-O(\eps^{4C/\sqrt{K}})\geq 1/2$ and the result follows. This completes the proof of \Cref{prop:perceptron}.
\end{proof}

\begin{Ualgorithm}
	\centering
	\fbox{\parbox{6in}{
			{\bf Input:} Access to at most $\opt^{1/\sqrt{K}}$-\new{corrupted} samples from a distribution $D$, $\opt,\eps,K>0$.\\
{\bf Output:} A $d$-degree PTF $h$ and interval $I$ so that: $\pr_{(\x, y) \sim D}[h(\x) \neq y\mid I]
\leq O_{d,K}(\opt_I^{1-1/\sqrt{K}}) + \epsilon\,,$ and a list of unclassified regions $Q$.\\
{\bf Parameters:} $c>0$ sufficiently small absolute constant.
\begin{enumerate}
\item Let $Z=\{\eps,\ldots,\opt^{O(1/\sqrt{K})}\}$. 
\item For each $\eps'\in Z$, do
\begin{enumerate}
    \item Run \Cref{alg:perceptron} for $\log(1/\eta)$ times with parameters $\eps',K$ and get $T=\{(p_1,I_1)\ldots(p_{\log(1/\eta)},I_{\log(1/\eta)})\}$.
    \item For each $(p,I)\in T$, if  $\Pr_{(\x,y)\sim D}[p(\x)\neq y\mid \x\in I]\leq O(\eps'^{1-8/\sqrt{K}})$, then keep $(p,I)$ and stop the search.
\end{enumerate}
\item Use \Cref{prop:set-cutter} and output a list $Q$ of unclassified regions with respect to  $(p,I)$.
\item Return $(p,I),Q$.
\end{enumerate}}}
\vspace{0.2cm}
\caption{Partial-Classifier} \label{alg:loop}
\end{Ualgorithm}

As outlined in our discussion, our algorithm partitions the unclassified regions into more manageable sub-regions. 
In particular, the partition ensures that the resulting distributions conditioned on each of the sub-regions  
satisfy good (anti-)concentration properties. 
To facilitate this, we introduce the following definition, which characterizes the attributes of each subregion. 
Specifically, given a polynomial transformation $\vec q$ and a rectangle $R$ we want the normal distribution conditional on $\vec q(\x) \in R$ to have both good anti-concentration and concentration behavior, while $\vec q $ is also level-k
super non-singular.
\begin{definition}[(Anti-)concetrated Polynomial Slab]\label{def:region-properties}
Let $\vec q:\R^n\mapsto \R^\ell$ and a rectangle $R\in\R^\ell$. We say that a tuple $(\vec q, R)$ is an $(\eps,k,K,d,\ell,\zeta)$-(anti-)concentrated if the following properties are satisfied:
\begin{enumerate}
    \item Let $\mathcal P=\{\x\in \R^n: q(\x)\in R\}$, then for any polynomial $g:\R^n \mapsto \R$ of degree-$d$, for $\eps < t < \eps^{2/K}$,
it holds
\begin{align*}
\Pr_{ \vec x \sim \normal(\vec 0, \vec I) \mid   {\mathcal P} }
\lp[ 
\abs{g(\vec x)} < t \; \snorm{ \normal(\vec 0, \vec I) \mid   {\mathcal P} , L_2}{g}
\rp] \leq t^{1/(2d)} \, ,
\end{align*}
and  for all $t > 0$, it holds
\begin{align*}
\Pr_{ \vec x \sim \normal(\vec 0, \vec I) \mid   {\mathcal P} } \lp[ \abs{g(\vec x)} > t \snorm{\normal(\vec 0, \vec I) \mid   {\mathcal P}, L_2}{g}
\rp] \leq O_{d,K}\lp( \eps^{-1} t^{- K  } \rp)\;.
\end{align*}
\item The mass of $\mathcal P$ is lower bound by $\zeta$, i.e., $\Pr_{ \vec x \sim \normal(\vec 0, \vec I)  }[q(\x)\in R]\geq \zeta$.
\item $\vec q$ is level-$k$ super non-singular 
with respect to the tuple $(\eps, d, K, M=3)$.
\end{enumerate}
\end{definition}

Our subsequent proposition demonstrates that \Cref{alg:loop} classifies a region that fulfills the (anti-)concentration property of \Cref{def:region-properties}. \Cref{alg:loop} employs \Cref{alg:perceptron} to categorize the majority of points effectively. Subsequently, it decomposes any remaining unclassified region into a list of smaller subregions, each inheriting the  (anti-)concentration property
of \Cref{def:region-properties}.

\begin{proposition}\label{prop:combine-perceptron-superpoly}
Let $n, k, d,\ell,K \in \mathbb Z_+$, and $\eps ,\opt,\eta\in (0, 1)$. Let $D$ be a distribution on $\R^n \times \{\pm 1\}$ such that (i) $D$ is at most $\opt$-corrupted with respect an unknown $d$-degree PTF, (ii) the $\x$-marginal of $\D$ follows standard $n$-dimensional normal.
Let $(\vec q,R)$ be $(\opt,k,K,d,\ell,\zeta)$-(anti-)concentrated. Let $D_Q$ be the distribution conditioned on $\vec q(\vec x) \in R $.
There exists an algorithm (\Cref{alg:loop}) that draws $N =O(n^d)\poly_{K,d}(d,1/\eps)\log(1/\eta)$ at most $\opt^{1/\sqrt{K}}$-corrupted samples from $D_Q$, runs in sample-polynomial time
and with probability at least $1-\eta$ computes:
\begin{enumerate}
    \item A set of $m$-dimensional vector-valued polynomial $\vec {q'}: \R^n \mapsto \R^m$ where $m=O_{d,\ell,K}(1)$ and a set of $m$-dimensional axis-aligned rectangles $\{ R^{(1)},\ldots R^{(t)} \}$ with $t\leq \opt^{-2m}$ satisfying
\begin{enumerate}
    \item For each $R^{(i)}$, the tuple $(\vec q',R^{(i)})$ is  $(\opt,k-1,K,d,m,\zeta)$-(anti-)concentrated.
    \item 
    For the union of the sets $ S := \bigcup_{i} \{ \vec x: \vec q'(\x) \in R^{(i)} \}$, it holds that 
    \[
    \Pr_{ \vec x \sim \normal(\vec 0, \vec I)  }
    \lp[ \vec x  \in S \mid  \vec q(\vec x) \in R\rp] < \opt^{c/(2d)}\,.
    \]
\end{enumerate}
    \item An at most $d$-degree polynomial $p$ and a region $I=\{q(\x)\in R\cap S^c\}$ such that
\[
\pr_{(\x, y) \sim D_Q}[\sgn(p(\x)) \neq y \mid \x \in I ]
\leq C_{d,K}\opt_I^{1-O(1/\sqrt{K})}+\opt+\eps\,,
\]
for some absolute constant $C_{d,K}>0$.
\end{enumerate}
\end{proposition}
\begin{proof}
From \Cref{def:region-properties}, for any $\rho\in(\opt,\opt^{2/K})$, we have that for any polynomial $p:\R^n \mapsto \R$, it holds
\begin{equation}
    \Pr_{ \vec x \sim D_Q  }
\lp[ 
p(\vec x) < \rho \;
\|p(\vec x)\|_{D_Q,L_2}\rp]
\leq \rho^{1/(2d)}.\label{eq:anticonc-2}
\end{equation}

In what follows for a tensor $\vec A \in \R^{k_1 \otimes \cdots \otimes k_m}$ we define $\vec A^{\flat}$ its flattening to a vector of $\R^{k_1\cdots k_m}$.
Let $\vec z=((\x,1)^{\otimes d})^{\flat}$ and denote as $D_{\otimes}$ the distribution of $\vec z$. Notice that the $\vec z$ contains a basis for all polynomials of degree at most $d$, so the dimension of $\vec z$ is $(n+1)^d$. Then for any vector $\vec v\in \R^{(n+1)^d}$, we have that $\vec v\cdot \vec z$ is an at most $d$-degree polynomial, therefore we have that for any $\vec v\in \R^{(n+1)^d}$ it holds that
$$
\Pr_{ \vec z \sim D_{\otimes}  }
\lp[ 
|\vec v\cdot\vec z| < \rho \;
\|\vec v\cdot\vec z\|_{D_\otimes,L_2}\rp]
\leq  \rho^{1/(2d)} .
$$
We first show that we can transform $\vec z$ so that the covariance of $\vec z$ is equal to identity.
Note that $\E[\vec z \vec z^{\top}]$ is Hermitian matrix, so it can be written as $\E[\vec z \vec z^{\top}]=\vec P \vec D \vec P^{\top}$,
for some diagonal matrix $\vec D$ and orthonormal matrix $\vec P$. Let $\vec T=\vec P \vec D^{-1/2}$ and let $\vec z'=\vec T \vec z$.
It holds that $\E[\vec z' (\vec z')^{\top}]=\vec I$. Furthermore, note that if $\vec v\cdot\vec z$ is a polynomial before the transformation, then $(\vec T^{-1}\vec v)\cdot \vec z'=\vec v\cdot \vec z$. Therefore all the polynomials can be expressed in the new space. To calculate the $\vec P,\vec D$, we can run standard Gram–Schmidt orthonormalization and get a $\gamma$-approximation (in $L_2$ sense) of these matrices with $\poly(d/\gamma)$ samples and time.

Denote by $\mathcal C$ the at most $\opt^{1/\sqrt{K}}$-corrupted oracle to the distribution $D_Q$. Before we apply \Cref{alg:perceptron} to find a new hypothesis, we need to guess the corruption rate of $\mathcal C$. For this purpose, we construct a list $Z=\{\eps,2\eps,\ldots, \opt^{1/\sqrt{K}}\}$ containing all the possible corruption levels. If $\opt_Q$ is the true corruption level of $\mathcal C$, then there exists $\opt'\in Z$, such that $\opt_Q\leq\opt'\leq \opt_Q+2\eps$.
Therefore, for the correct noise rate, from \Cref{prop:perceptron}  using \Cref{alg:perceptron} with probability at least $2/3$, we get a polynomial $p:\R^n\mapsto \R$ and a set $I=\{|p(\x)|\geq 2\opt_Q^{4/\sqrt{K}}\}$ so that
\[
\pr_{(\x, y) \sim D_Q}[\sgn(p(\x)) \neq y \mid \x \in I ]
\leq C_{K,d}(\opt_I+\eps)^{1-O(1/\sqrt{K})}\,.
\]
By running the same procedure $\log(1/\eta)$ and testing the result, we can amplify the probability of this procedure to $1-\eta$. Note that from \Cref{eq:anticonc-2}, we have that 
\begin{equation}
    \pr[\x\in I\mid \vec q(\x)\in R]\geq 1-(\opt_Q)^{2/(d\sqrt{K})}\geq 1-(\opt)^{2/(d K)}\;,\label{eq:proof32}
\end{equation} where we used that $\opt_Q\leq \opt^{1/\sqrt{K}}$.
To classify the rest of the points in the set $\{q(\x)\in R,\x\not\in I\}$, we need to split the region into smaller sets using \Cref{prop:set-cutter}.
From \Cref{prop:set-cutter}, given the polynomial $p$ and the current polynomial vector $q$ for some $m=O_{d,K,\ell}(1)$, we get the following: a new $m$-dimensional vector-valued polynomial $\vec {q'}: \R^n \mapsto \R^m$ and a set of $m$-dimensional axis-aligned rectangles $\{ R^{(i)} \}$ of size at most $\opt^{-2m}$ satisfy so that the sets for each $i$, it holds that $(\vec q',R^{(i)})$ is $(\opt,k-1,K,d,m,\eps^{3+5md}\zeta)$-(anti-)concentrated.
Moreover, let $ S := \bigcup_{i} \{ \vec x: \vec q'(\x) \in R^{(i)} \}$ , we have that $\pr[\x\in S,\x\in I\mid q(\x)\in R]\leq \opt^2$ and $\pr[\x\not\in S,\x\not\in I\mid \vec q(\x)\in R]\leq \opt^2$. Define $J=(I\cup S^c)-S$. Note that for the region $J$ it holds that $  \pr[\x\in J\mid \vec q(\x)\in R]\geq  \pr[\x\in I\mid \vec q(\x)\in R]-\opt^2$ and therefore, he have that
$\pr_{(\x, y) \sim D_Q}[\sgn(p(\x)) \neq y \mid \x \in J ]
\leq C_{K,d}(\opt_I+\eps)^{1-O(1/\sqrt{K})} +\opt\,.$ and that $ \pr[\x\in J\mid \vec q(\x)\in R]\geq  1-2(\opt)^{2/(d K)}$.


\end{proof}

\begin{Ualgorithm}
	\centering
	\fbox{\parbox{6in}{
			{\bf Input:}  Access to at most $\opt$-corrupted samples from a distribution $D$ over $\R^n\times\{\pm 1\}$, noise rate $\opt>0$, parameters $\eps,K>0$\\
{\bf Output:} A decision list of PTFs $h$ so that: $\pr_{(\x, y) \sim D}[h(\x) \neq y]
\leq O_{d,K}(\opt^{1-O(1/\sqrt{K})}) + \epsilon$.
\begin{enumerate}
\item Run \Cref{alg:loop} on $D$ and get $h=\{(p,I)\}$, $Q$.  \label{alg:end-step1}
\item \label{alg:while-cond-final}While $Q$ is not empty, do \label{alg:while-loop}
\begin{enumerate}
    \item $(\vec q,R)\gets \mathrm{pop}(Q)$. \label{alg:pop}
    \item Let $D_Q$ be the distribution $D$ conditioned on $\vec q(\x)\in R$.
    \item Run \Cref{alg:loop} on $D_Q$ and get $\{(p,I)\}$, $Q'$. \label{alg:partial-hypothesis}
    \item Let $Q=Q\cup Q'$ and $h=h\cup \{p,I\}$.
\end{enumerate}
     \item Return $h  $.
\end{enumerate}}}
\vspace{0.2cm}
\caption{Combined Algorithm} \label{alg:combined}
\end{Ualgorithm}

\subsection{Proof of \texorpdfstring{\Cref{thm:final-thm}}{}}
Let $\mathcal C$ be the $\opt$-corrupted oracle access to the distribution $D$. For any event $Q$, we denote $\mathcal C_Q$ the oracle access to the distribution $D_Q$ i.e., the distribution $D$ conditioned on the event $Q$. We can simulate $\mathcal C_Q$ from $\mathcal C$ by accepting only points that satisfy the event $Q$. We denote as $\opt_Q$ the corruption rate of the oracle $\mathcal C_Q$. Let $k$ be a sufficiently large multiply of $d$ and $C_K=O(1/\sqrt{K})$. We can assume without loss of generality that $\opt$ is less than an absolute constant $C'=O_{d,K}(1)$ Otherwise, the constant hypothesis will have error $O(C')$. Moreover, if $\eps>\opt$ then we can take $\opt=\eps$ as the error guarantee of the algorithm will be $O(\eps^{1-O(1/\sqrt{K})})$. Otherwise, we can assume that $\opt/2<\eps<2\opt$ as we can find a $2$-approximation of $\opt$ by re-running the algorithm $1/\eps$ times.
The distribution is standard $d$-dimensional normal, hence, it satisfies the assumptions of \Cref{prop:combine-perceptron-superpoly} for $\vec q(\x)=\vec 1$ and $R=\R^n$. Therefore, from Step \eqref{alg:end-step1}, we get a polynomial $p$ and a region $I$ so that 
$\pr_{(\x, y) \sim D}[\sign(p(\x))\neq y\mid \x\in I]
\leq C'(\opt+\eps)^{1-C_K}\,.$
First, note that our algorithm in Step \eqref{alg:while-loop}, maintains a list of unclassified regions $Q_1,\ldots, Q_\xi$. In order to show that our algorithm works, we need to show that in each iteration, we decrease the unclassified region and we output a classifier that classifies accurately most of the points. We first show that we can remove the regions with noise rates more than $\opt^{C_K}$ as their total mass contributes to the error at most $\opt^{1-C_K}$. Denote with $\Xi_1,\ldots,\Xi_l$ the regions in $Q$ so that 
$
\pr_{(\x, y) \sim D}[p^*(\x) \neq y\mid \Xi_i ]
\geq (\opt)^{C_K}.$
We show that $\pr_{(\x, y) \sim D}[\x\in \cup_{i=1}^l \Xi_i ]\leq \opt^{1-C_K}$.
\begin{claim} \label{clm:excess-noise}
    It holds that $\pr_{(\x, y) \sim D}[\x\in \cup_{i=1}^l \Xi_i ]\leq \opt^{1-C_K}$.
\end{claim}
\begin{proof}
    Let $A\cup_{i=1}^l\Xi_i=\R^n$. We have that
    \begin{align*}
        \opt&=\pr_{(\x, y) \sim D}[p^*(\x) \neq y ]\\
        &=\pr_{(\x, y) \sim D}[p^*(\x) \neq y \mid \x\in A]\pr_{\x\sim D_\x}[\x\in A]+\sum_{i=1}^l\pr_{(\x, y) \sim D}[p^*(\x) \neq y \mid \x\in \Xi_i]\pr_{\x\sim D_\x}[\x\in \Xi_i]
        \\&\geq \sum_{i=1}^l\pr_{(\x, y) \sim D}[p^*(\x) \neq y \mid \x\in \Xi_i]\pr_{\x\sim D_\x}[\x\in \Xi_i]
        \geq \opt^{C_K}\pr_{\x\sim D_\x}[\x\in \cup_{i=1}^l \Xi_i ]\;.
    \end{align*}
    By rearranging the above inequality, the result follows.
\end{proof}


Next, we show that in each iteration, we remove a large mass and we classify optimally the points of this mass. Let $(\vec q,R)$ be the element, we pop from $Q$ at Step \eqref{alg:pop}. For some $k\in \Z,\ell>0$, the $(\vec q,R)$ is $(\eps,k,K,d,\ell,\eps)$-(anti-)concentrated. For this region one of two events can happen: Either the corrupted oracle $\mathcal C_Q$ has a corruption rate more than $\opt^{C_K}$ in which case Step \eqref{alg:partial-hypothesis} gives a constant hypothesis in this region, or we are given 
a partial classifier $(p,I)$ and a new set of regions $\{\vec q',R^{(i)}\}$ that are $(\eps,k-1,K,d,m,\eps^{3+5md})$-(anti-)concentrated, for $m=O_{\ell,K}(1)$. For the partial hypothesis, we have from \Cref{prop:combine-perceptron-superpoly}, that 
\[
\pr_{(\x, y) \sim D_Q}[\sgn(p(\x)) \neq y \mid \x \in I ]
\leq C_{K,d}'\opt_I^{1-C_K}+\opt+\eps\,, 
\]
 Note that $\Pr[\x\in I\mid q(\x)\in R]\geq 1-\opt_{I}^{c/(2d)}$, therefore in each step we remove at least $\pr[q(\x)\in R](1-\opt_{I}^{c/(2d)})$ mass.

     Let $S_{i}$ be the unclassified regions on depth $i$. This means that set $S_i$ is a union of sets $R^{(i)}$ that are constructed after $i$ applications of Polynomial-Set-Partitioner. We show that after $t=O(d C_K)$ steps, we have classified all but $O(\opt)$ mass. Note that after $t$ iterations, we have that $\Pr_{ \vec x \sim \normal(\vec 0, \vec I)  }
    \lp[ \vec x \in S_{t}\rp]\leq \opt^{C_Kct/(2d)}$, hence for $t=O(d C_K)$ this is of order $\opt$. Therefore, we can drop the remaining sets.
    To analyze the correctness of our algorithm, we denote the following, let $\{(p_1,I_1),\ldots,(p_T,I_T)\}$ be the decision list classifier and let $\Xi$ be the union of regions discarded because the corruption rate was more than $\opt^{C_K} $. We denote as $h(\x)$ the classifier that if $\x\in I_i$ for some $i$ then $h(\x)=\sign(p_i(\x))$.
    We have that
    \begin{align*}
   \Pr_{ \vec x \sim \normal(\vec 0, \vec I)  }
    \lp[ h(\x)\neq y\rp]&\leq  \sum_{i=1}^T   \Pr_{ \vec x \sim \normal(\vec 0, \vec I)  }
    \lp[\sign(p_i(\x))\neq y\mid \x \in I_i\rp]\Pr_{ \vec x \sim \normal(\vec 0, \vec I)  }
    \lp[\x \in I_i\rp]
    \\&+ \Pr_{ \vec x \sim \normal(\vec 0, \vec I)  }
    \lp[\x \in \Xi\rp]+\Pr_{ \vec x \sim \normal(\vec 0, \vec I)  }\lp[\x \not\in  \cup_{i}I_i\rp]\;.
    \end{align*}
    We have shown that $ \Pr_{ \vec x \sim \normal(\vec 0, \vec I)  }
    \lp[\sign(p_i(\x))\neq y\mid \x \in I_i\rp]\leq C_{K,d}'(\opt_{I_i})^{1-C_K}+\opt+\eps$.
    Therefore we have that 
    \begin{align*}
         \sum_{i=1}^T   \Pr_{ \vec x \sim \normal(\vec 0, \vec I)  }
    \lp[\sign(p_i(\x))\neq y\mid \x \in I_i\rp]\Pr_{ \vec x \sim \normal(\vec 0, \vec I)  }
    \lp[\x \in I_i\rp]&\leq C_{K,d}'\sum_{i=1}^T (\opt_{I_i}^{1-C_K}+\opt+\eps)\Pr_{ \vec x \sim \normal(\vec 0, \vec I)  }
    \lp[\x \in I_i\rp]
    \\&\leq C_{K,d}'\opt^{1-C_K}+\eps\;,
    \end{align*}
    where the last inequality follows from the power means inequality. Furthermore, from \Cref{clm:excess-noise}, we have shown that $ \Pr_{ \vec x \sim \normal(\vec 0, \vec I)  }
    \lp[\x \in \Xi\rp]\leq \opt^{1-C_K}$. Finally, we have shown that $\Pr_{ \vec x \sim \normal(\vec 0, \vec I)  }\lp[\x \not\in  \cup_{i}I_i\rp]=O(\opt)$. Therefore,
    \[
    \Pr_{ (\vec x,y) \sim D  }
    \lp[ h(\x)\neq y\rp]\leq C_{K,d}'(\opt)^{1-C_K}+\eps\;.
    \]
This completes the proof of \Cref{thm:final-thm}.
 \bibliographystyle{alpha}
\bibliography{mydb}
\appendix

\end{document}